\documentclass[twocolumn,traditabstract]{aa} 
\usepackage{graphicx}
\usepackage{txfonts}
\usepackage{natbib}

\newcommand{\Checkmark}{Y}

\newcommand{\kms}{\hbox{km s$^{-1}$}}

\begin{document}

\title{Disks and Outflows in CO Rovibrational
Emission\\
 from Embedded, Low-Mass
Young Stellar Objects
\thanks{This work is based on observations collected at the European Southern 
Observatory Very Large Telescope under program ID 179.C-0151.}}

\author{Gregory J. Herczeg\inst{1}, Joanna M. Brown\inst{1},
Ewine F. van Dishoeck\inst{1,2}, \& Klaus M. Pontoppidan\inst{3}}

\institute{$^1$Max-Planck-Institut f\"ur extraterrestriche Physik, Postfach 1312, 85741 Garching, Germany; gregoryh@mpe.mpg.de;\\
$^2$Sterrewacht Leiden, Leiden University, P.O. Box 9513, 2300 RA Leiden, The Netherlands;\\
$^3$Space Telescope Science Institute, 3700 San Martin Drive, Baltimore, MD 21218, USA}

 \date{Received December 1, 2010; accepted April 2011}
 \authorrunning{Herczeg et al.}
  \titlerunning{CO Emission from Embedded Objects}

\abstract{Young circumstellar disks that are still embedded in dense molecular envelopes may differ from their older counterparts, but are historically difficult to study because 
emission from a disk can be confused with envelope or outflow emission.  CO fundamental emission is a potentially powerful probe of the disk/wind structure within a few AU of young protostars.
In this paper, we present high spectral ($R=90,000$) and spatial ($\sim 0\farcs3$) resolution {\it VLT}/CRIRES M-band spectra of 18 low-mass young stellar objects (YSOs) with dense envelopes in nearby star-froming regions to explore the utility of CO fundamental ($\Delta v=1$) 4.6  $\mu$m emission as a probe of very young disks.   CO fundamental emission is detected from 14 of the YSOs in our sample.  
The emission line profiles show a range of strengths and shapes, but can generally be classified into a broad, warm component and a narrow, cool component.
The broad CO emission is detected more frequently from YSOs with bolometric luminosities of $<15$ $L_{\odot}$ than those with $>15$  $L_\odot$.  The broad emission shares many of the same properties as CO fundamental emission seen from more mature disks around classical T Tauri stars (CTTSs) and is similarly attributed to the warm ($\sim 1000$ K) inner AU of the disk.   CO emission from the inner disk is not detected from most YSOs with a high bolometric luminosity.
Instead, the CO emission from those objects is produced in cooler ($\sim 320$ K), narrow lines in $^{12}$CO and in rarer isotopologues.  From some objects, the narrow lines are blueshifted by up to $\sim 10$ \kms, indicating a slow wind or outflow origin.  For other sources the lines are located at the systemic velocity of the star and likely arise in the disk.
 For a few YSOs, spatially-extended CO and H$_2$ S(9) emission is detected up to $\sim 2^{\prime\prime}$ from the central source and is attributed to interactions between the wind and surrounding molecular material.  Warm CO absorption is detected in the wind of six objects with velocities up to 100 \kms, often in discrete velocity components.  That the wind is partially molecular where it is launched favors ejection in a disk wind rather than a coronal or chromospheric wind.}



\maketitle

%
%

\section{INTRODUCTION}
Circumstellar disks play a central role in the growth of young stars and in the formation of planetary systems \citep[e.g.][]{Shu94,Lissauer1993}.  Disks channel material onto the star and provide an environment for grains to grow into planets.   While the star+disk system is still enshrouded in the envelope, the star builds up most of its mass, likely accreted from the disk.  Meanwhile, the disk mass can be replenished by the surrounding envelope.  Eventually the envelope and later the disk disappear, leaving only the star, any planets or planetesimals that may have formed, and remnant debris.  

During the Stage 0 phase of young stellar object (YSO) evolution, the dense envelope surrounds the star and any disk that may be present.  Most of the luminosity from the central protostar is reprocessed by cold dust and escapes as far-IR emission \citep[e.g.][]{Andre93,Robitaille06,Crapsi08}.  At the Stage 1 phase of YSO evolution, the envelope and disk masses are similar.  Any disk likely forms during the Stage 0 phase.  \citet{Evans09} measured population fractions and converted them to timescales to estimate that the Stage 0 stage typically lasts 0.1-0.16 Myr and that Stage I stage typically lasts $\sim 0.5$ Myr.  Fast disk formation also has theoretical support in hydrodynamic models of collapsing cold cores \citep{Bate2010,Machida2010}.

When the protostar is still embedded in an optically-thick envelope, the disk is difficult to study because emission from the disk, if present, can be confused with emission from the envelope.  Observations with high spatial resolution can break this degeneracy by spatially discriminating between the location of compact disk emission and extended envelope emission \citep[e.g.][]{Pad99,Eis05,Beckford2008,Tobin2010}.  
For example, sub-mm interferometry has been used to separate a compact emission source, as expected for a disk, and an extended emission source, as expected for an envelope \citep[e.g.][]{Keene90,Wilner96,Looney00,Brown2000,Jorg05,Jorg07,Lom08,Jorg09}.  A wide range of disk-to-envelope mass ratios have been measured for low-mass young stellar objects, as would be expected for a sample that includes systems at a range of evolutionary phases \citep{Lom08,Jorg09}.
  In most cases the disk+envelope mass is much smaller than the stellar mass, indicating that the main accretion phase has ended.  Disk masses are similar for objects in Stage 0 and Stage I stages \citep{Jorg09}, which suggests that once the disk grows to a certain size, the mass that the disk loses to the star is replaced by accreting mass from the envelope.
The presence of disks around Stage I objects is supported by observations of Keplerian emission profiles in cold rotational lines of HCO$^+$ \citep[e.g.][]{Hog00,Saito01,Brinch07,Lom08}.

These (sub)-mm observations trace molecular emission from the disk at large radii, but little is known about the inner AU of young disks that are still embedded in envelopes.  If the region is similar to the inner disks around classical T Tauri
stars (CTTSs), the chemical and temperature structure can be probed by
the same line diagnostics used to study CTTS inner disks \citep[see reviews by][]{Naj07,Dull07}.
Unfortunately, many of these techniques are difficult to apply to embedded objects because emission from the disk, envelope, outflow, and cloud can be confused and because embedded objects are highly reddened.  For example, in {\it Spitzer} IRS spectral surveys of young stars, emission in mid-IR gas lines is detected more frequently and is more luminous from embedded objects than from classical T Tauri stars \citep{Lah07,Flac09,Lah10}.  However, follow-up of potential disk tracers, including [O I], [Ne II] and H$_2$ emission at high spectral resolution \citep{Har95,Her06,vanBoek09,Pascucci09}, at high spatial resolution \citep{Wal03,Saucedo2003,Beck08}, or in large samples \citep{Guedel2010}, indicate that even for CTTSs, outflow emission often dominates over any disk emission.  Emission from 
disks and outflows can be even more confused for Stage 1 objects because they typically drive more powerful outflows than CTTSs.  Indeed, emission in H$_2$ 1-0 S(1) and 2-1 S(1) lines from embedded objects shows significant spatial extent but with small central velocities, indicating that the gas is kinematically quiescent \citep{Greene2010}.   H$_2$O emission in the near-IR and sub-mm has also been attributed to disks around embedded objects \citep{Car04,Jorg10}.

Fundamental-band ($\Delta v=1$) CO emission in the M-band has been demonstrated to be a powerful probe of warm ($\sim$ 500-1000 K) gas in terrestrial planet forming regions of disks around CTTSs \citep{Naj03,Rettig2004,Goto06,Sal07,Salyk09}.  For a Keplerian disk, the spectral width of the emission, when combined with a known disk inclination, constrains the location of the inner radius where the CO gas is to $\sim 0.05-0.1$ AU \citep{Naj03}, near where CTTS disks are expected to be truncated by the stellar magnetosphere \citep{Kon91,Shu94,Joh07,Gregory08}.  Spectro-astrometry of CO fundamental emission from three CTTS disks with inner dust holes confirms that the CO emission is produced in a disk and demonstrates that molecular gas is located within the dust hole (Pontoppidan et al.~2008; see also Pontoppidan et al.~2011).    CO fundamental emission is also commonly detected in disks around Herbig AeBe stars \citep{Bla04,Bri07}.  The Herbig AeBe disks with inner dust holes also lack warm CO emission within those dust holes, in contrast to CTTSs \citep{Goto06,vanderPlas2009,Bri09}.  Several Herbig AeBe stars also show bright emission from highly-excited vibrational levels, up to $v=7$, likely indicative of strong UV emission from the stellar photosphere \citep{Bri09,vanderPlas2009}

Several studies have shown that warm CO is also present around embedded objects.  In one case, GSS 30, CO fundamental emission is seen up to $2^{\prime\prime}$ (240 AU projected distance) from the central source.    In a larger sample, \citet{Pon03} detected bright CO emission from 18 of 44 objects observed with {\it VLT}/ISAAC at moderate resolution ($R=\lambda/\delta\lambda=10^4$), but did not analyze this emission because, when spectrally unresolved, multiple absorption and emission components can introduce spurious velocity shifts and cause misleading non-detections \citep[see also][]{Pon05}.  \citet{Brittain05} attributed CO emission from the embedded object HL Tau to the disk, based on similarity of the emission profile seen at high resolution to the profiles from CTTSs that lack envelopes.
Early studies by  \citet{Chandler1993} and \citet{Naj96} found vibrationally-excited CO overtone ($v=2-0$) emission from two embedded objects, WL 16 and L1551 IRS 5.

Now that CO $v=1-0$ emission has been used to study large samples of disks around CTTSs and Herbig AeBe stars, applying the same technique to younger YSOs could reveal the presence and structure of young disks that are difficult to detect with other gas tracers.  Embedded objects should have larger accretion rates than classical T Tauri stars (CTTSs), which may introduce differences between young disks around embedded objects and more mature disks around CTTSs. 
Any warm ($500-2000$ K) molecular gas in the inner disks of Stage I YSOs should produce bright and broad CO emission lines.
CO fundamental emission from warm gas has the potential to be a clean diagnostic of disks during the embedded phase of YSO evolution and is free from contamination by colder emission expected from both the outer disk and envelope.  Moreover, even if a compact, disk-like structure is present, as inferred from sub-mm interferometry,  resolved line profiles could indicate whether the gas has had time to settle into Keplerian rotation \citep[e.g.][]{Brinch09}.

In this paper, we explore the utility of CO fundamental emission as a possible probe of these young disks by analyzing high spectral resolution ($R=\lambda/\Delta \lambda\sim 90,000$) M-band observations of 18 YSOs observed with CRIRES on the {\it Very Large Telescope}.  CRIRES is uniquely capable of high-resolution M-band spectroscopy with relatively broad spectral coverage at high spatial resolution.  The data presented here were obtained as part of a large program to survey CO emission from YSOs at different stages of pre-main sequence evolution \citep{Pontoppidan2011b}.  Most observations in this paper were obtained under exceptional weather conditions.  In \S 2, we describe the observations and the selected sample.  In \S 3, we describe and analyze the observed line profiles, including separating the emission into broad and narrow components, analyzing spatially-extended CO and H$_2$ emission, and finding CO absorption in winds of six objects.  In \S 4, we attribute the components to different regions in the YSO and discuss the implications of our results on disk structure and wind launch regions.

\section{OBSERVATIONS}

\subsection{Target Selection}
The selected targets are Stage I$^1$ YSOs
 that are still embedded in circumstellar envelopes and that are members of nearby star-forming regions.  Most of the targets in our sample were chosen based on fundamental CO emission that was previously detected in $R\sim10,000$ ISAAC spectra \citep{Pon03,Pon05}.  Most targets are low-mass YSOs based on their luminosity.  One object, SVS 20 S, may have a luminosity consistent with intermediate-mass YSOs.  A few other objects may evolve into Herbig AeBe stars.
\footnotetext[1]{For simplicity, we use the physical terminology ``Stage'' rather than the observational terminology ``Class'' throughout the paper, although in some contexts ``Class'' would be the more appropriate term.}

\begin{table}
\caption{Evidence for envelopes within our sample}
\label{tab:embedded}
\begin{tabular}{lccccc}
\hline
Source            & mm-int.$^a$ & Compact$^b$  & Outflow$^c$  & Ref.$^d$\\
\hline
TMC 1A           &   \Checkmark   & \Checkmark & \Checkmark & 1,2,3,7 \\
GSS 30    &  \Checkmark        & \Checkmark  & n &  1,2,4\\
WL 12             &  \Checkmark       & \Checkmark  &  s & 1,2,4\\
Elias 29           & \Checkmark & \Checkmark    &  \Checkmark   & 1,2,4,8\\
IRS 43 S+N          & \Checkmark    & \Checkmark   &  s & 1,2,4\\
IRS 44 E+W         &     --        & \Checkmark  & \Checkmark   & 1,4 \\
IRS 63              & \Checkmark & \Checkmark    &   \Checkmark &  1,2,4\\
L1551 IRS 5 AB$^e$&  \Checkmark    & \Checkmark  &  \Checkmark   &  3,7,13\\
HH 100 IRS      &      --       & \Checkmark  & \Checkmark    & 5,9\\
WL 6                &     --        & \Checkmark      &  n & 1,4 \\
Elias 32           &      --       & \Checkmark  &   -- &  4\\
CrA IRS 2         &      --       &     --        &   --     &  -- \\
HL Tau            &   \Checkmark  & \Checkmark  & \Checkmark  & 3,10,12 \\
SVS 20 S+N            &    --        &  \Checkmark &  --   & 7\\
RNO 91           &      --      & \Checkmark & \Checkmark &  6,11\\
\hline
\multicolumn{5}{l}{$^a$Envelope detected in mm-interferometry.}\\
\multicolumn{5}{l}{$^b$Presence of cold, compact gas centered at object position.}\\
\multicolumn{5}{l}{$^c$Refers to cold molecular outflows detected in the (sub)-mm.}\\
\multicolumn{5}{l}{(s):  Molecular outflow is detected in only a single lobe.}\\
\multicolumn{5}{l}{(n):  No molecular outflow detected.}\\
\multicolumn{5}{l}{$^d$The references are examples of each phenonemon and}\\
\multicolumn{5}{l}{~~~~~ are not intended to be complete.}\\
\multicolumn{5}{l}{$^e$Unresolved and hereafter referred to as L1551 IRS 5}\\
\multicolumn{5}{l}{1:  \citet{Bontemps1996}, 2: \citet{Jorg09}}\\
\multicolumn{5}{l}{3:  \citet{Saito01}, 4:  \citet{TvK09}}\\
\multicolumn{5}{l}{5: \citet{TvK09b}, 6:  \citet{Arce06}}\\
\multicolumn{5}{l}{7:  \citet{Gregersen00}, 8: \citet{Lom08}}\\
\multicolumn{5}{l}{9: \citet{Groppi2007}, 10:  \citet{Wilner96}}\\
\multicolumn{5}{l}{11:  \citet{Chen2009}, 12:  \citet{Cabrit1996}}\\
\multicolumn{5}{l}{13:  \citet{Wu2009}}\\
\end{tabular}
\end{table}

Because SEDs of heavily-extincted stars can be similar to SEDs of stars embedded in circumstellar envelopes, the presence of an envelope for objects in our sample is typically confirmed by the presence of at least one of the following three criteria:  (1) a clearly extended component in the visibility curves of (sub)-mm emission \citep{Lom08,Jorg09}, (2) a compact structure of cold dense gas, such as HCO$^+$, at the same position as the target \citep{Saito01,Groppi2007,Lom08,TvK09}, or (3) a cold molecular outflow seen close to the YSO in both a red- and blue-shifted lobe \citep[e.g.][]{Bontemps1996,Arce06}.  Table~\ref{tab:embedded} summarizes the evidence for our classification.  One object, CrA IRS 2, is less-well studied and not confirmed as a Stage 1 object by these criteria, but is likely embedded in an envelope \citep{Nutter2005}.
For the purposes of this paper, both objects in spatially-resolved multiple systems with envelopes are assumed to be embedded (see Appendix A for more details).

The classification of these targets as embedded is consistent with most previous classifications, but can differ from classifications that are based on SEDs alone.  For example, \citet{McClure2010} used 12-5  $\mu$m colors to classify WL 12, IRS 43, and IRS 44 as YSOs with envelopes and WL 6, Elias 29, and IRS 63 as YSOs with disks but no significant envelope.  When classifying the latter three objects as disks, \citet{McClure2010} explain the discrepancy with previous work by suggesting that the envelopes are tenuous and close to disappearing.  This interpretion may apply to  WL 6 and IRS 63 \citep{TvK09}.  However, the sub-mm visibility of Elias 29 shows no indication of any compact structure, which means that the envelope mass must be much larger than the mass of the undetected disk \citep{Lom08}.  The disks for these three sources may dominate their near-IR spectra, but the sub-mm interferometry requires the presence of an envelope.

\begin{table*}
\caption{Sample Properties}
\label{tab:sample.tab}
\begin{tabular}{|lcccccccccc|}
\hline
Target         &  d  &  $PA_{out}^a$ & $A_V$ &   $T_{phot}$ & $L_{bol}$ & $v_{lsr}$   & $v_{abs}^b$ &\multicolumn{2}{c}{Flux$^c$}  &  \\
                   & pc &  $^\circ$ & mag     &     K             & $L_\odot$  & \kms\  & \kms\  &4.8 $\mu$m   & Method  & Ref \\
\hline
HH 100 IRS  & 130 &-- & 30    & 4060 &  15  & 5.9  &  5.8 & 11.3 & ISO & 2,8\\ 
CrA IRS 2     & 130 &  -- & 20    & 4900 & 12  &  (5.9) & 6.1 & 5.40 & ISO & 8,15 \\
WL 6            & 120  & $\sim 40$ & 9.8 &  --     & 2.4 &  3.5  & 4.5 & 1.4 & IRAC& 1,23\\
WL 12          & 120  & -- & 9.8 &  4000 & 2.4 &  3.5   &  7.6 & 1.1 & IRAC&1, 10\\
IRS 63         & 120  & $\sim 240$ & 9.8 &   --     & 3.0 &  3.5  &  2.0 & 1.3 & IRAC&1,24\\
IRS 43 S+N     & 120  &  25 & 9.8 &  4400  & 5.5   &  3.5 & 4.1 & 1.6 & IRAC &1,10,17\\
IRS 44 E+W     &120   &  $\sim0$ & 9.8 &  4400  & 25 &  3.5   &  5.8 & 2.8 & IRAC & 1,10,20\\
Elias 32        & 120  & -- & 9.8 & 3400    & 1.0 &  3.5  & 5.7 & 0.39 & IRAC &1,10\\
Elias 29        & 120  & $\sim 135$ & 9.8 &      --   & 38  &  3.5  & 5.5 &  24.3 & ISO & 1,21\\
GSS 30          & 120  & $\sim 45$ & 9.8 &     --     & 13  &  3.5  & 7.5 & 23 & IRAC &1,20\\
HL Tau          & 160  &  $\sim 45$ & 7.4 &  4350 & 6.6 & 7  &  8.2 & 5.49 & ISO &3,4,9,22\\
RNO 91         & 120   & 155 &-- & -- & 2.3 & 0.8   & 0.5 & 1.8 &   16 & 5,12,18\\
SVS20 S    &  415  & --  & 30   &5900   &142  & 1.9  & 8.3 & 3.5  & IRAC  & 10,14,7 \\
SVS20 N     &  415  & -- & 4  & 3270 & 0.27 &  1.9 &  (8.3) & 1.1  & IRAC &13,7 \\
TMC 1A          & 160 & $-10$   & --     &  --     & 2.8 & 5.6  &  5.3 & 1.0 & IRAC &6, 3,7, 25\\
L1551 IRS 5 AB & 160  &  256 & 28  &  3300 & 23 & 6.9  &  4.7 & 3.94 & ISO &6, 3,7,10,11,19\\
\hline
\multicolumn{4}{l}{$^a$Approximate position angle of outflow} & 
\multicolumn{7}{l}{$^b$$v_{lsr}$ of $^{13}$CO and C$^{18}$O absorption in our CRIRES spectra}\\
\multicolumn{3}{l}{$^c10^{-14}$ erg cm$^{-2}$ s$^{-1}$ \AA$^{-1}$} & \multicolumn{6}{l}{() indicates assumed value from nearby source.}\\
\multicolumn{11}{l}{Distances:  130 pc to CrA \citep{deZeeuw1999}, 120 pc to Oph \citep{Loinard2008}, 415 pc to Serpens}\\
\multicolumn{1}{l}{ } & \multicolumn{10}{l}{\citep{Dzib2010}, and 160 pc to Taurus \citep{Torres2009}}\\
\multicolumn{11}{l}{1: \citet{Evans09}, 2:  \citet{TvK09b} 3:\citet{Fur08} 4: \citet{Hayashi1993}}\\
\multicolumn{11}{l}{5:  \citet{Lohr2007};  6:  \citet{Yang2002} 7: \citet{Gregersen00} 8: \citep{Nis05}}\\
\multicolumn{11}{l}{9: \citet{Whi04};  10: \citet{Dopp05}; 11: \citet{Prato2009}; 12: \citet{Chen2009}}\\
\multicolumn{11}{l}{13: \citet{Oliveira09}; 14: \citet{Ciardi05};  15: \citet{Meyer2009}; 16:  \citet{Boogert2008}}\\
\multicolumn{11}{l}{17: \citet{Grosso2001}; 18: \citet{Arce06};  19: \citet{Pyo2009}; 20: \citet{Allen2002}; 21: \citet{Ybarra2006}}\\
\multicolumn{11}{l}{22: \citet{Beckwith89}; 23: \citet{Gomez2003}; 24: \citet{Zhang2009}; 25: \citet{Chandler1996}}\\
\multicolumn{11}{l}{}\\
\multicolumn{11}{l}{}\\
\multicolumn{11}{l}{}\\
\end{tabular}
\end{table*}

\begin{table*}
\caption{Observation Log}
\label{tab:obslog.tab}
{\scriptsize
\begin{tabular}{lcccccccccc}
\hline
Target  & Alt. Names & RA$^a$ & DEC$^a$ & UT Date  & $t_{exp}$ (s) & airmass & FWHM$^b$ &  $v_{bary}$ (\kms) & $\lambda$ settings &S/N$^d$\\ 
\hline
\hline
IRS 43 & YLW 15  & 16:27:27 & -24:40:51 & 2008-08-06  &960$^{l}$&  1.06  & 0.32 & -17   & 4716, 4730, 4868 & 40\\
IRS 43  & YLW 15  & 16:27:27 & -24:40:51 & 2008-08-07  & 1200 &  1.02  & 0.25 & -17   & 4831  & 40\\
IRS 44$^d$  & YLW 16a&  16:27:28 & -24:39:34 & 2008-04-27  & 720& 1.41  &  0.56   &  27   & 4716  & 15\\
IRS 44  & YLW 16a &  16:27:28  & -24:39:34 & 2008-04-30  & 720$^m$& 1.26  &  0.54  &  25   & 4730, 4833 & 10\\
IRS 44$^e$  & YLW 16a &  16:27:28 & -24:39:34 & 2008-05-01&  1200 & 1.61  &  0.86   &  25   & 4946 & 9 \\
IRS 44$^e$  & YLW 16a &  16:27:28 & -24:39:34 & 2008-08-06 &  1200 & 1.31  & 0.33   & -17   & 4868 & 20\\
IRS 44$^e$  & YLW 16a &  16:27:28 & -24:39:34 & 2008-08-07 & 960  & 1.38  & 0.32 & -17   & 4716, 4730 & 30\\
IRS 63  & GWAYL 4          &  16:31:36 & -24:01:29 & 2007-04-26  & 600 & 1.02  &  0.37 &  28   & 4716, 4730, 4833, 4929 & 32\\
WL 12   & YLW 2   & 16:26:44 & -24:34:48 & 2007-09-02  & 960 &  1.22  & 0.31 & -20   & 4716, 4730 & 48\\
WL 12   & YLW 2   & 16:26:44 & -24:34:48 & 2007-09-04  & 960 &  1.25  & 0.31 & -20   & 4833 & 45\\
WL 12   &YLW 2   & 16:26:44 & -24:34:48 & 2010-03-03  & 2280 &  1.05 & 0.31 &40  & 4946 &60 \\ 
WL 6    & YLW 14           & 16:27:22 & -24:29:53 & 2007-04-27  & 480 & 1.20  & 0.29 &  27   & 4716, 4730, 4833 & 30\\
HH 100 IRS & V710 CrA   & 19:01:51 & -36:58:10& 2007-09-03  & 600 & 1.27  & 0.43 & -17   & 4716, 4730 & 170\\
HH 100 IRS & V710 CrA      & 19:01:51 & -36:58:10 & 2007-09-06  & 480 & 1.43  & 0.26 & -18   & 4833 & 140 \\
HH 100 IRS & V710 CrA   & 19:01:51 & -36:58:10 & 2008-08-03  & 480 &  1.11  & 0.44 &  -6   & 4710, 4730, 4868,4946 & 100\\
Elias 32  & YLW 17A & 16:27:28 & -24:27:20 & 2008-05-03  & 1200 & 1.19  & 0.32 &  24   & 4716, 4730, 4833 & 10\\
Elias 32$^e$& YLW 17A        & 16:27:28 & -24:27:20& 2008-08-09  & 960 & 1.07  &   0.45   &  17     & 4716 & 3\\
Elias 32$^f$& YLW 17A& 16:27:28 & -24:27:20& 2008-08-10  & 960  &   1.12      & 0.52 &    18   & 4730 & 7.5\\
Elias 29$^e$&  WL 15 & 16:27:09 & -24:37:19 & 2008-08-09  & 360 & 1.37  & 0.59 & -17   & 4716, 4730, 4868 & 130 \\
CrA IRS 2& CHLT 1  & 19:01:42 & -36:58:32 & 2007-04-26  &480&  1.02  & 0.37 &  34   & 4716,4730,4833,4929 & 110\\
RNO 91 & HBC 650             &  16:34:29 & -15:47:02 & 2007-04-24    &  600 & 1.05   & 0.28  &  30 & 4716,4730,4929,4946 & 55\\
RNO 91 & HBC 650             &  16:34:29 & -15:47:02 &  2007-04-25    &  1200 & 1.10   & 0.26   & 30 & 4770,4780 & 60\\
HL Tau$^g$  & HBC 49                & 04:31:38 & 18:13:58 & 2007-10-12  & 600$^{n}$&   1.61    &  0.24 & 10  & 4716,4730,4868 & 120\\
HL Tau$^h$ & HBC 49                & 04:31:38 & 18:13:58 & 2010-02-19   &  1080 & 1.71   & 0.22  &  42 & 4716,4730,4800,4820 & 150\\
SVS 20$^i$ & [EC92] 90 & 18:29:58 & 01:14:06 & 2007-04-24     &  1200$^{o}$ & 1.11  & 0.42 & 41 & 4716,4730,4929,4946 &60 \\ 
TMC 1A$^j$     &  IRAS 04365+2535 & 04:39:35 & 25:41:45 & 2010-02-10     & 1200$^p$  & 1.67 &   0.54   &  -38    & 4716, 4730, 4946 & 30 \\
GSS 30$^h$ & Elias 21 & 16:26:21 & -24:23:06 &  2010-03-04       & 600$^q$   & 1.15 & 0.30    & 40 & 4716, 4730,5115 & 58\\
L1551 IRS 5$^k$ & HBC 393  & 04:31:34 & 18:08:05 &  2007-10-17         &  840  & 1.50 & 0.82 & -4 & 4716,4730 & 17\\
\hline
\multicolumn{11}{l}{$^a$2MASS position (J2000) from \citep{Skr06}.}\\
\multicolumn{11}{l}{$^b$In arcsec, includes both the seeing and any real spatial extension in the continuum.}\\
\multicolumn{11}{l}{$^c$Approximate S/N per pixel in a single setting, typically measured in regions with an average telluric absorption of 10-15\%.}\\
\multicolumn{11}{l}{Position angle of 0$^\circ$ except where noted:  $^d$PA=70;  $^e$PA=100;  $^f$PA=340; $^g$PA=-150; $^h$PA=45; $^i$PA=5; $^j$PA=160; $^k$PA=75; }\\
\multicolumn{11}{l}{~~~{\bf Listed position angle is approximate for cases where the slit was aligned with the parallactic angle.}}\\
\multicolumn{11}{l}{Exposure times same for all settings except where noted:  $^l$4868:  1440s;  $^m$4833:  960s $^n$4868:  960s $^o$4868:  4929/4946:  500; $^q$4946: 1080; $^q$5115:  1080s}\\
\multicolumn{11}{l}{}\\
\multicolumn{11}{l}{}\\
\multicolumn{11}{l}{}\\
\end{tabular}}
\end{table*}

Table \ref{tab:sample.tab} lists selected properties of our targets from the literature.  The stellar properties, including the effective temperature and luminosity of the photospheric emission and the extinction,  are uncertain because of the difficulty in detecting photospheric features \citep[e.g.][]{Whi04,Dopp05} and in accurately measuring the extinction.  In some cases, the $A_V$ and $L_{bol}$ are obtained from different sources.  The bolometric luminosities that are measured from the far-IR/sub-mm SEDs are not particularly sensitive to either extinction or spectral type.  All luminosities are corrected for the most recent distance measurement to the parent cloud.  In a few cases, significant discrepancies in derived stellar parameters exist in the literature$^{2}$.  
\footnotetext[2]{For IRS 63, the photospheric properties measured by \citet{Dopp05} are not adopted here because the $v_{lsr}=-26.5$ \kms\ of photospheric absorption features is highly anomalous relative to the other Stage I objects in the $\rho$ Oph molecular cloud and is inconsistent with the velocity of CO absorption by the envelope seen in our data.    We speculate that \citet{Dopp05} measured optically-thin absorption in an outflow and attributed the absorption to the photosphere.  Alternately, the slit rotation may have induced a large velocity offset, in which case the derived effective temperature would still be accurate (T. Greene, private communication).}

The sample likely includes a range of masses for the central object, with a general trend that more luminous objects are likely also more massive.  Unfortunately, the lack of accurate stellar properties prevents a reliable determination of mass from pre-main sequence evolutionary tracks.

The systemic velocity listed in Table~\ref{tab:sample.tab} is obtained from literature measurements of (sub)-mm lines of the molecular cloud or of the envelope.   All CO and H$_2$ velocities listed in the paper are relative to the measured velocity of $^{13}$CO and C$^{18}$O absorption in our M-band spectra.

\subsection{Observational Setup and Data Reduction}
We used CRIRES \citep{Kaeufl2008} on the VLT-UT1 to obtain high-resolution echelle spectra from 4.6--4.9  $\mu$m of 15 Stage I objects, three of which are binaries.  
Table~\ref{tab:obslog.tab} lists our observation log$^3$.
\footnotetext[3]{Data are available for download at http://http://www.stsci.edu/~pontoppi}

CRIRES has four 1024 x 512 pixel InSb detectors that each covers $\sim 0.02$  $\mu$m in a single integration.    Every object in our sample was observed at multiple wavelength settings to cover spectral gaps between chips and to observe lines with a wide range of rotational levels. Table~\ref{tab:obslog.tab} includes the wavelength setting used for each object.  Online Table A.1 lists the wavelength ranges covered for each setting.  
Each pixel covers $\sim 0.2$ \AA, or $\sim 1.25$ \kms, in the dispersion direction.  The $0\farcs19$-wide slit with $0\farcs086$ pixels in the dispersion direction leads to $R\sim90,000$, as measured from telluric absorption lines.

Observations were obtained with $10^{\prime\prime}$ nods in an ABBA pattern.  Each nod consists of 60s integrations.  Random dither patterns with a width of 1$^{\prime\prime}$ were used for each nod to distribute the counts over different pixels at each integration.  Total integration times for each setting are listed in Table~\ref{tab:obslog.tab}.

\begin{figure}
\includegraphics[width=90mm]{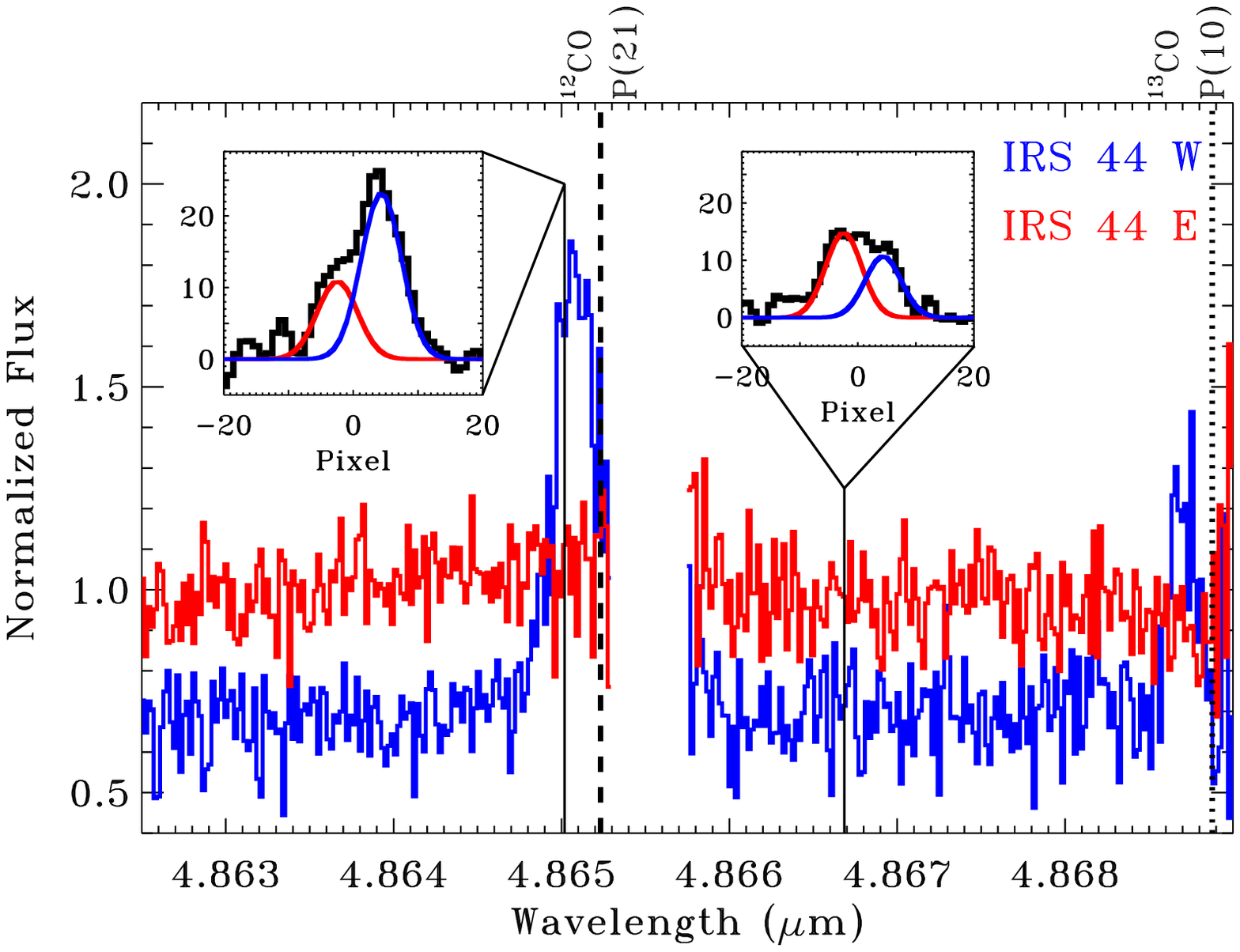}
\caption{Extraction of two components from the IRS 44 spectra.  The main plot shows the extracted counts versus wavelength for IRS 44 E (red) and IRS 44 W (blue) over a small wavelength range.  The two insets show the spatial profile of the counts in the cross-dispersion direction for a single pixel (left, within the CO P(21) line, right within the continuum).  Two Gaussian profiles were fit to the cross-dispersed emission profile and were subtracted to minimize contamination between components during the extraction.  CO emission is detected from the secondary, IRS 44 W (blue) but not from the primary, IRS 44 E (red).  The excised region is unreliable because of strong telluric CO absorption.}
\label{fig:irs44extract}
\end{figure}

Most sources in our sample were too faint for the MACAO adaptive optics system, which usually feeds CRIRES.  The AO loop was closed for HL Tau and RNO 91, though with a low Strehl ratio and poor correction.  For other objects, the M-band seeing during our observations was exceptional, typically between $0\farcs3$ and $0\farcs45$ as measured from the spatial profile of the M-band continuum emission in the cross-dispersion direction for each object.  Two objects, Elias 29 and L1551 IRS 5, have especially large seeing measurements, which indicates that either the seeing was poor or that the M-band continuum emission is spatially extended on $0\farcs5-1\farcs0$ scales.

Each object was acquired with the Ks filter.  After November 2008, the CRIRES control software implemented an automatic correction to the differential atmospheric dispersion between the K and M-bands.  Prior to November 2008, our observations were generally obtained at low airmasses or at parallactic angle to minimize the effect of atmospheric dispersion.   Because most of our observations did not use adaptive optics, the effect of atmospheric dispersion ($\sim 50$ mas at airmass of 1.7) is typically much less than the FWHM of the continuum in the cross-dispersion direction.  A real offset between the K-band and M-band continuum emission is also possible for embedded young stars.   In cases where the slit was aligned with the parallactic angle, the position angle did not change by more than $10^\circ$ in any observation.  All of our observations include strong continuum emission, and effects that may be attributed to different slit placements are likely minor.

\begin{figure}
\includegraphics[width=90mm]{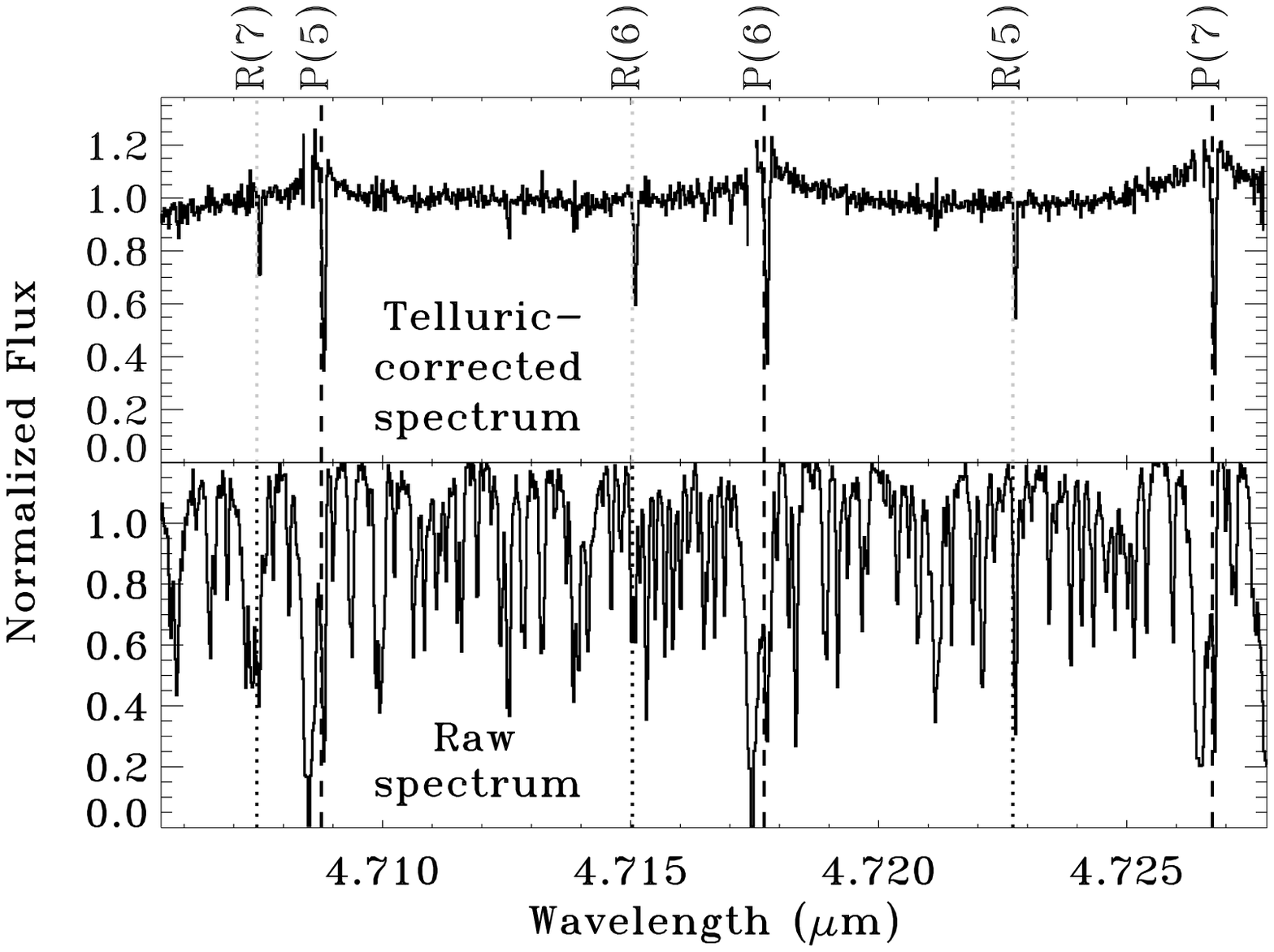}
\includegraphics[width=90mm]{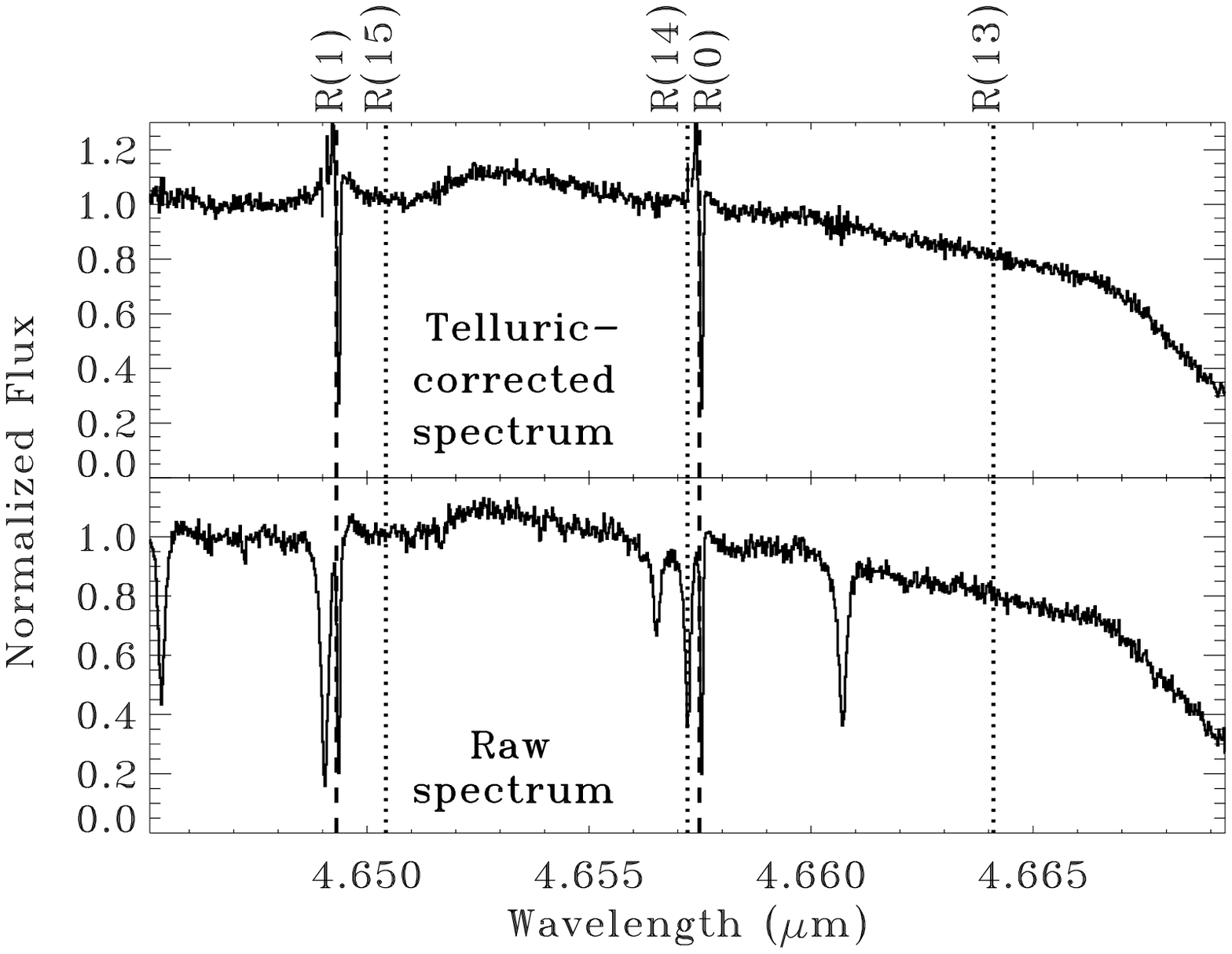}
\caption{The telluric correction for IRS 43 S+N from two chips in the $4716$ \AA\ setting.  The spectrum reaches $S/N\sim100$ per pixel in regions with few atmospheric lines (left) and $S/N\sim50$ per pixel in regions with many atmospheric lines (right).  $^{12}$CO lines are labeled in black and $^{13}$CO lines are labeled in red.}
\label{fig:telluric}
\end{figure}

A telluric standard at a similar airmass to our object spectrum was always observed in an adjacent observation and with an AO correction.  
Spectra from the object and the telluric standard were wavelength-calibrated using telluric absorption lines.  The science spectrum was then divided by the spectrum of the telluric standard.
Differences in spectral resolution between the telluric standard and the science target were always minor and, when necessary, were accounted for by convolving the spectrum of the telluric standard with a Gaussian profile to match the two resolutions.   
Figure~\ref{fig:telluric} demonstrates that high S/N ($\sim 50-100$ in the continuum) is obtained even in regions filled with telluric absorption lines.  The spectra were then shifted in velocity space to the local standard of rest.  Many of our observations were obtained on dates when the telluric absorption was shifted by large velocities from the local standard of rest to shift telluric CO absorption away from the central wavelength of CO lines for the objects.   The relative wavelength calibration is accurate to $\sim 1$ \kms.

Data were reduced following \citet{Pont08}.  The pixels on the detector are each $0\farcs086$ in the dispersion and cross-dispersion direction.  When coadding the 2D images, the light was shifted onto a common central pixel and resampled onto a finer ($0\farcs043$) scale.  Spectra were extracted from windows with a width that is twice the FWHM of the emission profile in the cross-dispersion direction.  A specialized extraction technique was applied to the $0\farcs3$ binary IRS 44 (see \S A1.1), which was marginally resolved on three nights when the seeing was exceptional ($0\farcs3$).
The spectra for the two components were extracted by fitting two Gaussian profiles to the cross-dispersed profile at each spectral bin in the 2D images (see Fig.~\ref{fig:irs44extract}).  The two Gaussian profiles both have the same FWHM.  The FWHM and positional offset between the two components are constant across the detector.  The Gaussian profile of one component was  subtracted from the data.  The counts were then extracted from a spectral window around the other component.  The extraction windows for the two components do not overlap to minimize contamination by the other component.

\begin{figure*}
\includegraphics[width=90mm]{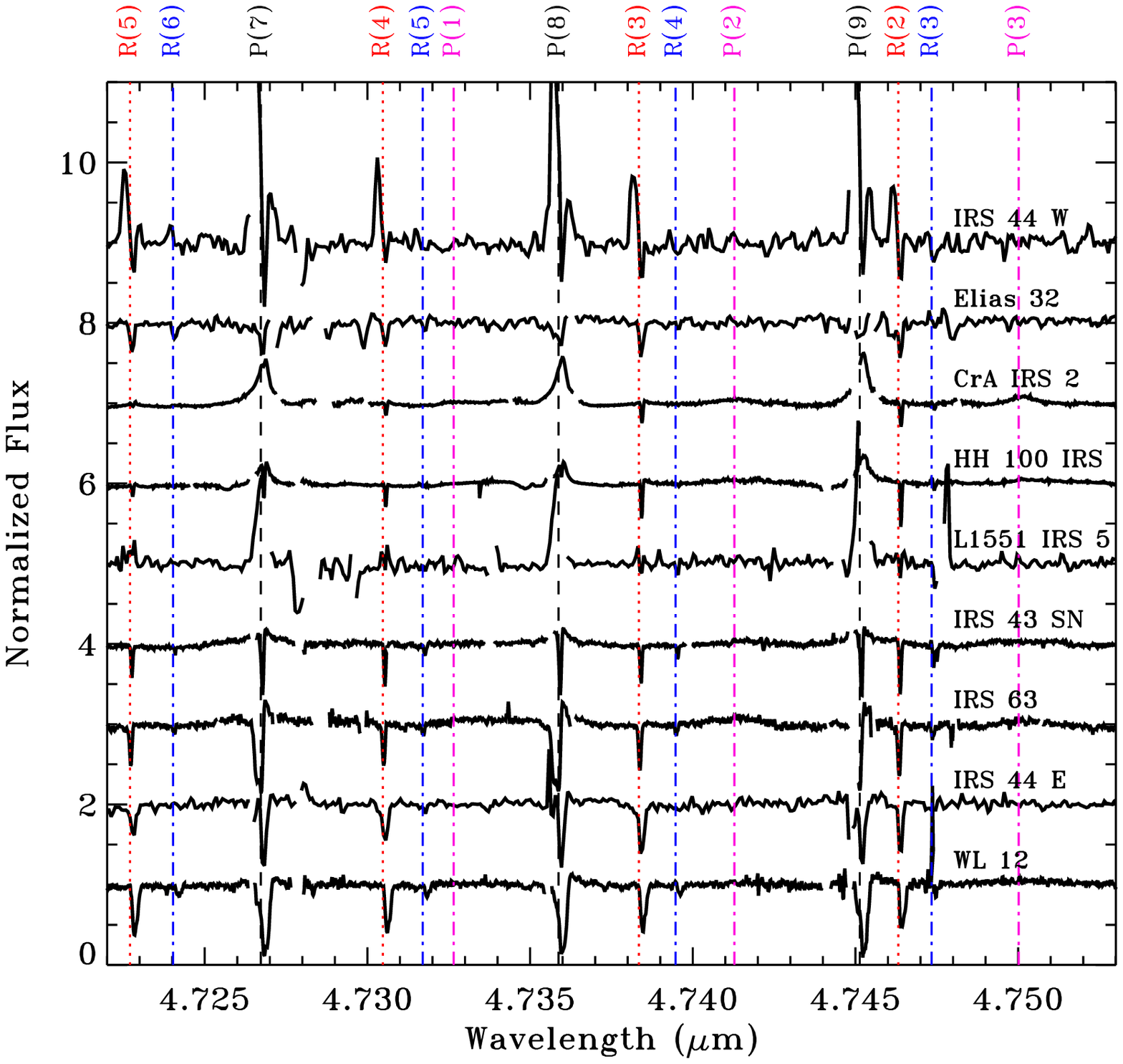}
\includegraphics[width=90mm]{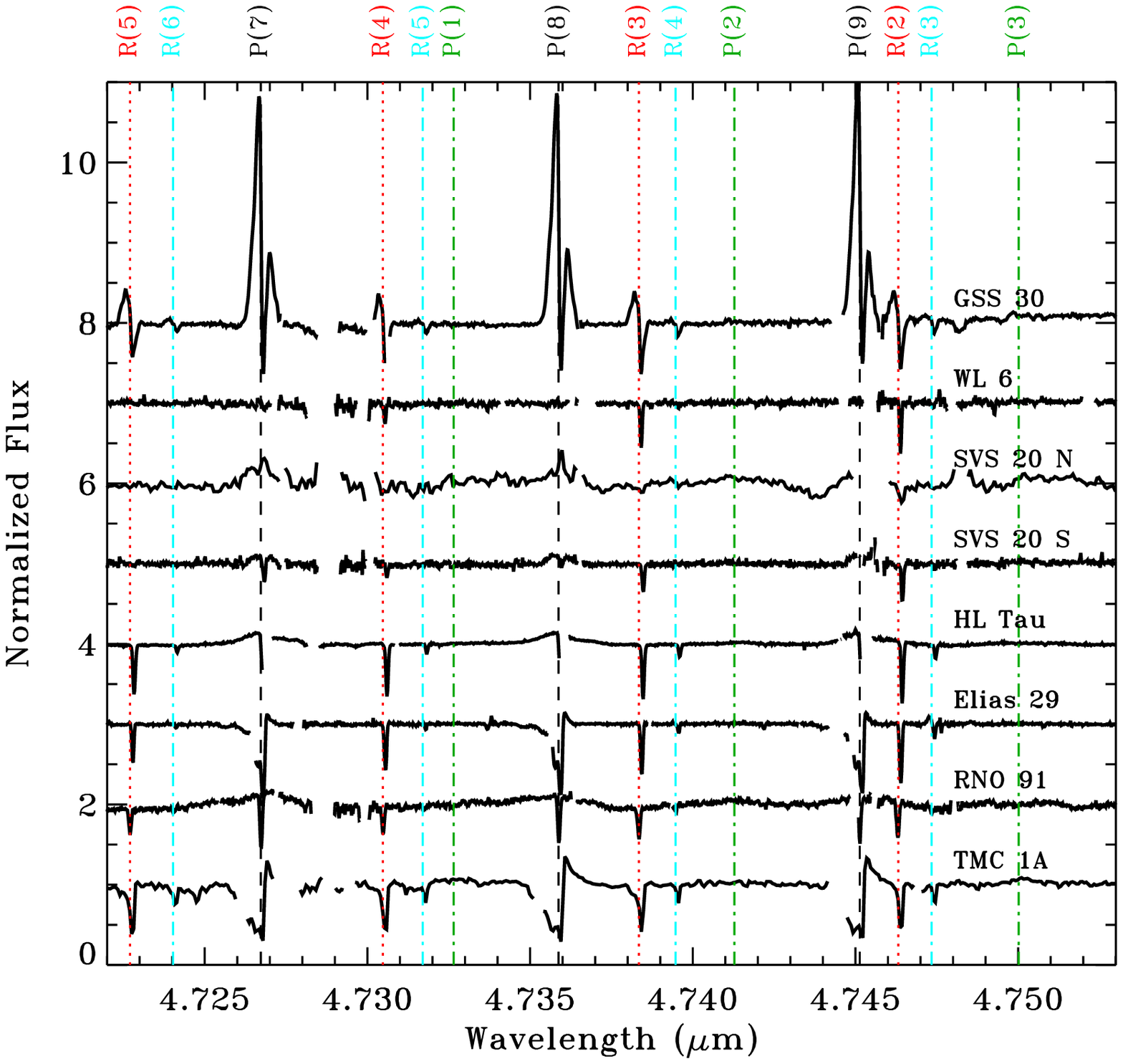}
\caption{A sample spectral region, with $^{12}$CO $v=1-0$ lines shown in black, $^{13}$CO $v=1-0$ shown in red, C$^{18}$O $v=1-0$ shown in blue, and $^{12}$CO $v=2-1$ shown in green.}
\label{fig:overview}
\end{figure*}

The M-band continuum flux is measured directly from ISO-SWS spectra for four objects \citep{Sloan2003} and from ground-based ISAAC spectra for one object \citep{Boogert2008}, and is inferred from linear fits to {\it Spitzer}/IRAC photometry for seven objects \citep{TvK09,Harvey07}.  For binaries, the continuum flux of each companion was calculated by using the M-band magnitude and the counts ratio in the continuum emission measured in our M-band observations.   Spatially-extended continuum emission for a given object would artificially increase the calculated M-band flux of the central source.  The CO line fluxes are calculated from equivalent width measurements and would be overestimated in cases where the M-band continuum is spatially extended.

Our CRIRES spectra yield a flat M-band continuum flux for every object, with a relative uncertainty of $\sim 30$\% between 4.5 and 5.0  $\mu$m.  In the four ISO spectra, the maximum relative flux difference between 4.5 and 5.0  $\mu$m is 30\%.  We also do not correct the absolute or relative CO emission line fluxes for extinction because of the wide range in calculated extinctions for any given object, the lack of methodological consistency in extinction calculations for the entire sample, and because extended CO emission may not suffer from the same extinction as the central source.  Applying extinctions that range from $A_V=10-37$ mag would increase CO line luminosities by factors of 1.2 to 3  \citep{Rieke85,Chapman09} relative to the measured luminosities reported in this paper.  The wavelength dependence of extinction in the M-band is negligible ($\sim 3$\% across the 4.6--5.0 $\mu$m spectral region for $A_V=20$ mag).

\subsection{Archival ISAAC Spectra}
In several subsections we compare our high-resolution CRIRES spectra to archival VLT/ISAAC M-band spectra of the same sources obtained in 2001--2002.  The ISAAC spectra span from 4.53--4.75  $\mu$m with $R=5000-10000$.  The spectra were obtained with a $0\farcs3$ slit width and generally with worse seeing than our CRIRES data, possibly because the embedded objects within the CRIRES program were preferentially observed on nights with good seeing.  As a consequence, the extracted ISAAC spectra usually sample larger areas on the sky.  The data were reduced and analyzed by \citet{Pon02} and \citet{Pon03}, including reporting detections and velocities for CO emission.

\subsection{Archival Near-IR Images of GSS 30}
To compare extended CO emission to continuum emission from GSS 30 (see \S 3.3), we downloaded an AO-fed K-band image from the {\it VLT} archive that was obtained with NACO on 19 September 2007 as part of {\it VLT} program 079.C-0502 (P.I. Chen).  Gaspard Duch\^{e}ne (private communication) provided us with an L-band AO image of GSS 30 obtained with {\it VLT}/NACO and published in \citet{Duc07}.

\section{DESCRIPTION OF M-BAND EMISSION FROM EMBEDDED OBJECTS}
M-band spectra cover many CO fundamental ($\Delta v=1$) transitions, H Pfund-$\beta$, H$_2$ S(9), and the 4.67 $\mu$m CO ice absorption band.   One spectral region of our CRIRES M-band spectra of embedded objects is shown in Figure~\ref{fig:overview}.  The M-band emission is dominated by a smooth continuum produced by warm dust, likely located in the disk close to the star \citep[e.g.,][]{Eis05,Eno09}.  No photospheric features are detected from any of our targets.  CO ice absorption \citep{Pon03} is detected towards most objects in our sample.  HL Tau lacks any CO ice absorption \citep[see also][]{Whittet89,Brittain05}, which suggests that much of the line-of-sight extinction occurs in a disk or flattened envelope in which little solid CO absorption is expected to be observed \citep{Pontoppidan05a}.
 Strangely, GSS 30 also shows only very weak ice absorption \citep[see also][]{Pon02}.  Absorption in gaseous $^{12}$CO and isotopologues is detected towards every object.  Narrow absorption lines are caused by the envelope and parent molecular cloud, while broad absorption arises in the wind.

\begin{figure}
\includegraphics[width=80mm]{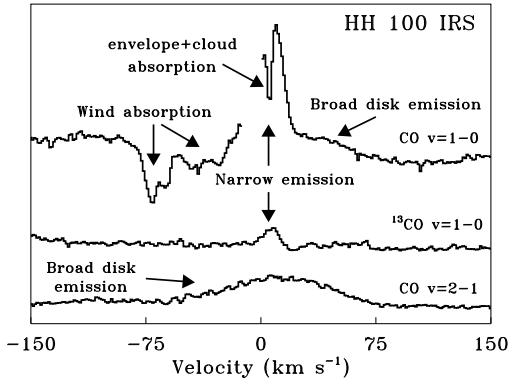}
\caption{Co-added profiles for CO $v=1-0$, $v=2-1$ and $^{13}$CO $v=1-0$ lines in the {\it VLT}/CRIRES spectrum of HH 100 IRS.  The different observed components are each identified with a physical region in the system.  For scaling, the peak of the $v=1-0$ line is at $\sim 1.38$, normalized to the adjacent continuum.}
\label{fig:hh100_components}
\end{figure}

\begin{figure*}[!htb]
\includegraphics[width=170mm]{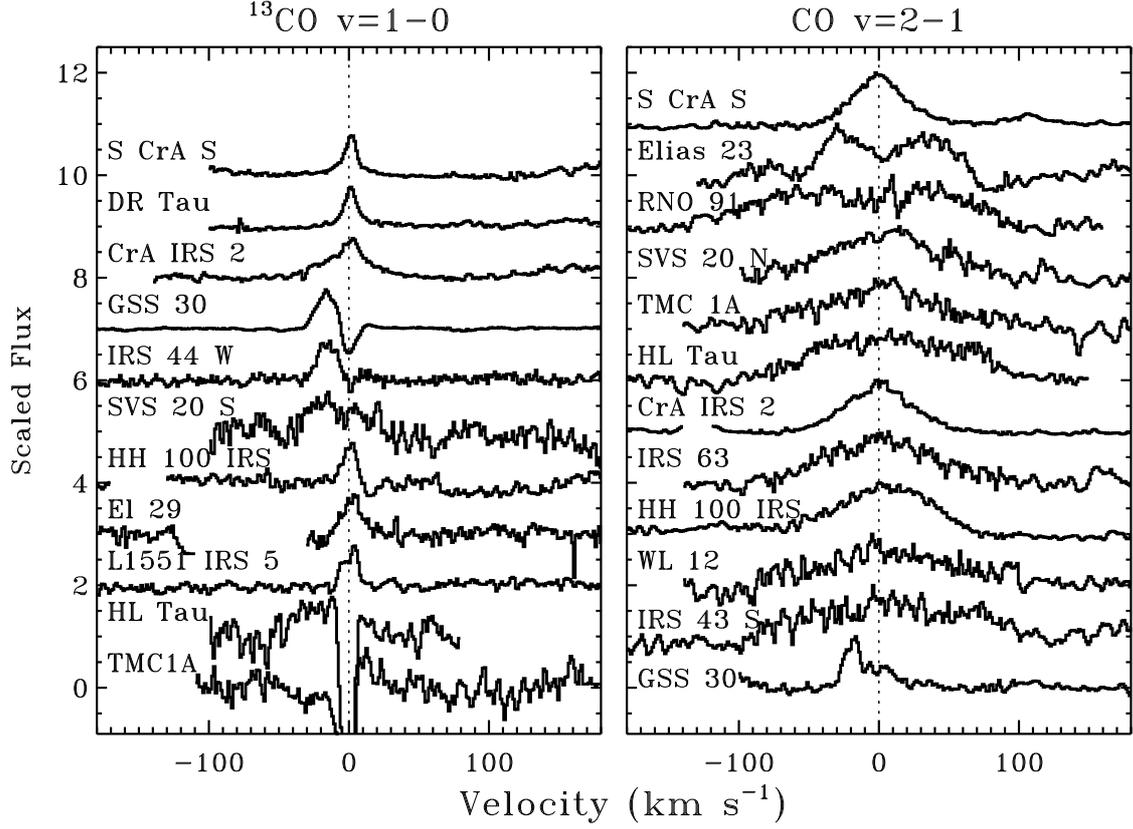}
\caption{Line profiles of $^{13}$CO (left) and CO $v=2-1$ (right) emission for detected sources in our sample.  For comparisons to CTTSs, the $v=2-1$ line profiles of Elias 23 and S CrA S and the $^{13}$CO line profiles of S CrA S and DR Tau are shown at the top.}
\label{fig:galleryx}
\end{figure*}

Emission in $^{12}$CO $v=1-0$ lines is detected from all but four embedded objects in our sample.  In general, the on-source CO emission can be described by a broad and a narrow component.  Broad CO emission, in both $v=1-0$ and in higher vibrational lines, is detected from 10 sources in our sample.  The narrow component, which has an optical depth sufficient to produce detectable emission in $^{13}$CO and C$^{18}$O lines, is detected from 9 sources.    These different components are more easily distinguished in the $v=2-1$ and isotopologue lines that are also covered in M-band spectra.
Figure~\ref{fig:hh100_components} shows how the broad component, seen in $v=2-1$ lines, and the narrow component, seen in $^{13}$CO lines, combine to create the $^{12}$CO $v=1-0$ emission line profiles for one object, HH 100 IRS.  Similarly for the other sources, the $^{13}$CO lines are typically narrower than the $v=2-1$ lines (Fig.~\ref{fig:galleryx}).
In addition to the on-source emission, very narrow CO emission is spatially-extended from two sources (GSS 30 and IRS 43, see \S 3.3).  Blueshifted CO absorption is detected from six sources (see \S 3.6).

\begin{figure*}[!htb]
\includegraphics[width=5.8cm]{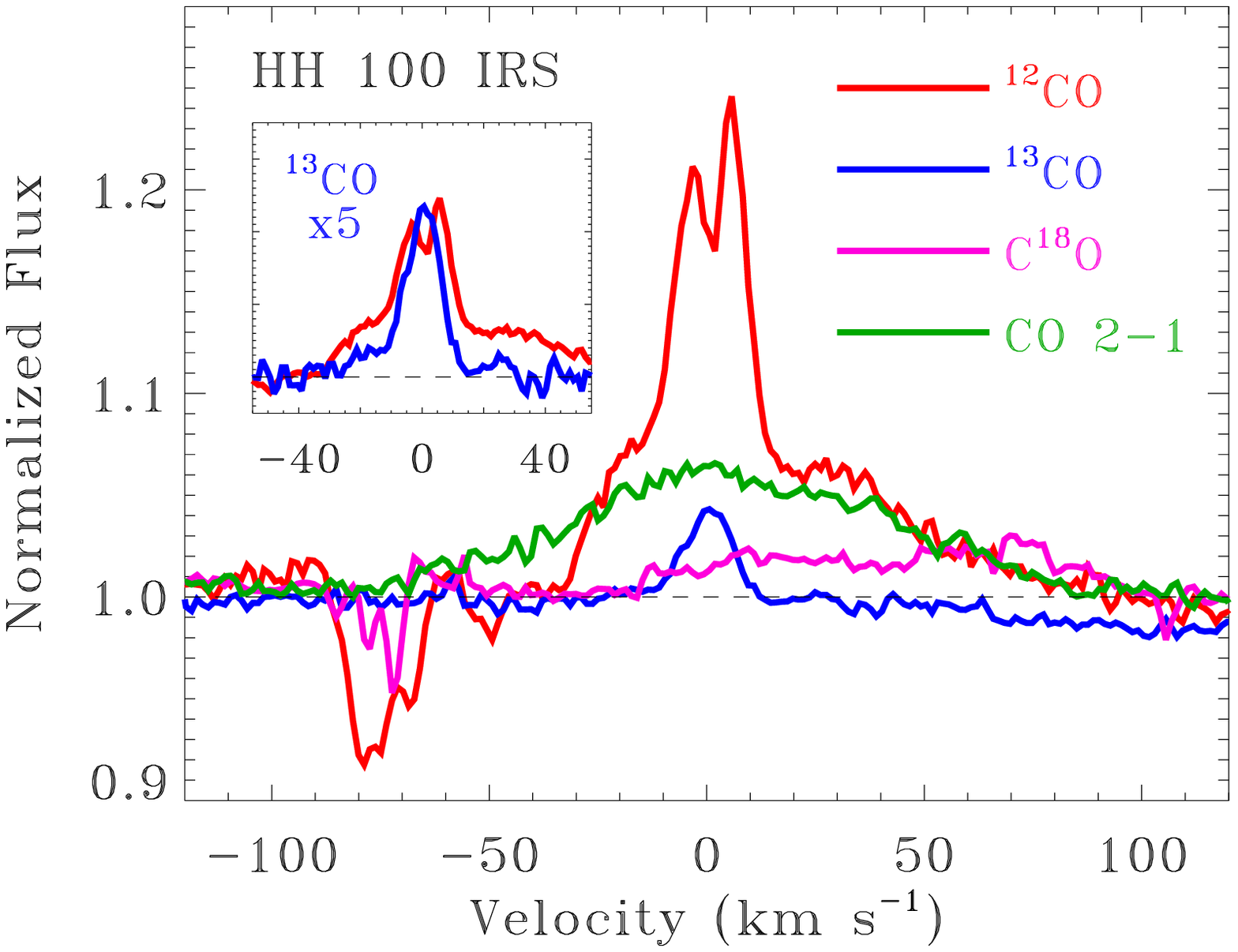}
\includegraphics[width=5.8cm]{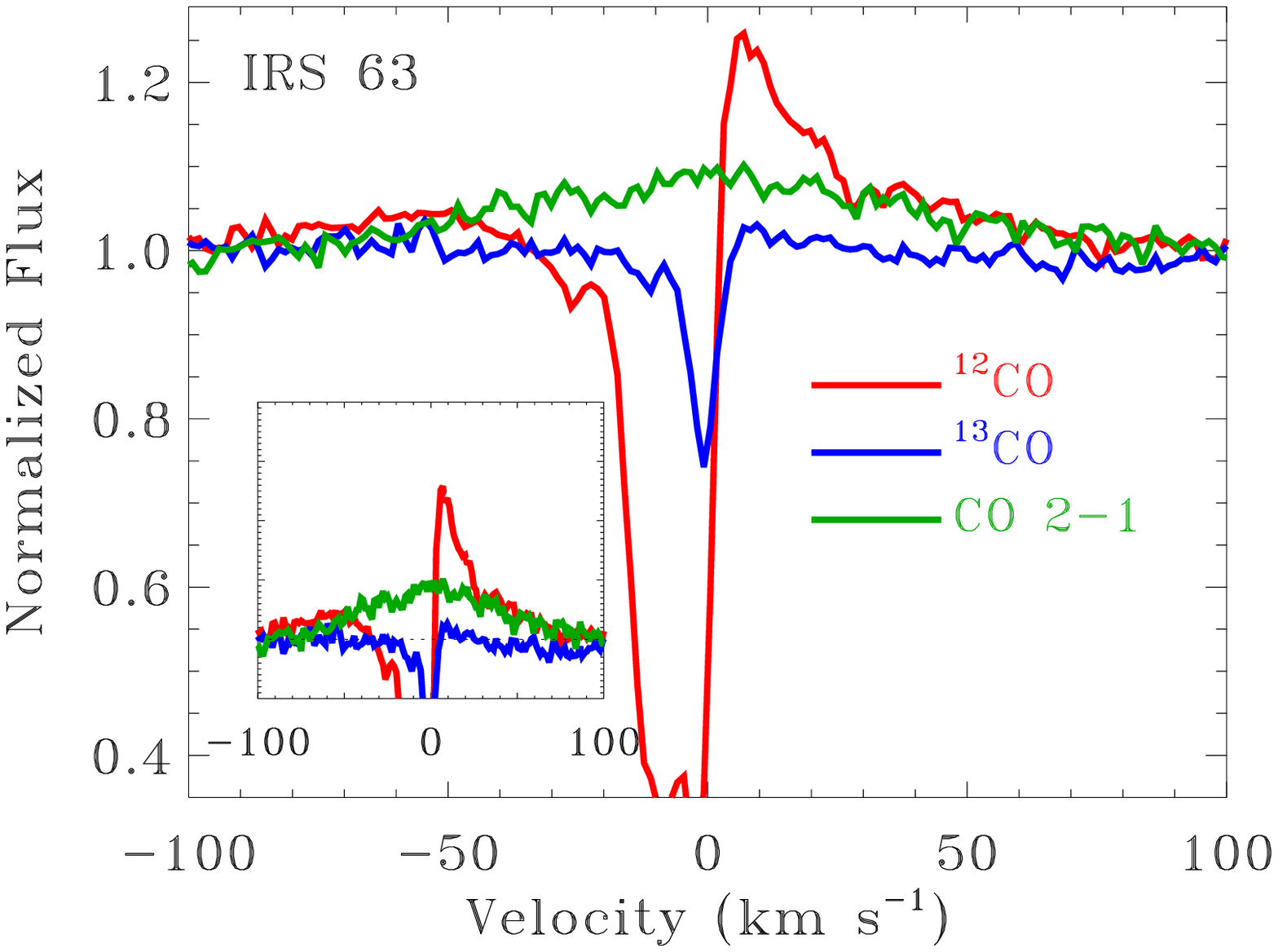}
\includegraphics[width=5.8cm]{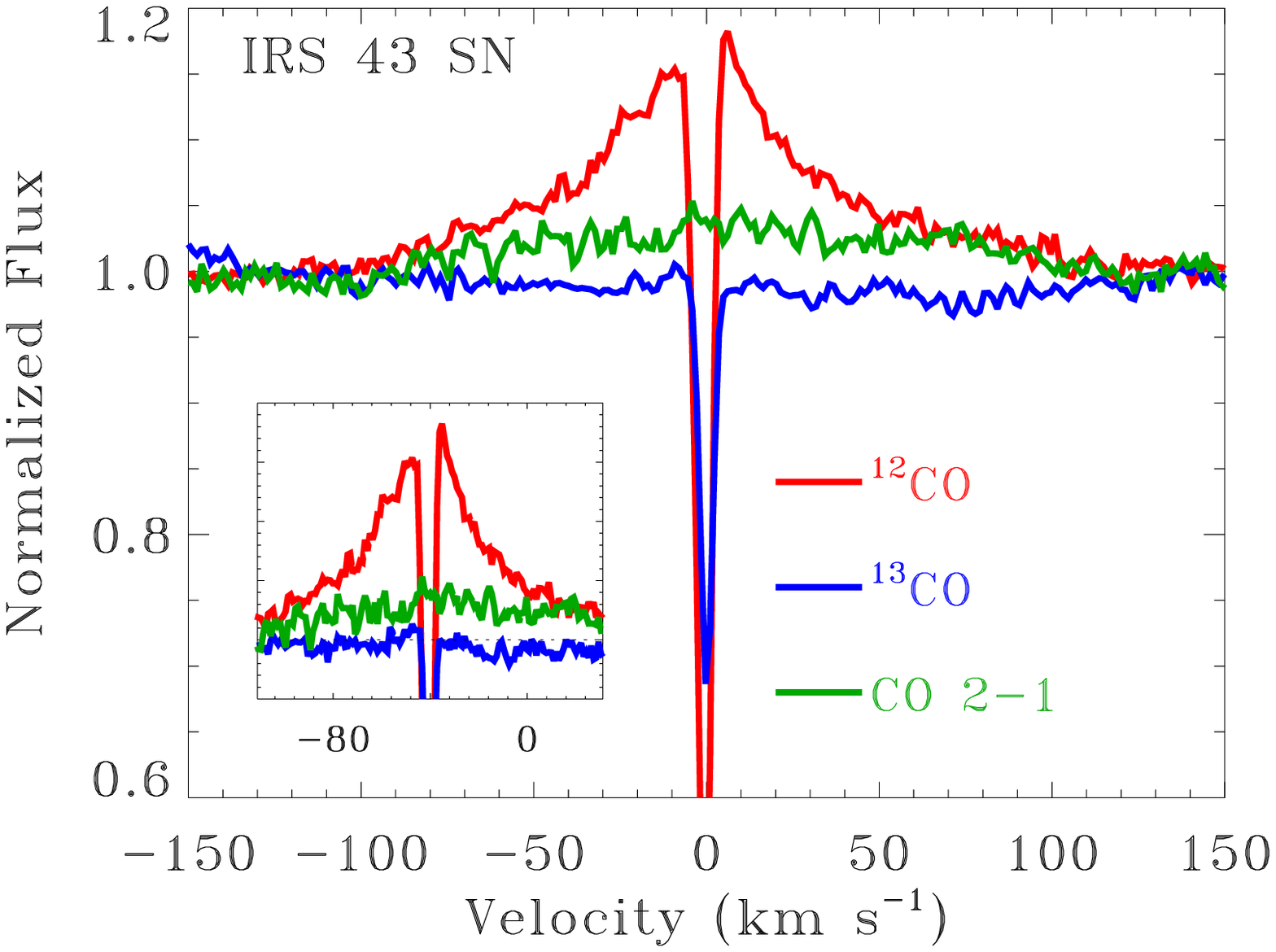}
\includegraphics[width=5.8cm]{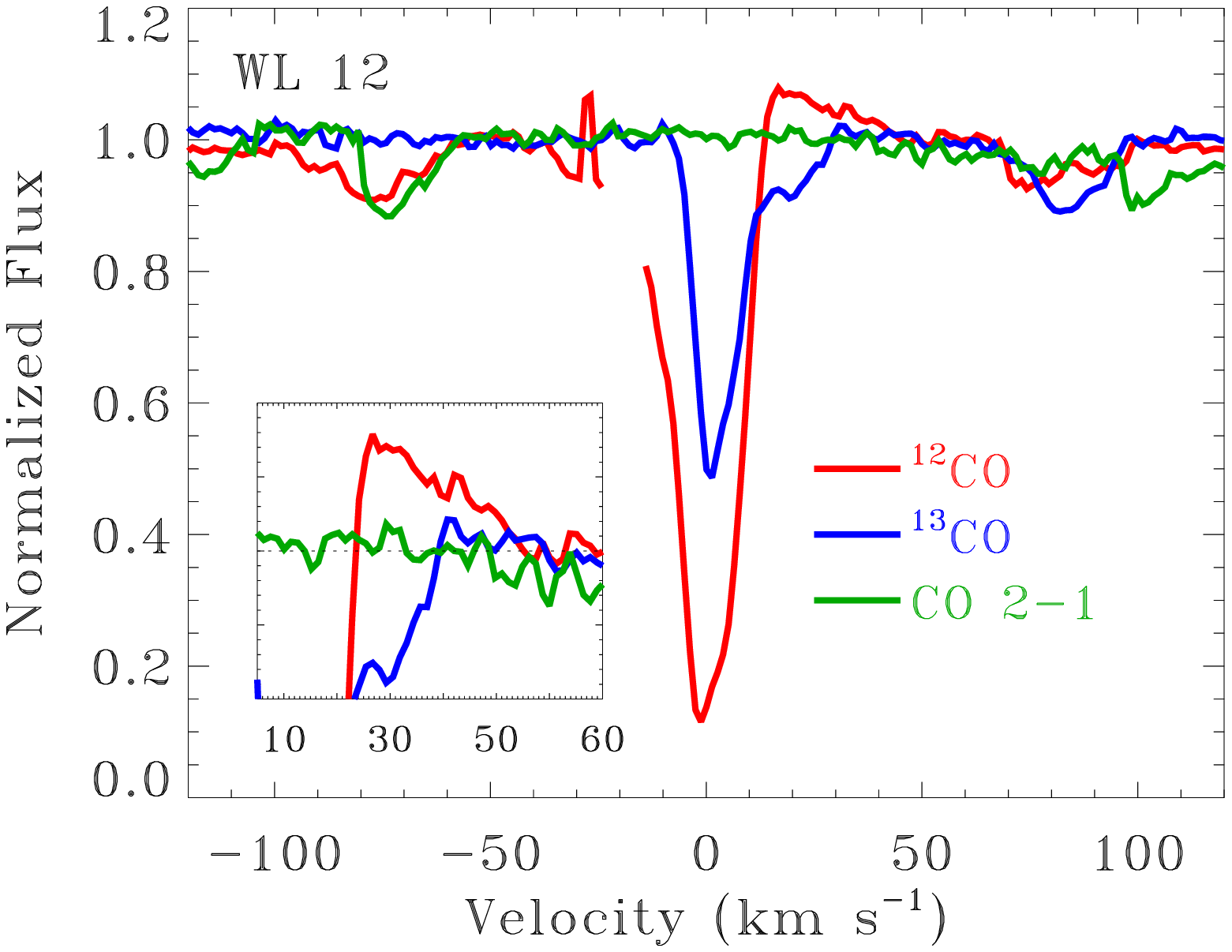}
\includegraphics[width=5.8cm]{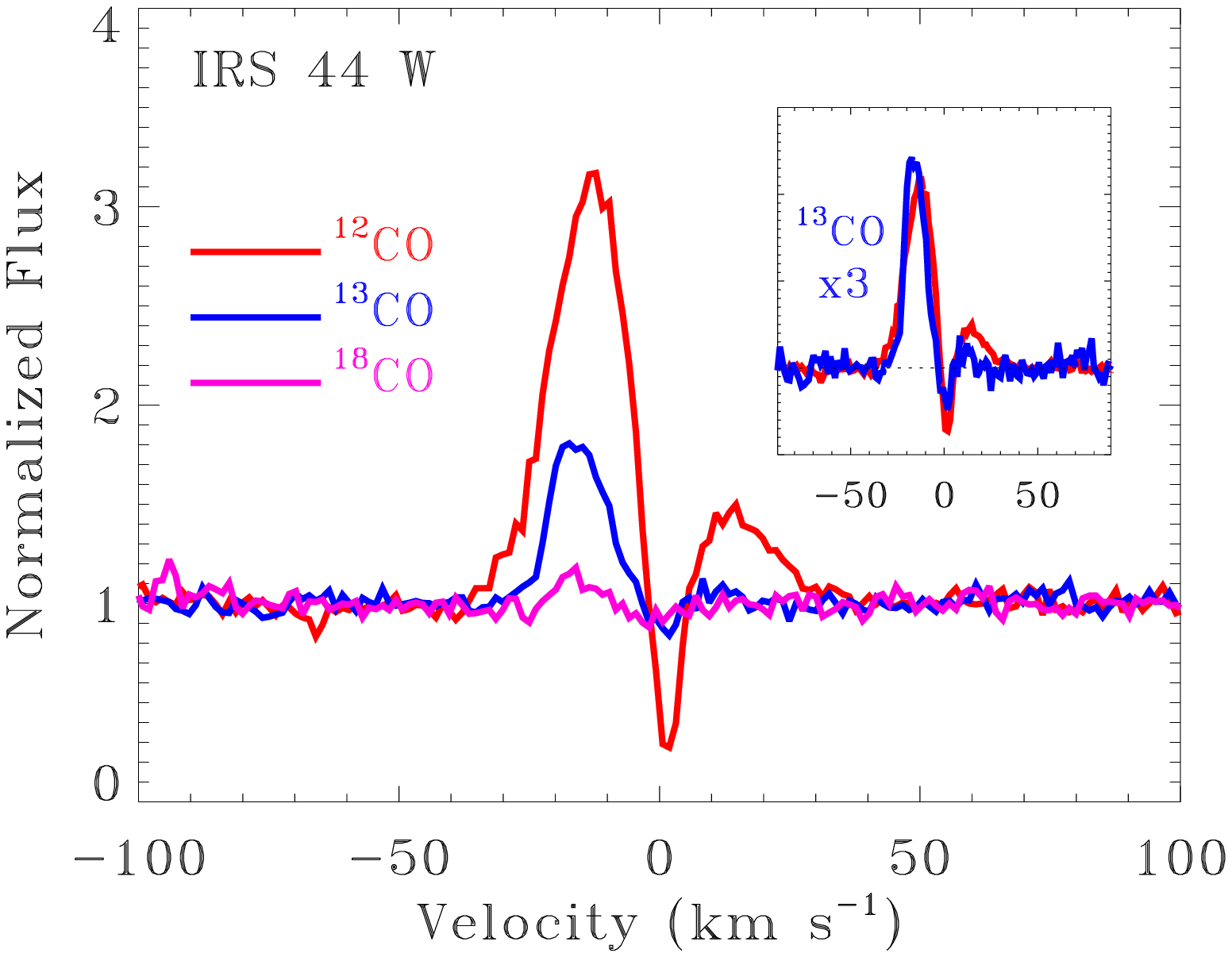}
\includegraphics[width=5.8cm]{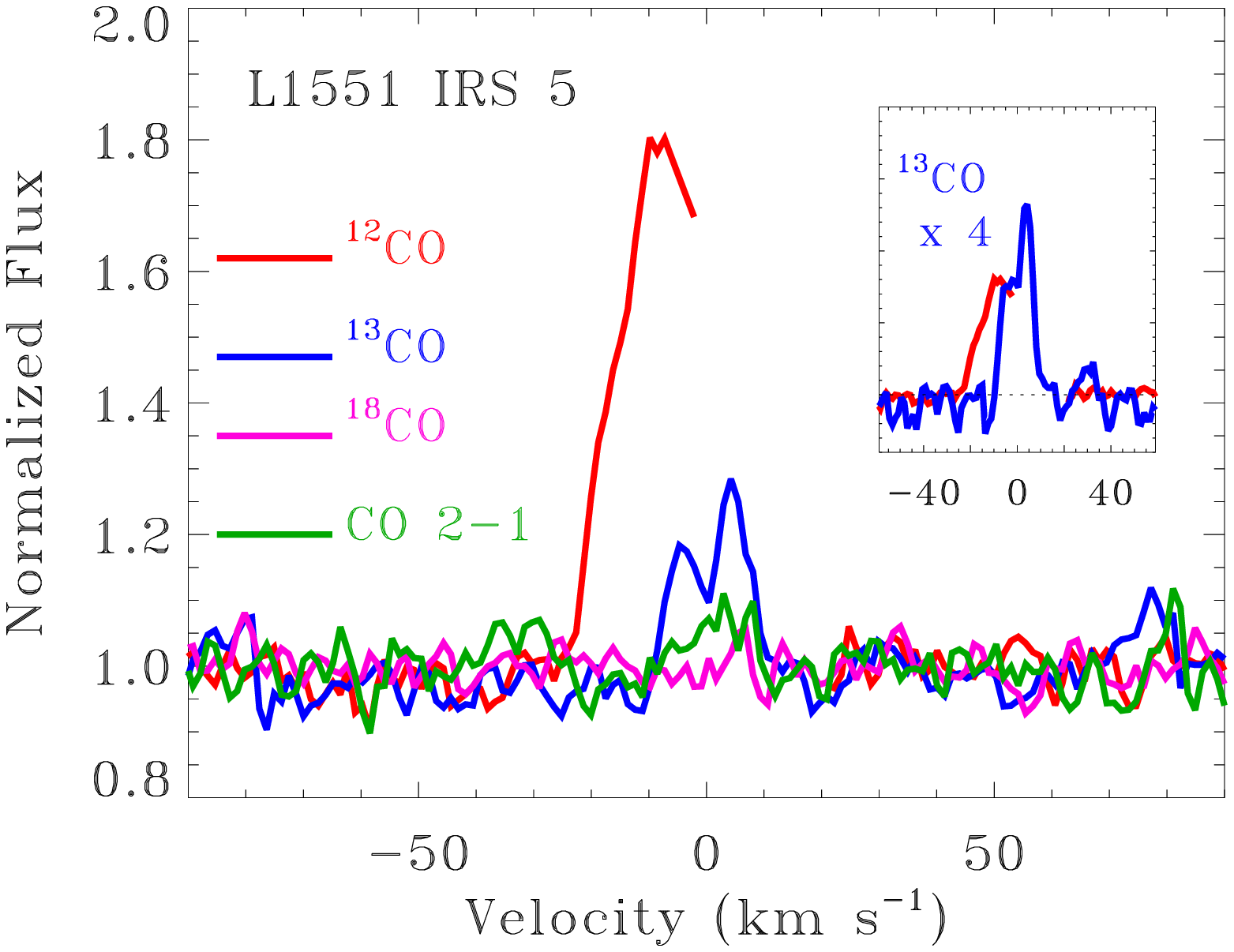}
\includegraphics[width=5.8cm]{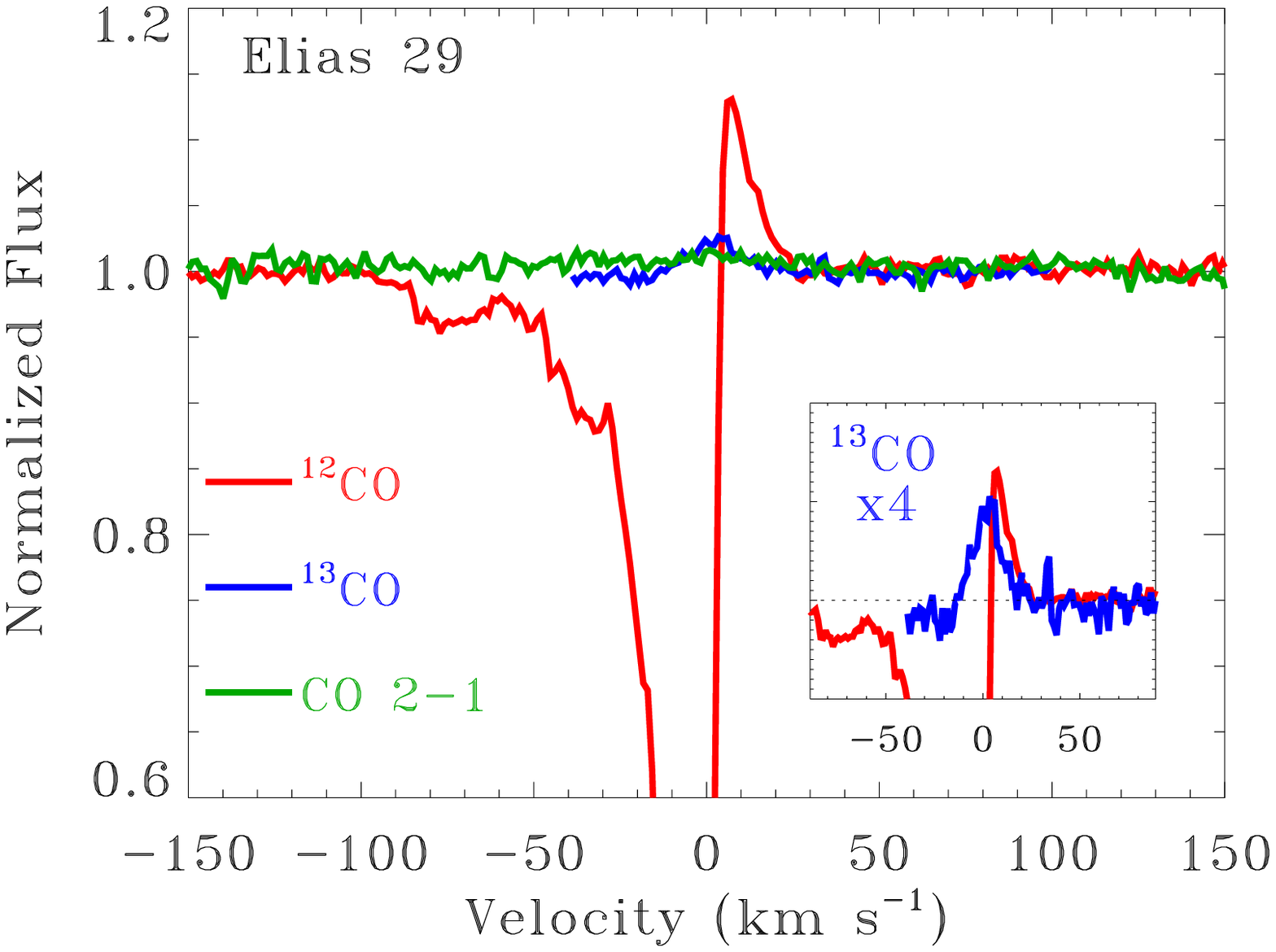}
\includegraphics[width=5.8cm]{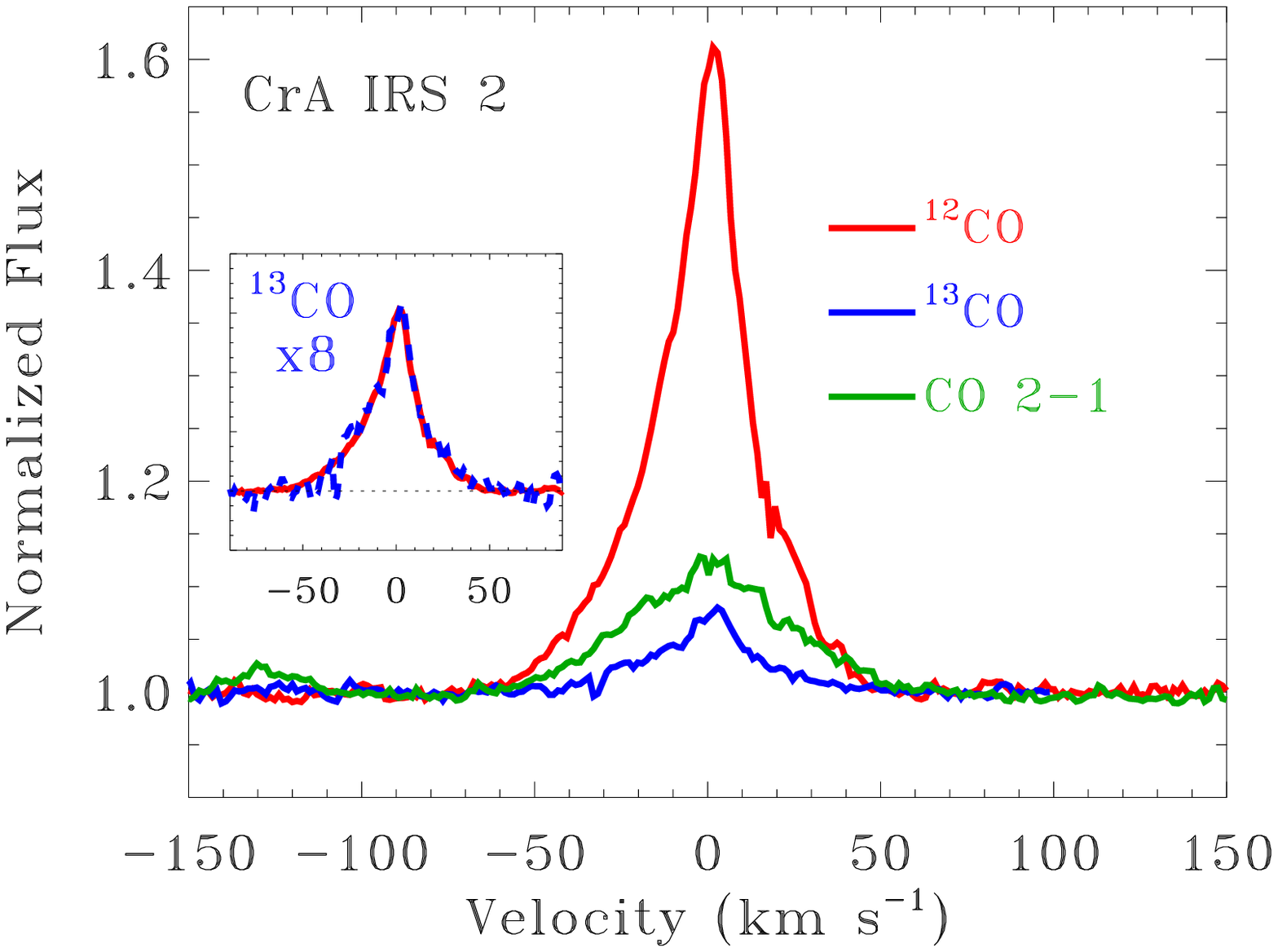}
\includegraphics[width=5.8cm]{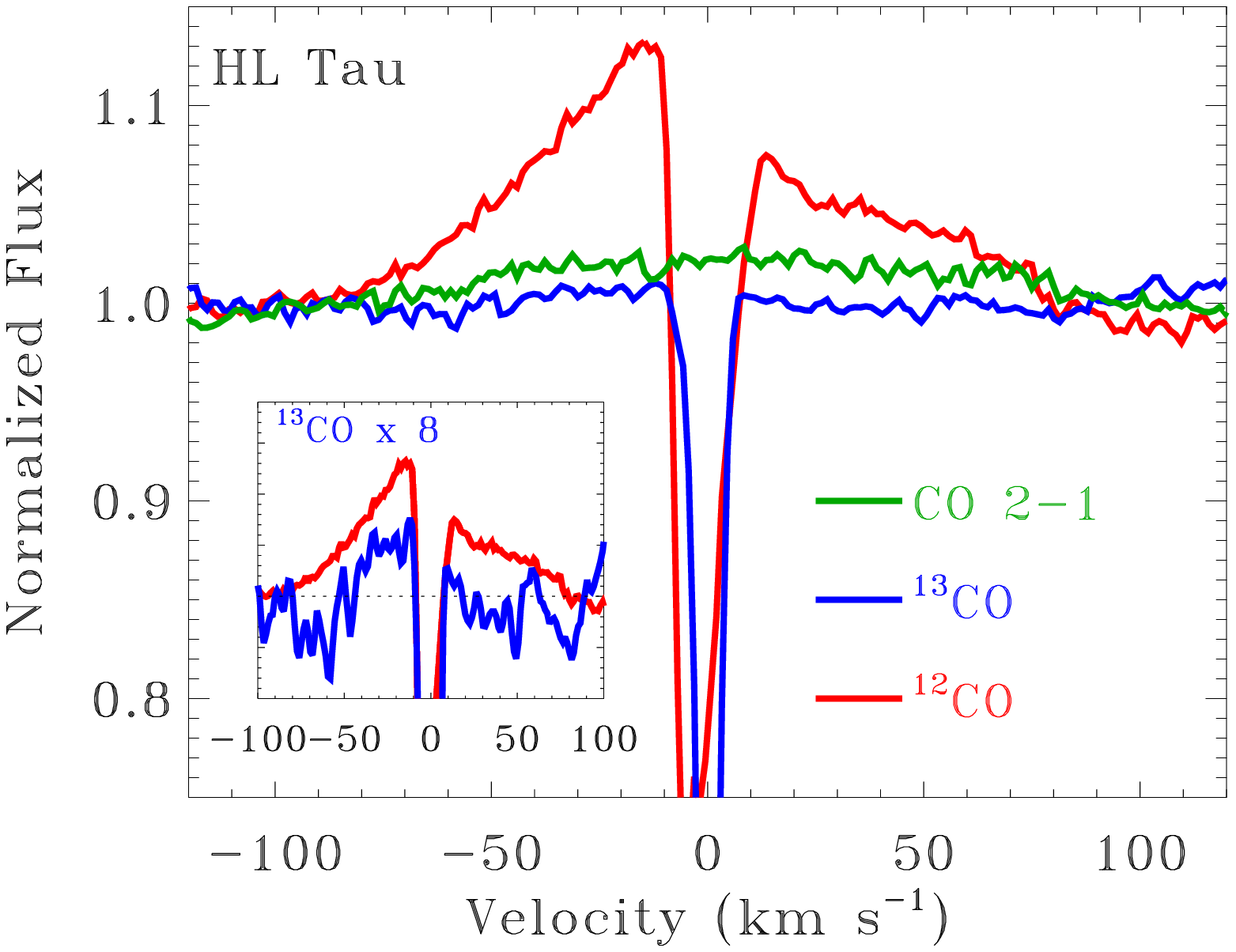}
\includegraphics[width=5.8cm]{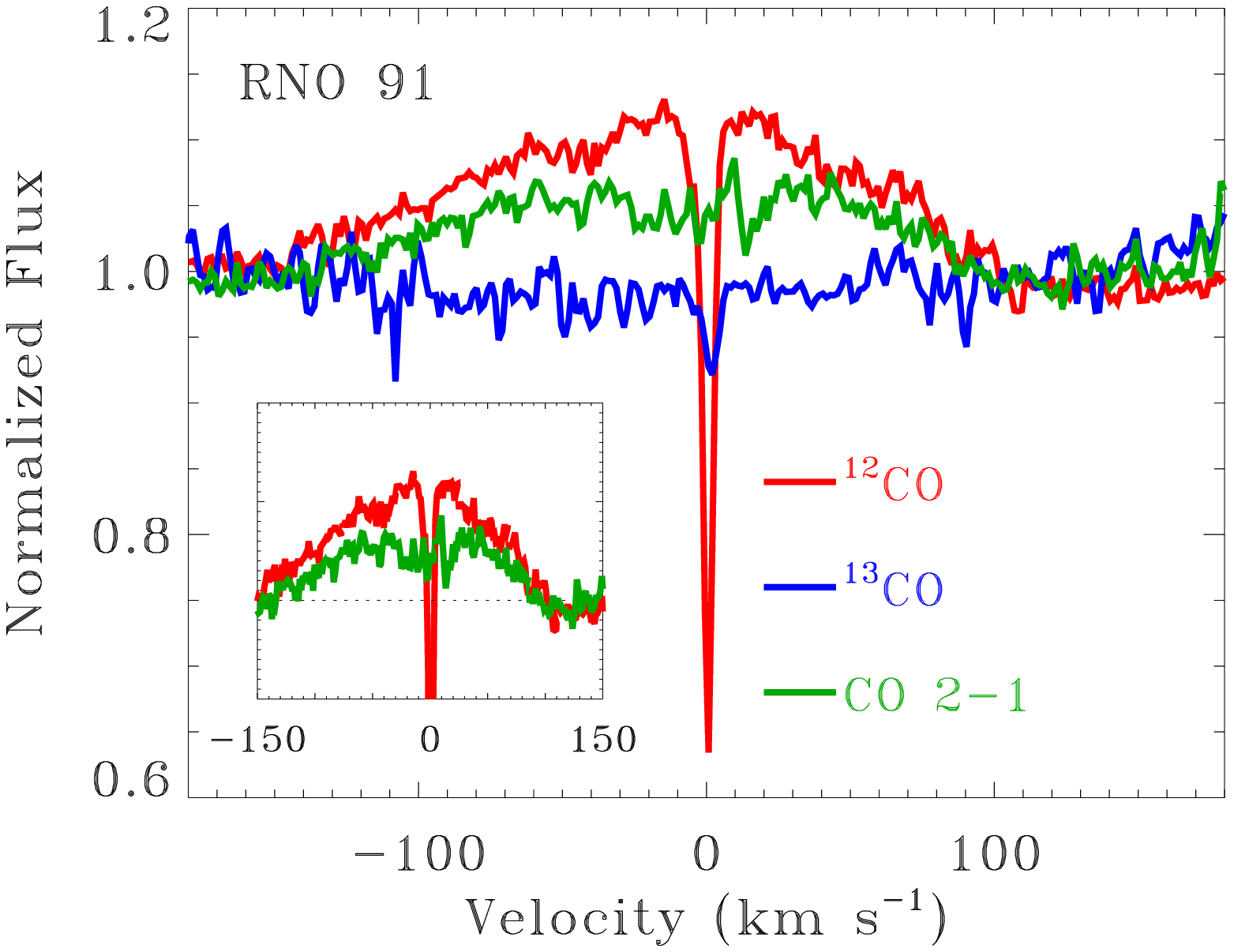}
\includegraphics[width=5.8cm]{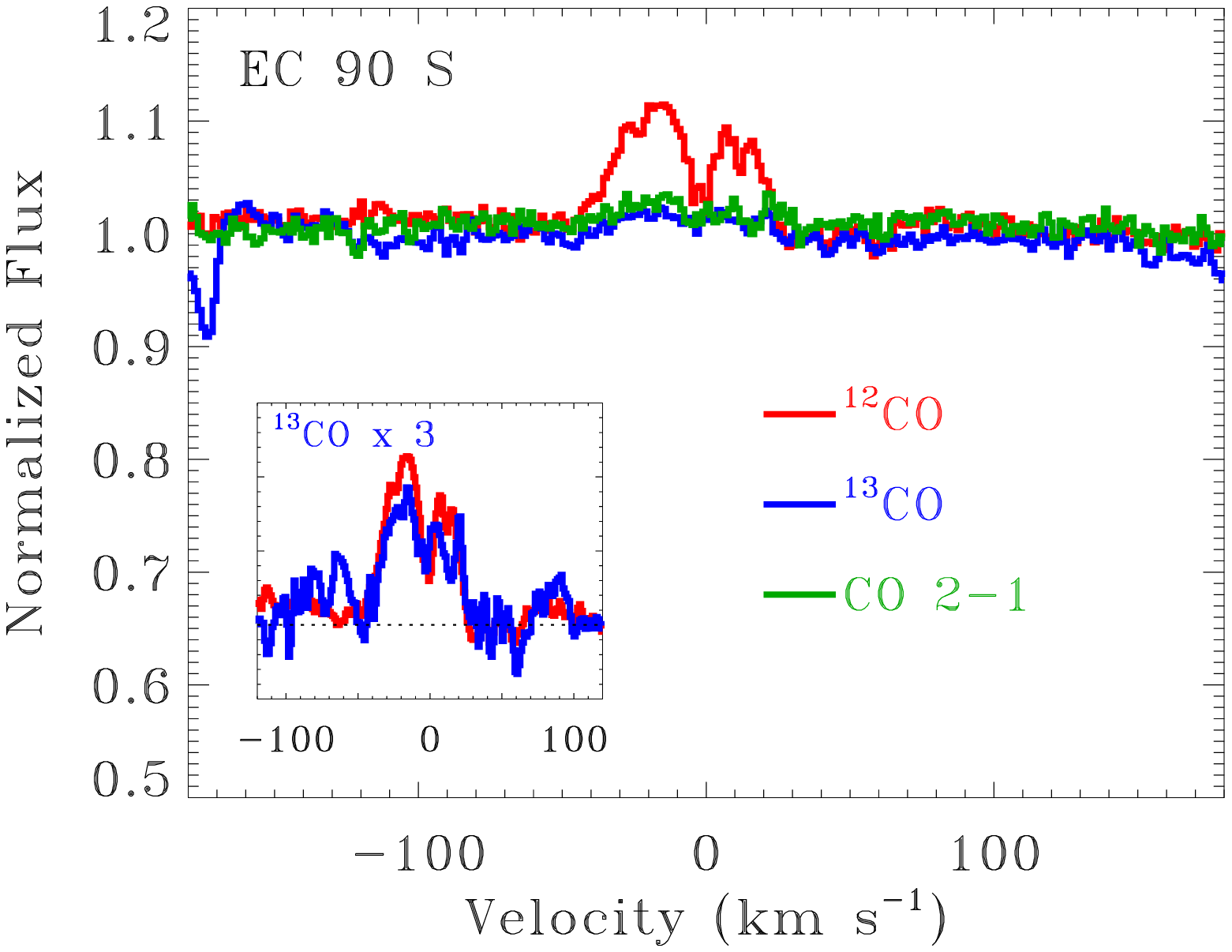}
\includegraphics[width=5.8cm]{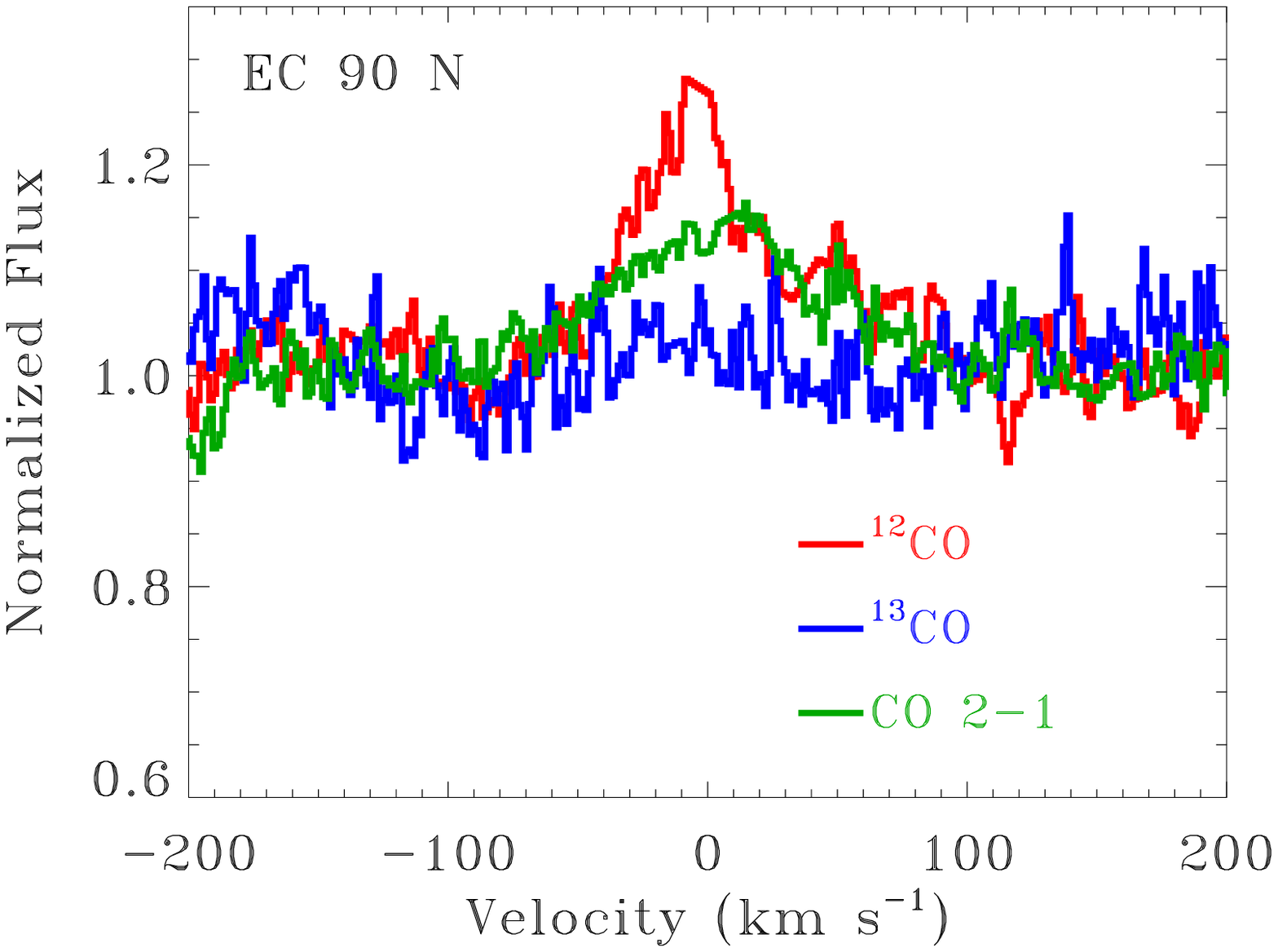}
\includegraphics[width=5.8cm]{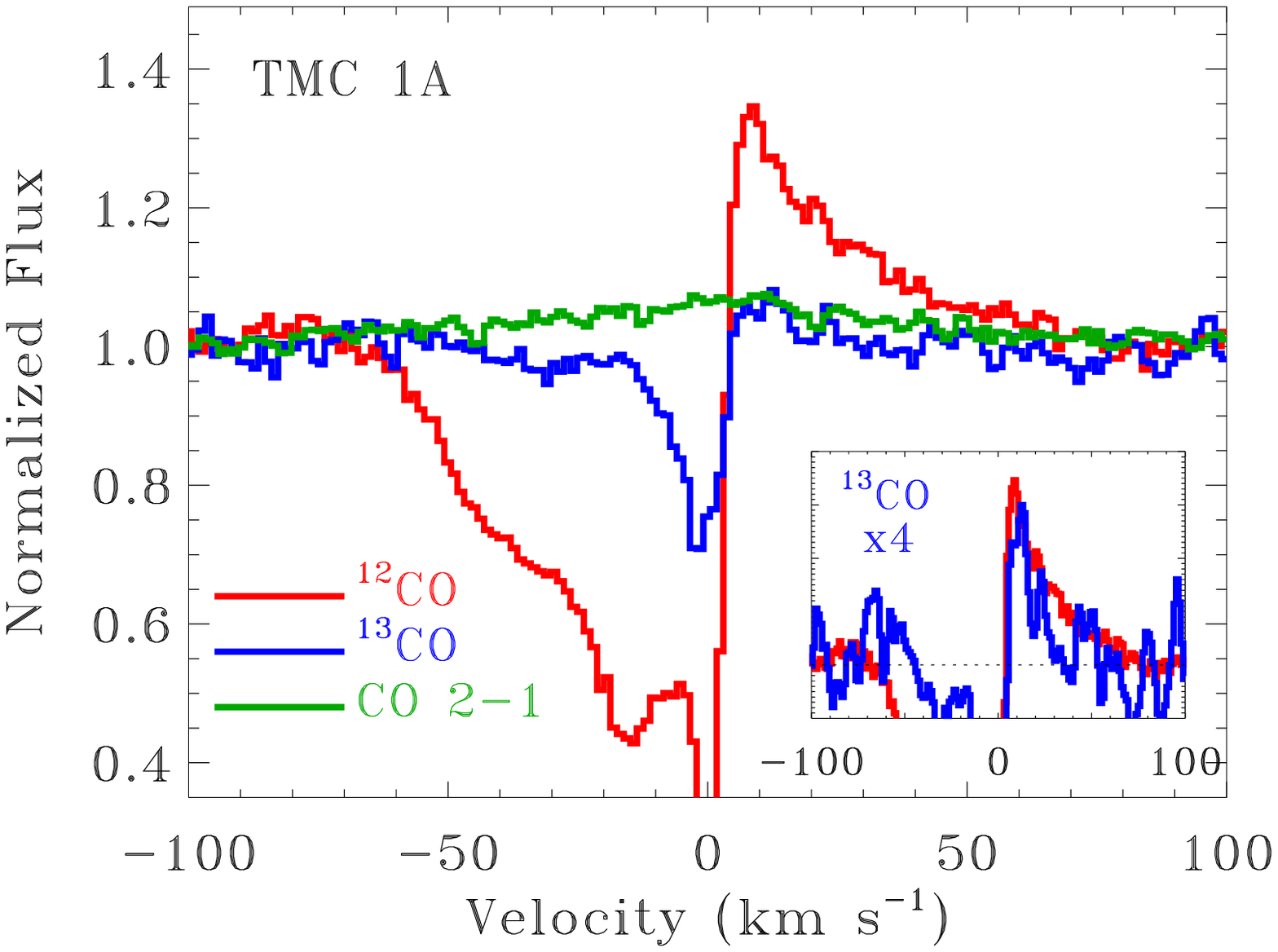}
\includegraphics[width=5.8cm]{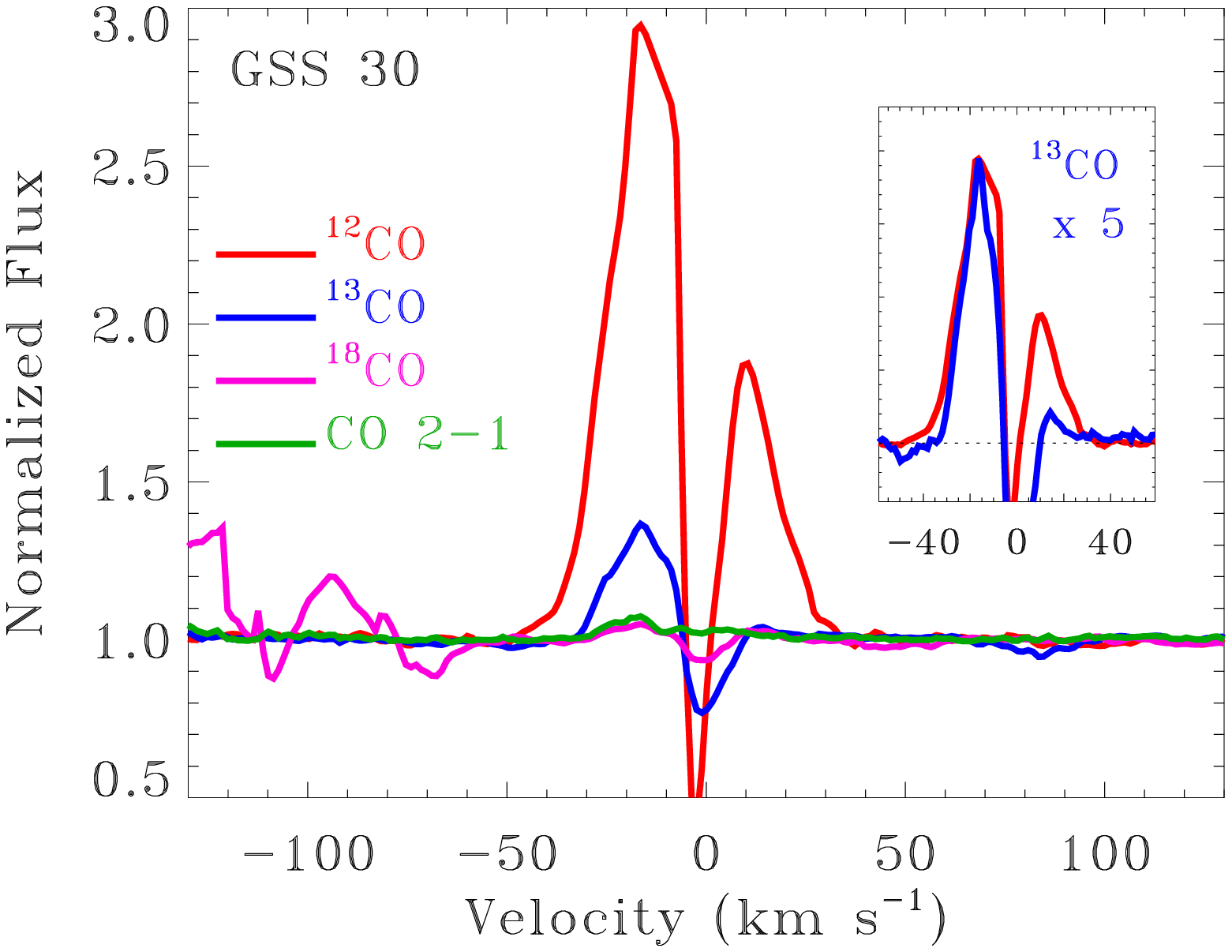}
\includegraphics[width=5.8cm]{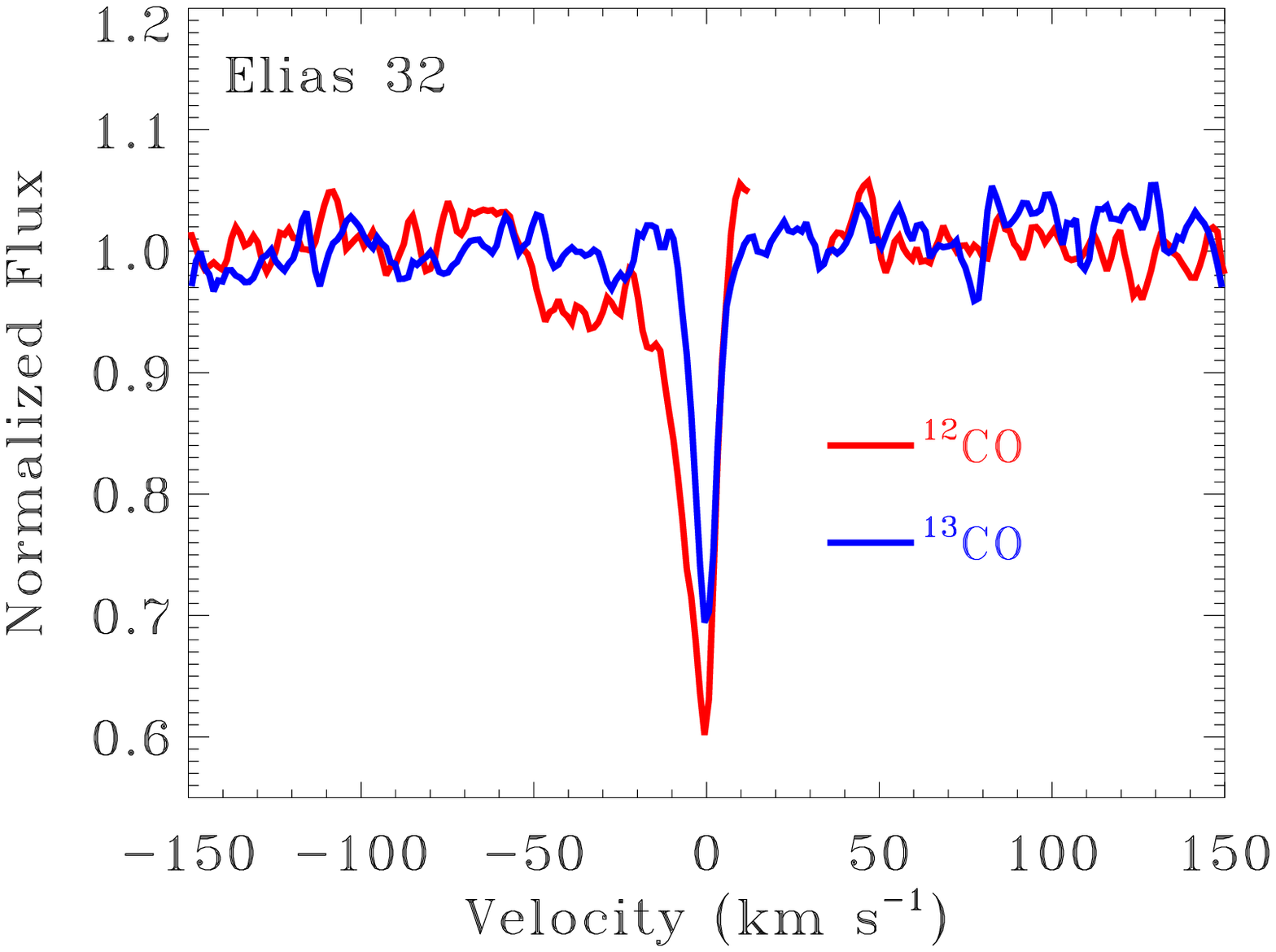}
\caption{Co-added line profiles from sources with detected CO emission.  In each plot the $^{12}$CO $v=1-0$ line is shown in red, the $^{13}$CO $v=1-0$ line is shown in blue, the C$^{18}$O $v=1-0$ line is shown in pink (IRS 44 W and GSS 30), and $^{12}$CO $v=2-1$ line is shown in green.  Inserts zoom in to compare the $^{12}$CO emission line profile to that of $^{13}$CO, C$^{18}$O, or $v=2-1$, depending on the source.}
\label{fig:gallery}
\end{figure*}

\begin{figure}
\includegraphics[width=85mm]{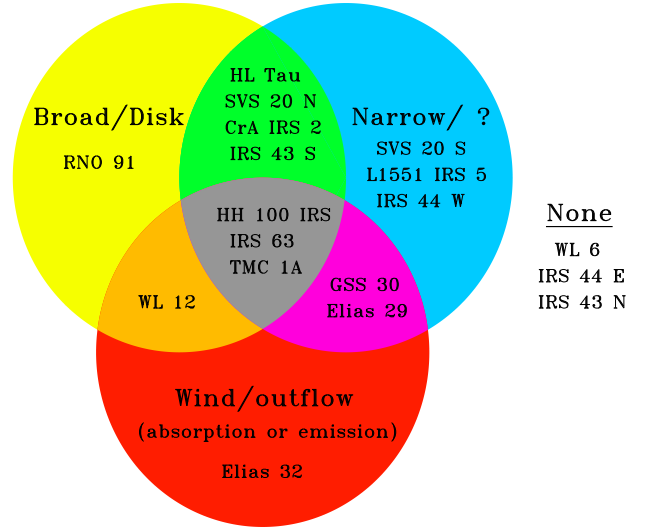}
\caption{A Venn diagram showing the presence of CO emission and absorption components from Table~\ref{tab:summco.tab}.}
\label{fig:covenn}
\end{figure}

Figure~\ref{fig:gallery} presents a gallery of co-added CO line profiles from all 
objects with detected CO emission.  Lines were co-added over all $J$ transitions 
that were observed and that have minimal or moderate contamination from telluric absorption.  
Some objects show strong emission in the narrow, optically-thick component, some objects show strong 
emission in broad, vibrationally excited emission, other objects show both, and still other objects show only
 weak or no emission in $^{12}$CO $v=1-0$ lines.  
Figure~\ref{fig:covenn} and Table~\ref{tab:summco.tab} (ordered by decreasing luminosity) summarize the presence and absence of each component 
for these objects.  Objects with a high bolometric 
luminosity, relative to the median in our sample, show emission in the narrow component but not the broad component.  Objects with a lower bolometric luminosity, relative to the median in our sample, emit in the broad 
component but not in the narrow component.  Hereafter, we loosely define high and low bolometric luminosities as greater than or less than $\sim 15$ $L_\odot$, respectively.

Table~\ref{tab:specprofs.tab} summarizes the properties of Gaussian profiles that were 
fit to the different isotopic and vibrational lines.  The fits were applied to lines coadded over all unblended 
rotational levels (Fig.~\ref{fig:gallery}).  The emission from $v^{\prime}>1$ is typically broad (FWHM of 50--160 \kms) and centered at the systemic velocity.  None of the lines show a double-peaked profile that is characteristic 
of Keplerian rotation in a disk with high inclination, although the co-added CO $v=2-1$ profiles from RNO 91 
tentatively show a weak dip in flux relative to a Gaussian profile.
On the other hand, 
the $^{13}$CO emission lines are narrow (FWHM between 10 to 50 \kms), with central velocities shifted between 0 to -15 \kms.  
The range of velocity centroids for the narrow component suggests that our crude classification scheme does not capture the full range of physical processes that produce this component.  

In most cases, the CO emission can be clearly distinguised as a broad or narrow component.
In a few spectra, the precise classification of CO components is unclear.  For example, the narrow component of CO emission from GSS 30 and IRS 44W each include emission from at least two regions based on distinct components in the line profiles.  These cases are discussed in more detail in the following subsections and in Appendix C.

Online Table A.2 presents the line fluxes and equivalent widths for the different components in several 
selected transitions.  Most $^{12}$CO $v=1-0$ and $^{13}$CO $v=1-0$ lines are partially-attenuated by absorption in the same transition.   The  listed line fluxes are what the  fluxes would be if no line-of-sight CO absorption occurred.  The fluxes are not corrected for extinction.  The absorbing CO gas in the circumstellar envelope or molecular cloud is often cold \citep[$\sim 10-100$ K,][]{Jorg02,Bergin2007} and only affects emission in low-$J$ transitions.
However, some absorption components are optically-thick for transitions with $J>30$.
In some cases, multiple CO absorption components are detected.  The listed fluxes are measured from Gaussian fits to the emission profile, ignoring any absorption component.   
The FWHM and central velocity are determined from fits to the 
summed profiles (Table~\ref{tab:specprofs.tab}).  The central velocities are listed 
relative to the velocity of $^{13}$CO and C$^{18}$O absorption listed in Table~\ref{tab:sample.tab}.  
Fitting profiles to the observed emission while ignoring the absorption components yields line fluxes measured consistently throughout a given spectrum.
In many cases the fit is applied only to a portion of the 
line profile to avoid absorption features, regions with optically-thick telluric absorption, 
contamination from emission in other components of the line, and emission in other lines.  
In each case the $1$-$\sigma$ uncertainty in flux is dominated by the uncertainty in the continuum level and does not include the uncertainty in the relative or absolute flux calibration. 

In the subsequent subsections, we describe separately the properties of the broad and narrow emission components, spatially-extended CO emission, H$_2$ emission, and CO wind absorption detected within our sample.  For each CO emission component, a model of an isothermal, plane-parallel slab of CO gas (see Appendix D) is used to calculate temperatures, column densities, and emitting areas, with results in Table~\ref{tab:props.tab}.

\begin{table}
\caption{Presence of CO emission components$^a$}
\label{tab:summco.tab}
{\scriptsize
\begin{tabular}{lcccccc}
\hline
Star & $L_{bol}$ &\multicolumn{2}{c}{Broad}      & \multicolumn{2}{c}{Narrow} &Wind \\
       & ($L_\odot$)&  $v=1-0$  &  $v=2-1$  &   $^{12}$CO  & $^{13}$CO  & abs.\\
\hline
SVS 20 S  &142 & n  & n  & \Checkmark  & \Checkmark & n   \\
Elias 29 & 41 & n  &  n  &  \Checkmark  & \Checkmark &  \Checkmark\\
L1551 IRS 5 & 23 & n & n & \Checkmark & \Checkmark  & n\\
IRS 44 E &  (18) & --  & n & n & --& n\\
GSS 30 &  14 & n? & n? & \Checkmark& \Checkmark & n\\
HH 100 IRS & 15& \Checkmark   &  \Checkmark   &  \Checkmark &  \Checkmark  & \Checkmark\\
CrA IRS 2& 12 &--  &  \Checkmark   &  \Checkmark& \Checkmark  &   n\\
 IRS 44 W & (9)  &   n & n & \Checkmark$^b$  & \Checkmark & n\\
HL Tau  & 6.6 & \Checkmark    &  \Checkmark &  \Checkmark & \Checkmark   &  n \\
IRS 43 S & 6.0 & \Checkmark  &  \Checkmark  &  \Checkmark? & n   & n\\
IRS 63   & 3.3 & \Checkmark  & \Checkmark  & \Checkmark & -- & \Checkmark \\
TMC 1A &  2.8 &\Checkmark& \Checkmark & \Checkmark & \Checkmark &  \Checkmark\\
WL 12$^d$ & 2.6 &   \Checkmark &--  & --& n & \Checkmark \\
WL 12$^e$ & 2.6 & --  & \Checkmark  & \Checkmark$^c$  &n & \Checkmark \\
WL 6     & 2.6 & n  & --   & n  & -- & n\\
RNO 91 & 2.5 & \Checkmark  & \Checkmark  & n  & n & n\\
Elias 32 & 1.1 & n  & n  & n & n & y\\
IRS 43 N  & -- & n  &  --  &  n  & --& --\\
SVS 20 N  &(0.27) & \Checkmark  & \Checkmark  & \Checkmark  & -- & n \\
\hline
\multicolumn{7}{l}{$^a$\Checkmark means present, n means not present}\\
\multicolumn{7}{l}{~~~~~ -- means not possible to determine}\\
\multicolumn{7}{l}{$^b$Two distinct narrow components}\\
\multicolumn{7}{l}{$^c$Weak emission tentatively identified as narrow component}\\
\multicolumn{3}{l}{$^d$Sep.~2007} & 
\multicolumn{3}{l}{$^e$Mar.~2010} \\
\end{tabular}}
\end{table}

\begin{table}
\caption{Average CO Emission Line Properties$^a$}
\label{tab:specprofs.tab}
\begin{tabular}{lccc}
\hline
Target   & Lines   & $v_{CO}^b$ (km s$^{-1}$) & FWHM (km s$^{-1}$) \\
\hline
\multicolumn{4}{c}{Broad Component}\\
\hline
\hline
HH 100 IRS$^c$   & $v=2-1$  &  3.2 (3.0)  & 80 (18) \\
HH 100 IRS$^c$   & $v=1-0$  &  (3.2)       & 92 (26) \\
RNO 91         & $v=1-0$  & -15.5 (4)  & 150 (40)   \\
RNO 91         & $v=2-1$   & -8  (4)     &  165 (40)   \\
IRS 43  S         &  $v=1-0$  & 0.7 (2)     &  120 (20)  \\
IRS 43  S        & $v=2-1$   &  13 (4)     &  146 (25) \\
IRS 63           & $v=2-1$    &  1.0 (1.0)  & 92 (16)   \\
IRS 63           & $v=1-0$     &  (1)$^d$  & 118 (30)  \\
WL 12$^h$        &  $v=2-1$  &  6.4 (2) &  98 (30) \\
HL Tau      & $v=2-1$    &  10 (3)   & 130 (40)  \\
HL Tau     & $v=1-0$    & -8.7 (1.0)   & 90 (15) \\
CrA IRS 2    & $v=2-1$ &  -0.2 (0.6) & 54 (4)  \\
TMC 1A      &  $v=2-1$  & 2.0 (2)    & 96 (20) \\
TMC 1A      &  $v=1-0$  & (2)$^d$   & (96) $^d$ \\
SVS 20 N   & $v=2-1$   & 3.5 (2) & 100 (30) \\
SVS 20 N   & $v=1-0$   & (3.5)$^d$ & 115 (40) \\
\hline
\multicolumn{4}{c}{Narrow Component}\\
\hline
\hline
HH 100 IRS$^c$   & $^{13}$CO  &  0.6 (0.4)   &  11 (3)   \\
HH 100 IRS$^c$   & $v=1-0$         &  1.3 (0.4)    &  15 (1.5)       \\
Elias 29        & $^{13}$CO   &   2.4 (1.5)  &  18 (7) \\
Elias 29$^e$             & $v=1-0$          &  (3.0) $^d$         &  19 (3) \\
IRS 63$^e$        & $v=1-0$            &  3 (10)        &  28 (13)   \\
HL Tau       &  $^{13}$CO   & -15 (6) & 42(14) \\
SVS 20 N      & $v=1-0$ & -8.8 (1) & 28 (5)                   \\
CrA IRS 2$^f$       & $^{13}$CO  &  0.0 (0.4)  &  34 (4)     \\
CrA IRS 2$^f$      & $v=1-0$         & 0.5 (0.3)&  16 (3)          \\
SVS 20 S      & $v=1-0$          & -8.0 (1.0) & 37 (7)         \\
SVS 20 S      & $^{13}$CO & (-8.0)$^d$  & 47 (23)          \\
IRS 44 W     & $^{13}$CO & 15.4 (0.4) & 14 (1)     \\
IRS 44 W       & C$^{18}$O & (15.4) $^d$ &8 (2)   \\
IRS 44 W   & $^{12}$CO & -1.0 (1.0) & 32 (4) \\
L1551 IRS 5 & $^{13}$CO & 0.8 (0.7) & 12 (2) \\
L1551 IRS 5 & $^{12}$CO & (0.8)$^d$  & 17 (2)   \\
L1551 IRS 5 & $v=2-1$ & (0.8)$^d$ & 13(2)   \\
GSS 30$^h$     & $v=2-1$   & -5.8 (2.0) & 42 (10)\\
GSS 30     & $v=2-1$    & -19.2 (0.5) & 10 (2)\\
GSS 30$^h$     & $^{12}$CO,$J>30$   & -10 (1) & 36 (3) \\
GSS 30 SW     &  $^{12}$CO   & -9.1 (1.0) & 6.3 (1.0)\\
GSS 30 NE     &  $^{12}$CO   & -8.6 (1.0) & 6.5 (1.0)\\
IRS 43 S$^h$           & $v=1-0$   &  -4 (1)     &  42 (7) \\
\hline
\multicolumn{4}{l}{$^a$Based on Gaussian profiles fit to coadded lines.}\\
\multicolumn{4}{l}{$^b$Relative to velocity of $^{13}$CO and C$^{18}$O absorption.}\\
\multicolumn{4}{l}{$^c$From spectrum obtained in August 2008}\\
\multicolumn{4}{l}{$^d$Forced value from same component in $v=2-1$ or $^{13}$CO transition}\\
\multicolumn{4}{l}{$^e$Fit only to red emission}\\
\multicolumn{4}{l}{$^f$Fit is not good because line is not Gaussian}\\
\multicolumn{4}{l}{$^g$March 2010}\\
\multicolumn{4}{l}{$^h$Tentative classification as narrow component}\\
\end{tabular}
\end{table}

\begin{table}
\caption{Physical Properties of CO Emission Region}
\label{tab:props.tab}
{\footnotesize \begin{tabular}{ccccccc}
\hline
Star        & Component & $T_{rot}$ (K)& $\log N$(CO)$^a$ & (Area)$^{0.5}$$^b$ \\
\hline
IRS 44 W  & Narrow$^c$            &  330$\pm 20$   &  19.15$\pm0.25$      &  3.6 \\
GSS 30    & Narrow$^c$            & $315\pm15$   &  $18.65\pm0.10$     &  18.1 \\
GSS 30    & Extended     & $250\pm30$  & $18.7\pm0.2$ & 3.6 \\
CrA IRS 2 & Broad      & $1100\pm200$ & --      & --\\
CrA IRS 2  &  Narrow     & $560\pm70$    & $19.1\pm0.1$ & 1.0\\
HH 100 IRS   &  Broad       & $1000\pm100$ & $17.8\pm0.2$ & 0.7 \\
\hline
\multicolumn{5}{l}{$^a$Column density $N$ in units of $\log$ cm$^{-2}$ here}\\
\multicolumn{5}{l}{~~~~and throughout paper, assumes $b=2.0$ \kms.}\\
\multicolumn{5}{l}{$^b$Square root of emitting area, in units of AU}\\
\multicolumn{5}{l}{$^c$Blueshifted emission}\\
\end{tabular}}
\end{table}

\subsection{Broad CO Emission From Warm Gas}
Broad emission in $v=2-1$ lines is clearly detected in 10 of 18 sources$^4$ (Figure~\ref{fig:galleryx}).  Emission in some $v=3-2$ and $4-3$ lines is detected from CrA IRS 2 and HH 100 IRS, which have high S/N.   Non-detections of lines from $v^{\prime}>2$ from other sources are generally not significant, assuming the same flux ratio in $v=3-2$ to $v=1-0$ lines as in CrA IRS 2 and HH 100 IRS.
\footnotetext[4]{Counting WL 12 as a detection, despite a non-detection in one of the two observations, and GSS 30 as a non-detection, despite some $v=2-1$ emission (see \S 3.2.1).}

In most cases, a component of similar shape and strength as the $v=2-1$ lines is seen in the $v=1-0$ lines.  The $v=1-0$ emission is often absorbed by CO in our line of sight to the emission region.  Broad CO wind absorption partially or totally obscures any CO $v=1-0$ emission on 
the blue side of the line profile of six sources (HH 100 IRS, IRS 63, Elias 29, TMC 1A, Elias 32, and WL 12, see \S3.7).  In most cases, the same broad component seen in $v=2-1$ lines is also identified 
in the $v=1-0$ lines based on the similarity of the emission line profiles on the red wing.  For HL Tau and CrA IRS 2, 
CO $v=1-0$ emission in a narrower emission component masks any possible broad emission that would have the same profile as the $v=2-1$ emission.  The broad emission component is not seen
in any $^{13}$CO lines within our sample.

The centroid of the $v=2-1$ emission is usually consistent with the systemic velocity.  
The centroids that deviate by $>3$ \kms\ (IRS 43 S, HL Tau, and WL 12) each have very low S/N in 
the coadded line profiles and as a result have unreliable central velocities.  In the cross-dispersion direction, the 
broad emission from HH 100 IRS, RNO 91, and CrA IRS 2 (the three cases with highest S/N in the 
broad lines and the highest spatial resolution) is centered at the same location (to 
within $\sim 0.2$ pix, or 1--2 AU at 120 pc) as the continuum emission and is not 
spatially extended (FWHM$\lesssim 1-1.5$ pix, or 6-8 AU at 120 pc).

To measure temperatures and column densities (Fig.~\ref{fig:rots.ps}), we select the two spectra, HH 100 IRS and CrA IRS 2, that 
have the highest S/N in broad line emission over a wide range of $J$ levels.  
The excitation diagram of CO emission from HH 100 IRS shows an upturn in $N/(2J+1)$ at low-$J$, 
characteristic of optically-thick emission from warm (500-1500 K) gas (see Appendix D).  The CO $v=2-1$ lines are more likely to be free of any optical depth effects.  No $^{13}$CO $v=1-0$ emission is detected, with an upper limit that is $\sim 10$ times weaker than the  $^{12}$CO $v=1-0$ emission.  A fit to the $v=2-1$ lines indicates a rotational temperature of $\sim 1000\pm100$ K.  At this temperature, the lack of detectable $^{13}$CO emission in this component requires a column density $\log N$(CO)$<18.0$.  In the excitation diagram, the curve for $v=1-0$ lines requires $\log N$(CO)$>17.5$.  The emission in $^{12}$CO $v=1-0$ lines with high-$J$ is underproduced at this temperature and column density, which may suggest that the rotational temperature derived from the $v=2-1$ lines is too low or that a second temperature is needed.  The $v=2-1$ lines are underproduced by $\sim 0.6$ dex, indicating that the vibrational temperature is warmer than the rotational temperature.  The flux ratio of $v=3-2$ and $2-1$ lines yields a vibrational temperature of $\sim2000$ K. The total emitting area is roughly (0.7 AU)$^2$.

For CrA IRS 2, the broad component in the $v=1-0$ lines is not measureable because it is mostly masked by much stronger emission in a narrower emission component (see Appendix C.3).  Broad emission is detected in $v=2-1$, $3-2$, and a few $4-3$ emission lines.  The $v=2-1$ and $3-2$ line flux ratios yield  a rotational temperature of $\sim 1000$ K, while the ratio of line fluxes in the different vibrational levels yields a vibrational temperature of $\sim 2500$ K.

\subsection{Narrow CO Emission From Optically-Thick Gas}

Emission in $^{13}$CO $v=1-0$ lines is detected from 9 of 18 embedded objects within our sample.   The same narrow component seen in $^{13}$CO emission is also seen in $^{12}$CO and C$^{18}$O lines, which indicates that the $^{12}$CO lines are optically-thick.  Some of these same objects also have a broad emission component.  In only one case, GSS 30, is CO $v=2-1$ emission detected in the narrow component, probably because of the high S/N and line-to-continuum contrast in that spectrum.   

The line profile of narrow emission is centered at the systemic velocity for most sources.  In addition to emission at the systemic velocity, IRS 44 and GSS 30 also show narrow components blueshifted by $\sim 10$ \kms.  Meanwhile, WL 12 (when the emission is detected, see variability in Fig.~\ref{fig:hhvar})
and SVS 20 S show only a blueshifted component.
In each case, the emission is relatively narrow, with FWHM between 10--50 \kms.  For most objects, the emission in this component  is centered at the same spatial location on the detector as the continuum emission and is generally not spatially extended.   In one case, IRS 44 W, some of the  $^{12}$CO is offset by $0\farcs07$ ($\sim 9$ AU) W of the star and is spatially extended in the slit by $0\farcs22$ ($\sim 28$ AU).   Some very extended emission is also detected from GSS 30 and IRS 43 and is described in \S 3.3.  In Appendix C, we discuss in detail the narrow emission from GSS 30, IRS 44, and CrA IRS 2, which each have high S/N CO spectra.

\begin{figure*}
\includegraphics[width=60mm]{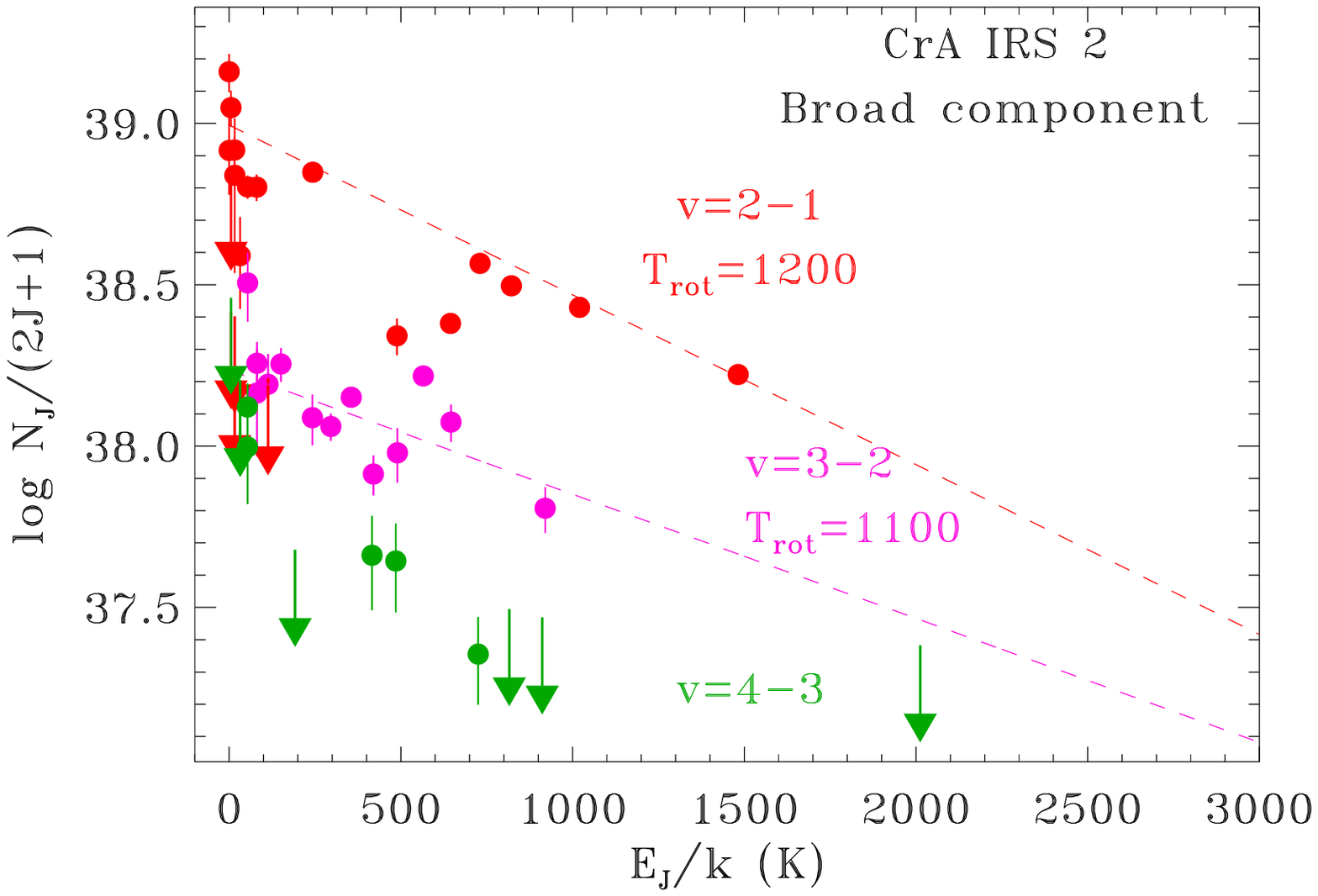}
\includegraphics[width=60mm]{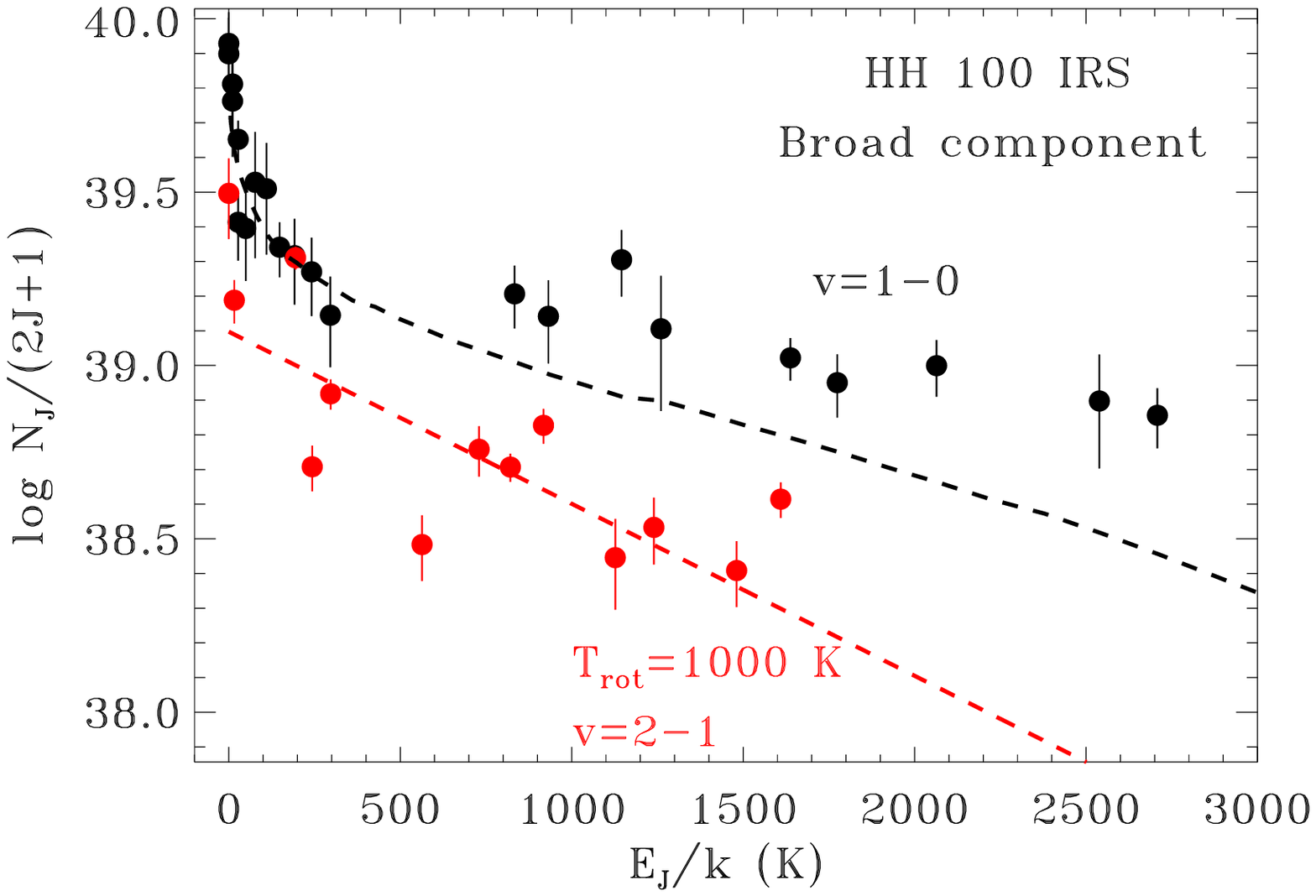}
\includegraphics[width=60mm]{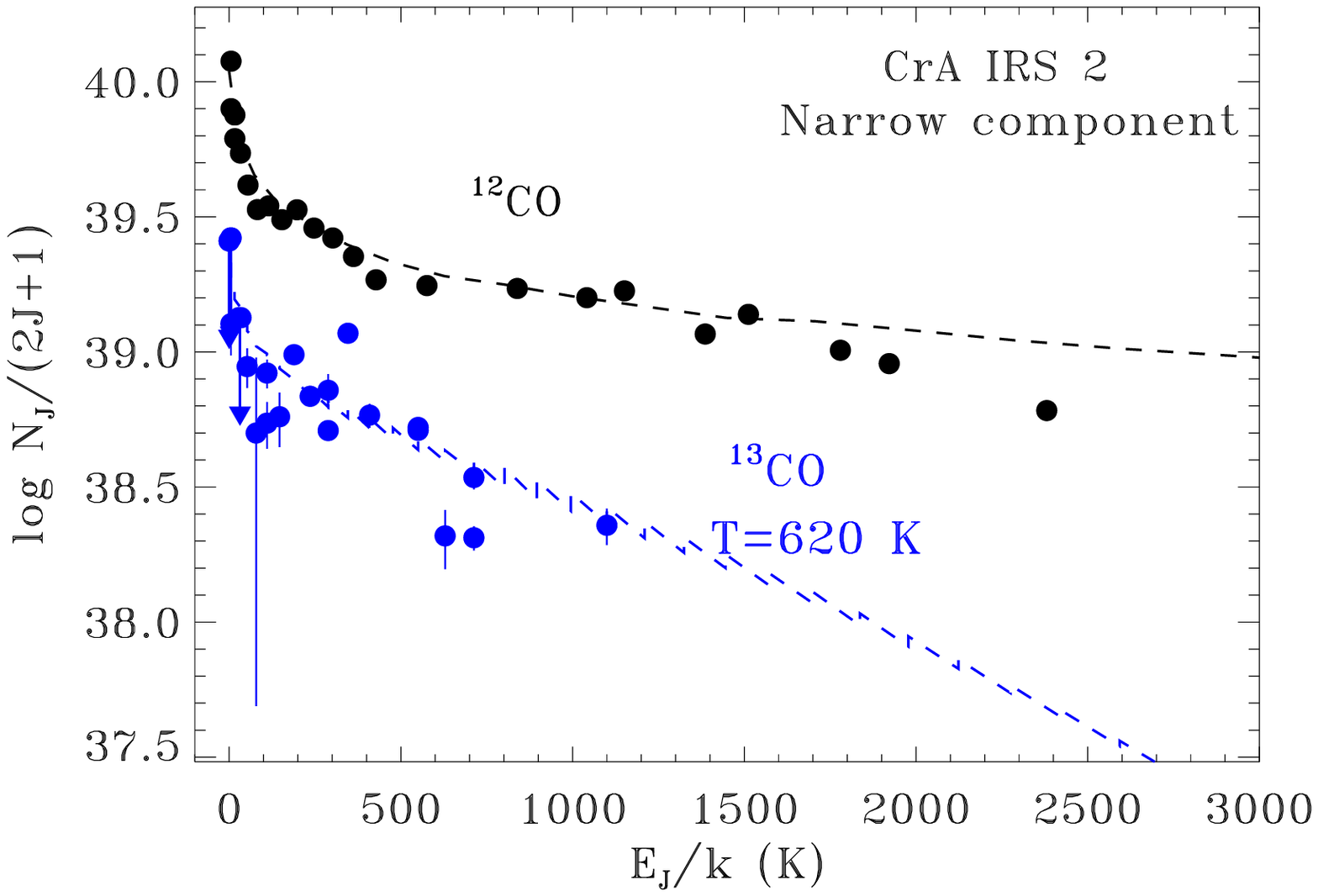}
\includegraphics[width=60mm]{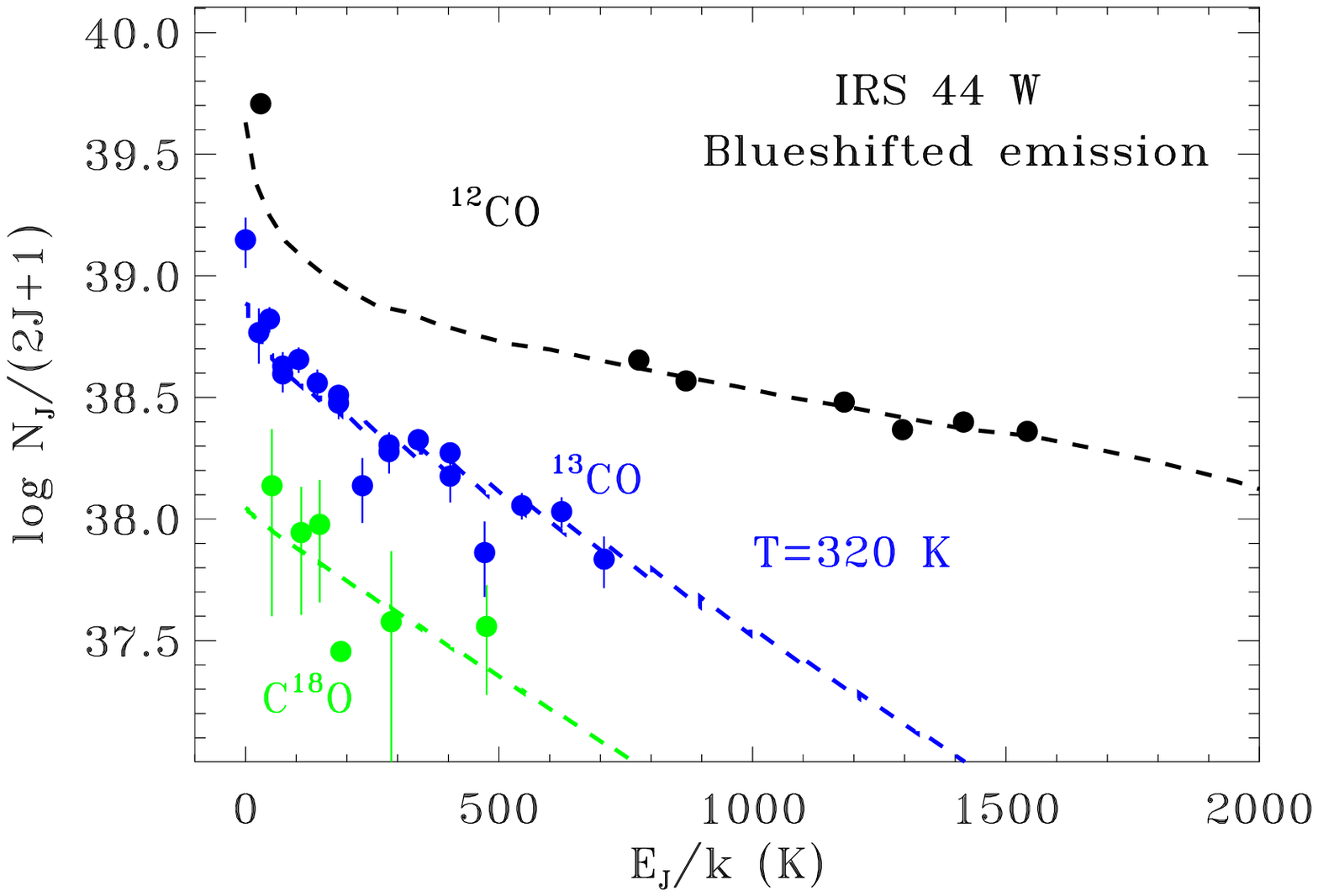}
\includegraphics[width=60mm]{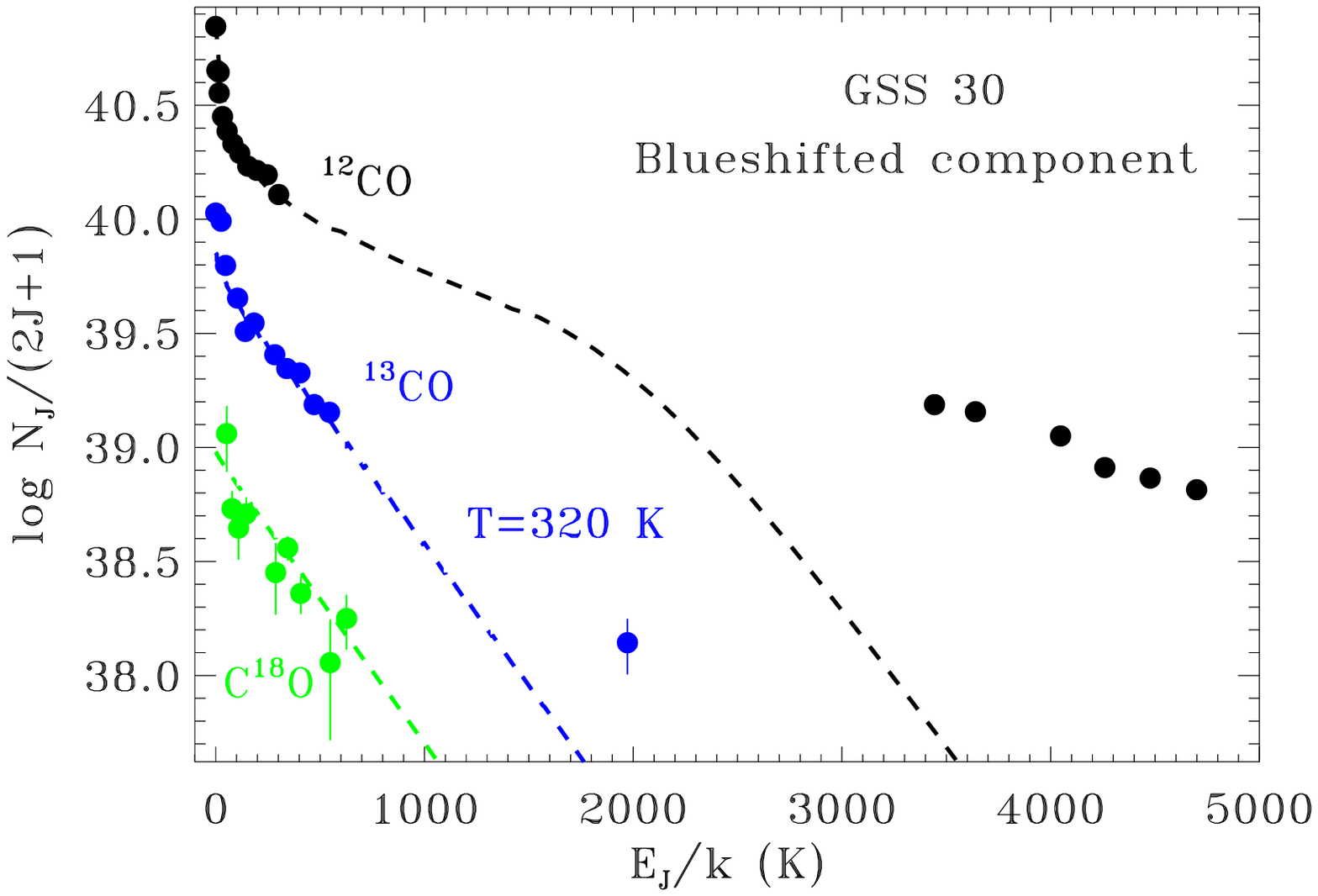}
\includegraphics[width=60mm]{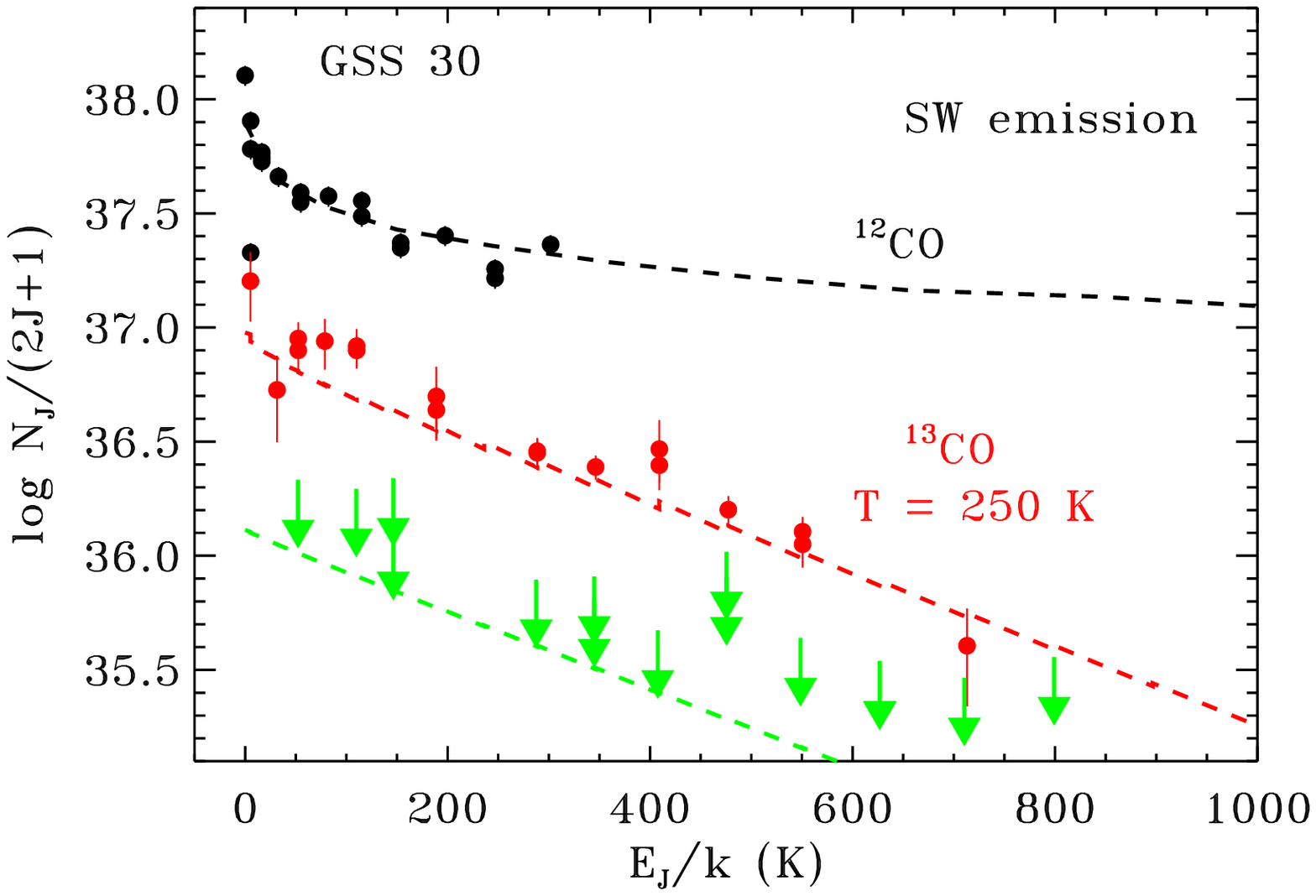}
\caption{Plots of apparent column density versus excitation for CO emission for the broad and narrow emission components and spatially-extended emission.  For the $^{12}$CO v=1-0 lines, the synthetic fluxes are shown only for the P transitions.  The dashed lines show the excitation expected from the $N$(CO) and $T$, based on models described in \S A.4.}
\label{fig:rots.ps}
\end{figure*}

Figure \ref{fig:rots.ps} shows the rotational diagram for narrow emission from GSS 30, IRS 44 W, and CrA IRS 2.  
Relative to the broad emission component, the narrow emission component is produced by gas that is cooler, with temperatures of 250--600 K, and more optically-thick, with column densities of $\log N$(CO)$\sim 19$.  However, as indicated by the range in velocity centroids, the narrow component may be produced by different processes for different objects.  
For IRS 44 W and GSS 30, the temperature and column density from the rotational diagrams apply only to the blueshifted emission because $^{13}$CO emission is not detected on the red side of the line profile.  

\begin{figure}
\includegraphics[width=90mm]{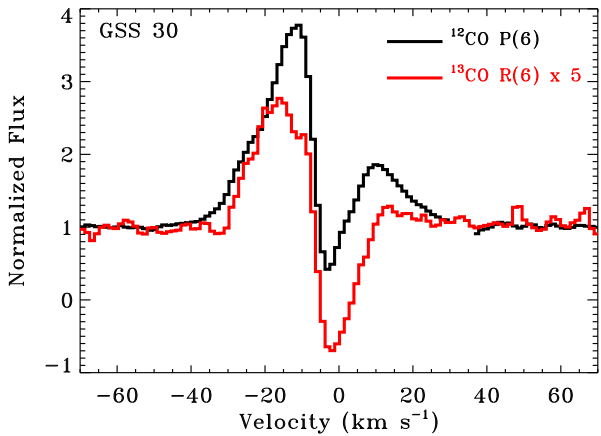}
\caption{A comparison of the $^{12}$CO P(6) and $^{13}$CO R(6) line profiles from GSS 30.  The two profiles should be similar, including absorption from the same lower level.  The absorption in the $^{13}$CO line is deeper and redder than that seen in the $^{12}$CO line, despite that the $^{13}$CO line cannot be more optically-thick than the $^{12}$CO line.  Instead, we infer the presence of redshifted absorption from infalling gas at 10--20 \kms.  The broader emission is produced far enough from the central source that it does not travel through the infalling gas and consequently does not suffer from absorption.}
\label{fig:gss30infall}
\end{figure}

Figure~\ref{fig:gss30infall} compares the $^{12}$CO P(6) line with the $^{13}$CO R(6) line for GSS 30.  The two lines should have similar emission and absorption properties, modulated by the different optical depths.  However, the $^{13}$CO absorption is actually broader on the red side of the line than the $^{12}$CO absorption.  The most likely explanation for this discrepancy is that the CO emission suffers less from CO absorption than does the continuum.  The stronger $^{12}$CO emission then fills in the absorption more than the weaker $^{13}$CO emission.  The absorption, detected out to +10 \kms, indicates the presence of infalling gas in our line of sight to GSS 30.

\subsection{Spatially Extended CO Emission}
The M-band spectra of both GSS 30 and IRS 43 show CO emission extended on arcsecond scales.  We concentrate here on the extended CO emission from GSS 30 because it is much brighter than that from IRS 43.  The slit position angle for the GSS 30 observation roughly aligns with the $\sim 45^\circ$ position angle of the outflow \citep[e.g.][]{Allen2002}.

\begin{figure}
\includegraphics[width=90mm]{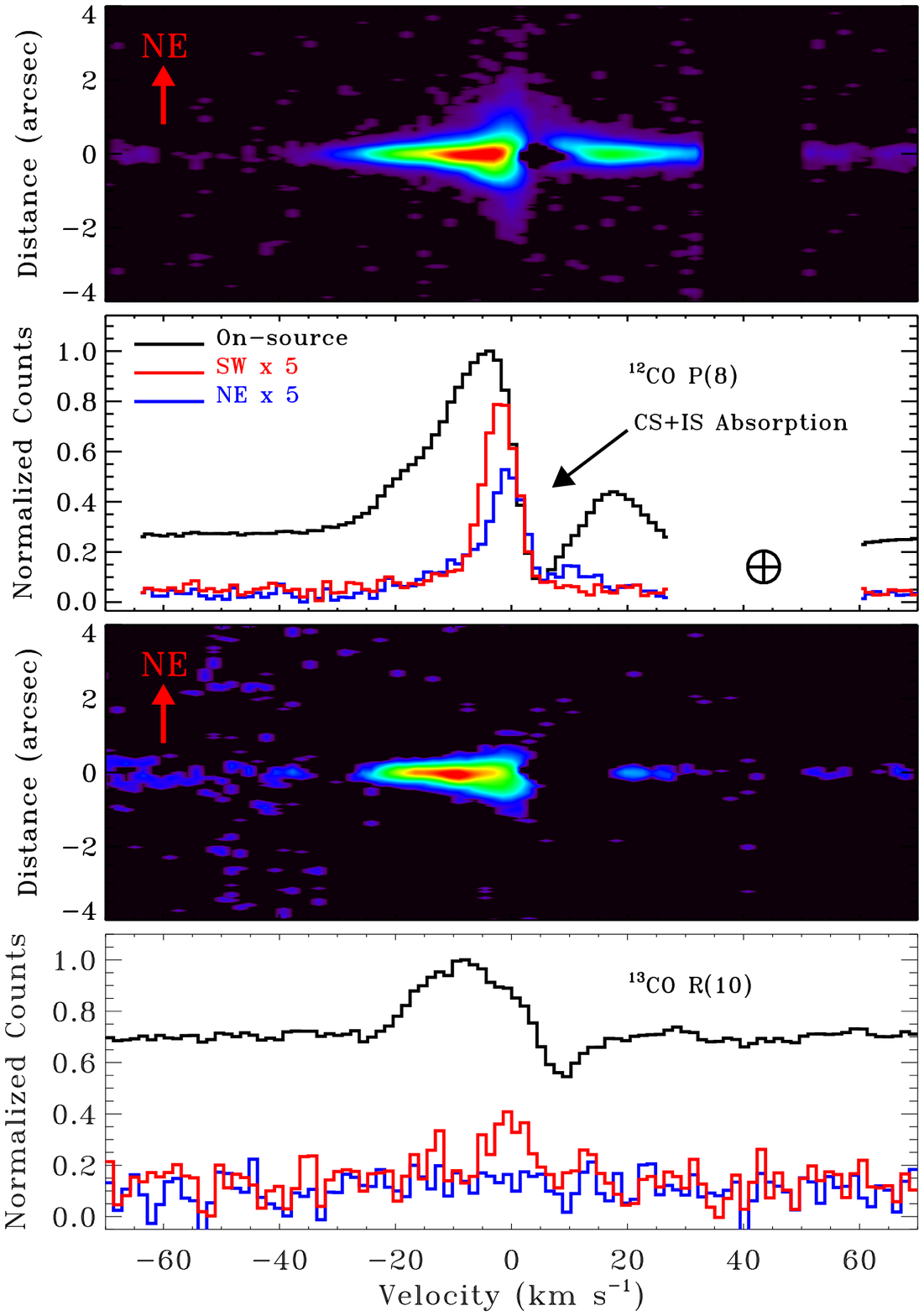}
\caption{Spectra around the $^{12}$CO $v=1-0$ P(8) and $^{13}$CO $v=1-0$ R(10) lines from GSS 30 seen on-source (black line), to the SW of the star (red line, multiplied by 5), and to the NE of the star (blue, multiplied by 5).  The off-source spectra are extracted from $0\farcs47-0\farcs93$ from the central source.  The 2D spectral images are obtained from coadding images over $^{12}$CO and $^{13}$CO lines.  Both CO and continuum emission are detected off-source in both directions along the slit.  The emission is strongest to the SW.  The off-source CO line emission is very narrow (FWHM$\sim 6.4$ \kms) and likely suffers from CO absorption in our line of sight.}
\label{fig:gss30_spatspecs}
\end{figure}

Figure \ref{fig:gss30_spatspecs} shows that emission in low-$J$ $^{12}$CO lines is detected out to $2\farcs2$ (265 AU at 120 pc) from the central source in both the NE and SW directions.  Emission in $^{13}$CO is also detected off-source in the SW direction and is barely detected to the NE.  Co-adding C$^{18}$O, CO $v=2-1$, and CO $v=1-0$ high-$J$ lines each yields marginal detections of emission to the SW and non-detections to the NE.  Most of the extracted emission is in narrow line profiles (FWHM~$\sim$~7 \kms) centered $2-3$ \kms\ blueward of the CO absorption.  A weak, broad (FWHM~$\sim20$~\kms) base is also detected in the line profile.  The CO absorption likely attenuates any emission on the red side of the $^{12}$CO line profiles.

\begin{figure*}
\includegraphics[width=60mm]{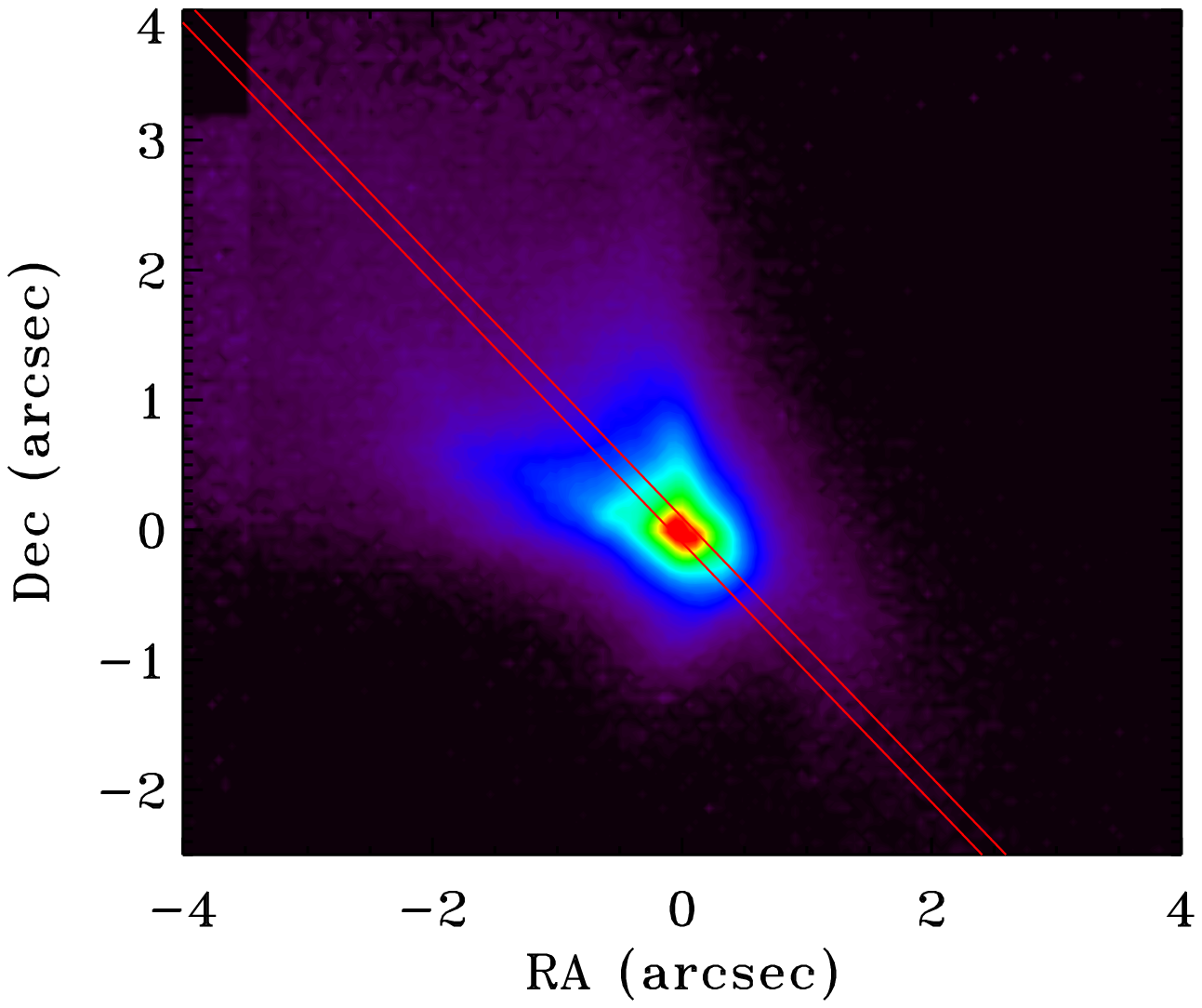}
\includegraphics[width=60mm]{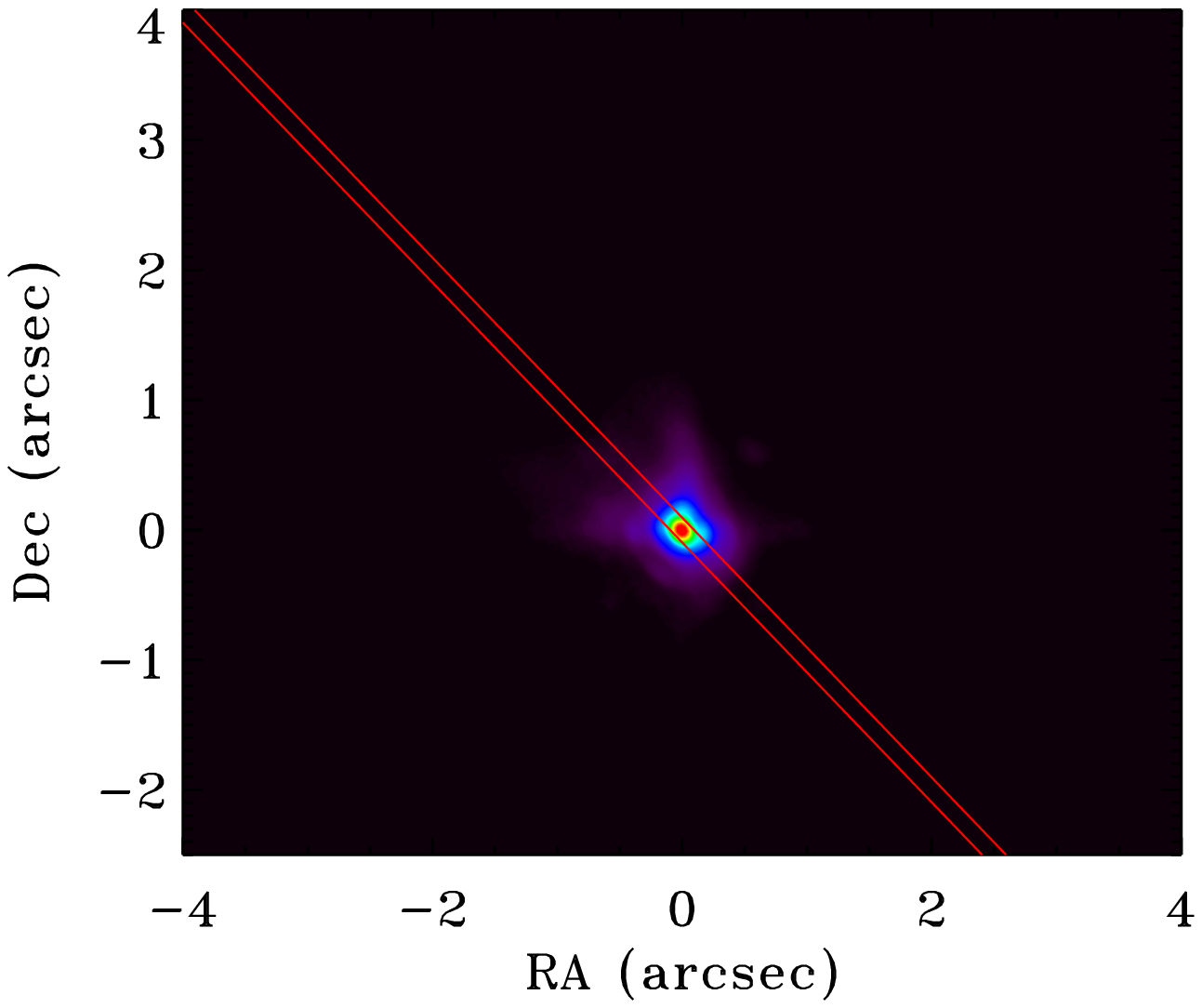}
\includegraphics[width=60mm]{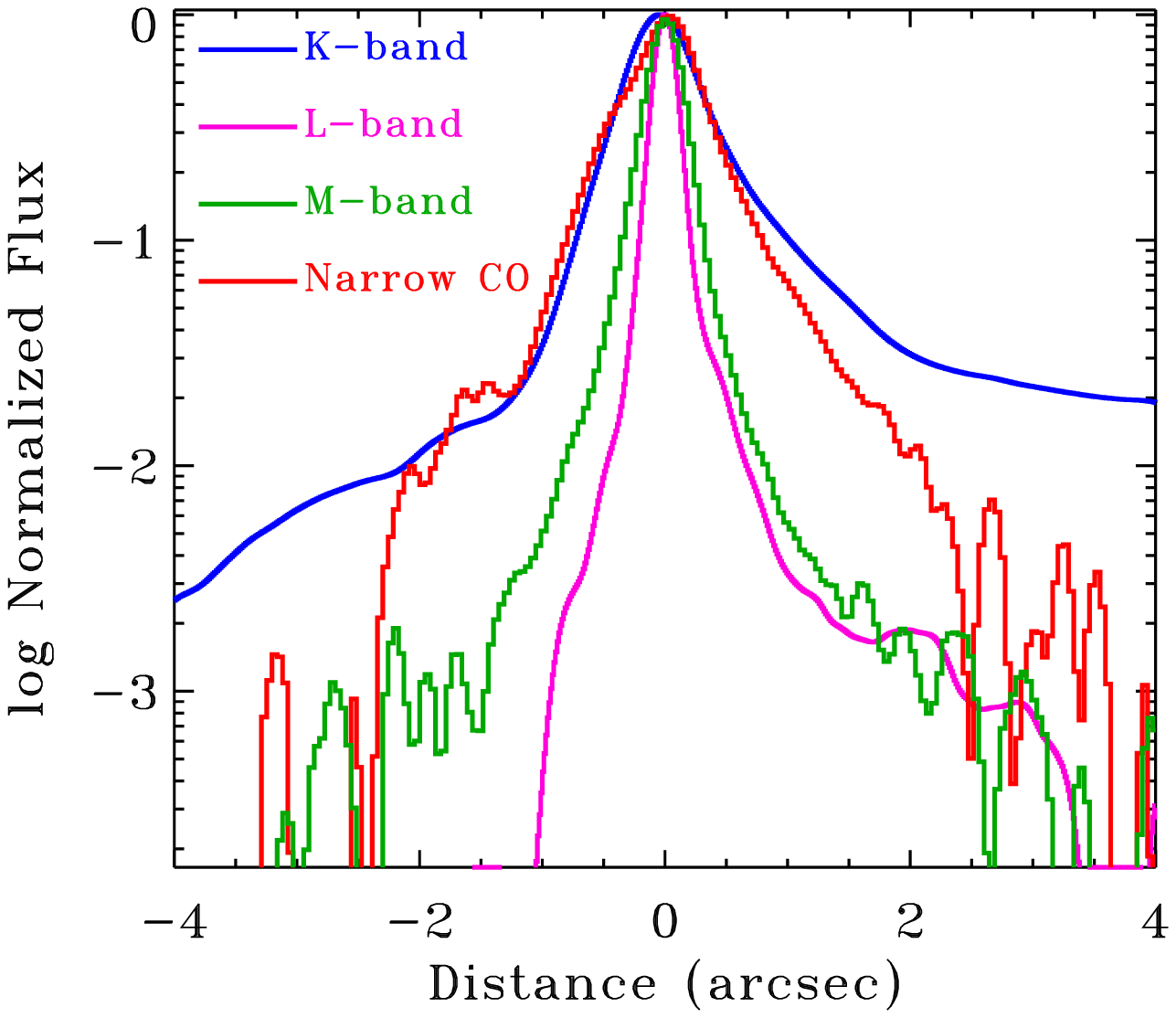}
\caption{{\bf Top:}  {\it VLT}/NACO K-band (left) and L-band (right, provided by Duch\^{e}ne) images of GSS 30, with the CRIRES slit in red. Nebulosity is detected to the NE and SW of the resolved central source.  {\bf Bottom:}  The cross-dispersion profile of K-band, M-band, and narrow CO emission from GSS 30 along the CRIRES slit.  The K-band emission is extracted from the NACO image, deconvolved to the $0\farcs3$ seeing during our CRIRES observation.  The M-band continuum and the on-source CO emission are unresolved.  That the M-band continuum emission is broader than the L-band continuum emission may be the result of broad wings on the CRIRES point-spread function, which we do not account for here.  The extent of the narrow CO emission is somewhat similar to the K-band nebulosity.}
\label{fig:gss30_slit}
\end{figure*}

Figure \ref{fig:gss30_slit} compares the extent of emission in CO and the M-band continuum to the spatial extent in archival L-band \citep{Duc07} and K-band images (3.6 and 2.2 $\mu$m, respectively) obtained with {\it VLT}/NACO, deconvolved to the approximate spatial resolution of our CRIRES observations.  The K-band continuum and CO line emission is much more spatially extended than the L- and M-band continuum emission.  The blue K-L color of the nebulosity suggests that the continuum emission is seen in reflected light.

Figure \ref{fig:rots.ps} shows the rotational diagram for CO line fluxes in the SW direction, extracted over a region from $0\farcs47$ to $0\farcs93$.  Each continuum pixel in this region has $\sim 2$\% the flux of the central pixel.  A linear fit to the $^{13}$CO emission yields a temperature of $250\pm30$ K.  The low $^{12}$CO/$^{13}$CO line flux ratio and the curved shape of the $^{12}$CO rotational diagram indicates that the $^{12}$CO emission is optically-thick, with $\log N($CO$)\sim18.7$.  With these parameters, the C$^{18}$O is just below our detection limit.  The line fluxes are reproduced with an emission area of (3.7 AU)$^2$, which is two orders of magnitude smaller than the approximate extraction region of $0\farcs43 \times 0\farcs 3$, or $\sim (43$ AU)$^2$ at 120 pc.  Lowering the Doppler $b$ parameter from 2.0 to 0.2 \kms\ decreases the column density to $\log N($CO$)\sim17.8$, thereby increasing the emission area to (11 AU)$^2$, which is still ten times smaller than the extraction region.  This discrepancy between emitting area and extraction area is somewhat surprising because the CO emission appears to be smoothly distributed in the slit and the spatially-extended K-band continuum emission is smoothly distributed in the K-band image.

\subsection{Non-Detections of CO Emission}

Four objects, WL 6, IRS 44 E, Elias 32, and IRS 43 N, are undetected in CO emission.  For these four objects, the presence of strong CO emission relative to the continuum can be definitively ruled-out in our CRIRES spectra.  However, the spectra are inconclusive regarding the presence of weak CO emission.  Each non-detection is discussed in detail in the following paragraphs.

No CO emission is detected in our CRIRES spectrum of WL 6.  The telluric absorption falls on the red side of the line.  In several other cases (WL 12, Elias 29, IRS 63), the CO $v=1-0$ emission is detected only on the red side of the line because interstellar and wind absorption both occur in the same transitions.  Some emission, with a strength that may be slightly above the continuum level, is detected between the telluric absorption and interstellar CO absorption.  This weak emission could be either CO or an artifact from the telluric correction.  If real, the weak CO emission would have blended with the interstellar/circumstellar absorption at low-resolution, leading to a non-detection in the ISAAC spectrum \citep{Pon03}.  The CRIRES spectrum is consistent with weak CO emission.

The non-detection of CO emission from IRS 44 E is complicated by the strong CO emission from IRS 44W.  Weak emission is seen from the deep absorption to +50 \kms, with a peak-to-continuum that reaches $\sim 0.05$.  However, this marginal detection is not considered significant because it may be an artifact  of the non-standard spectral extraction from this close binary (see \S 2.2).  Regardless of whether this emission is attributable to IRS 44 E, most of the CO emission from the IRS 44 system is produced by the W component.  Any CO emission from the E component is at least 5 times weaker than that from the W component.

Bright emission was previously detected in high-$J$ CO lines in an ISAAC spectrum of Elias 32 \citep{Pon03}, so our non-detections of CO emission from Elias 32 on three different nights$^5$ is surprising.
Since CRIRES would spectrally resolve CO emission from the narrow absorption line, the emission component should be detected at velocities that lack CO absorption.  The equivalent width required to explain the CO emission in the ISAAC spectrum is inconsistent with the non-detection of CO emission in the CRIRES spectrum of Elias 32.  Either the equivalent width of the CO emission is variable or the ISAAC spectrum detected mostly extended CO emission.  Some CO is detected in blueshifted absorption to Elias 32 (see \S 3.6).
\footnotetext[5]{We observed Elias 32 three times, with each spectrum having low S/N.  The highest S/N spectrum for Elias 32 was obtained on 3 May 2008, when the telluric CO absorption was located at +23 \kms.  As with WL 6, the non-detection of CO emission on that date is not significant.  From the spectrum obtained on 10 August 2008, we find a limit that the emission on the red side of the central absorption is less than 7\% above the continuum flux.  Weak CO emission from several other embedded objects would not have been detected with this S/N.}

The object IRS 43 N is a very faint companion to IRS 43 S.  The non-detection is not significant, although the equivalent width must be at least 3 times smaller than the equivalent width of CO emission seen from IRS 43 S.

\subsection{H$_2$ S(9) Emission}
\begin{figure}[!t]
\includegraphics[width=90mm]{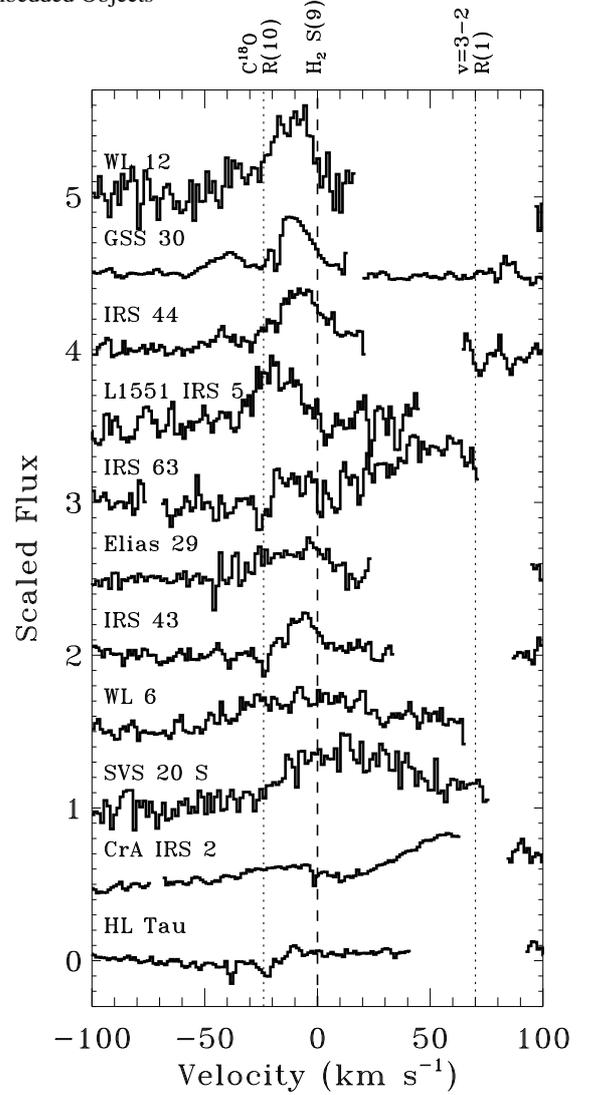}
\caption{H$_2$ line profiles from our sample, including non-detections from CrA IRS 2 and HL Tau.  The H$_2$ emission from SVS 20 S and WL 6 should be considered as tentative detections because the line is very broad.}
\label{fig:h2specs}
\end{figure}

\begin{figure}[t]
\includegraphics[width=40mm]{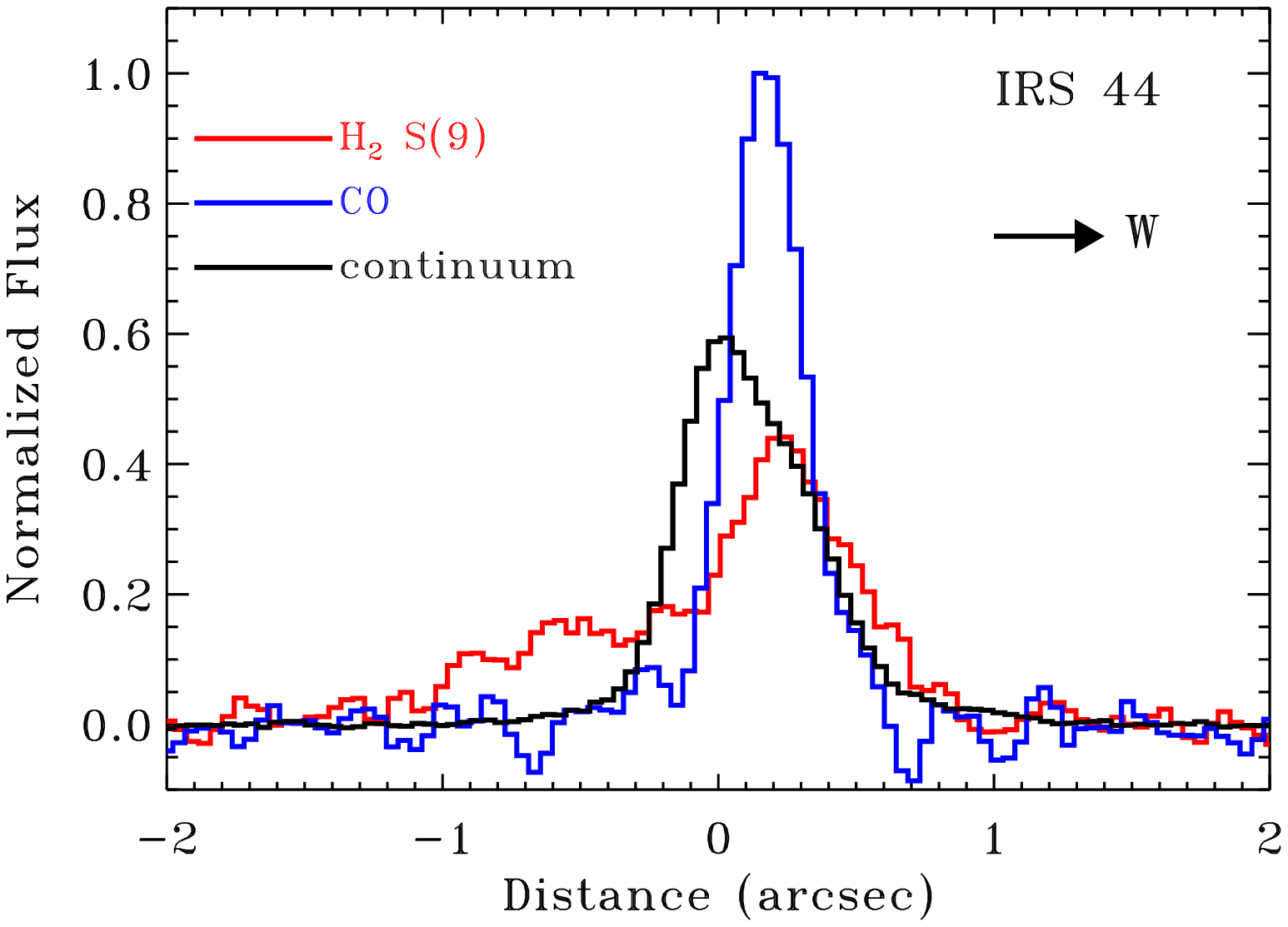}\includegraphics[width=40mm]{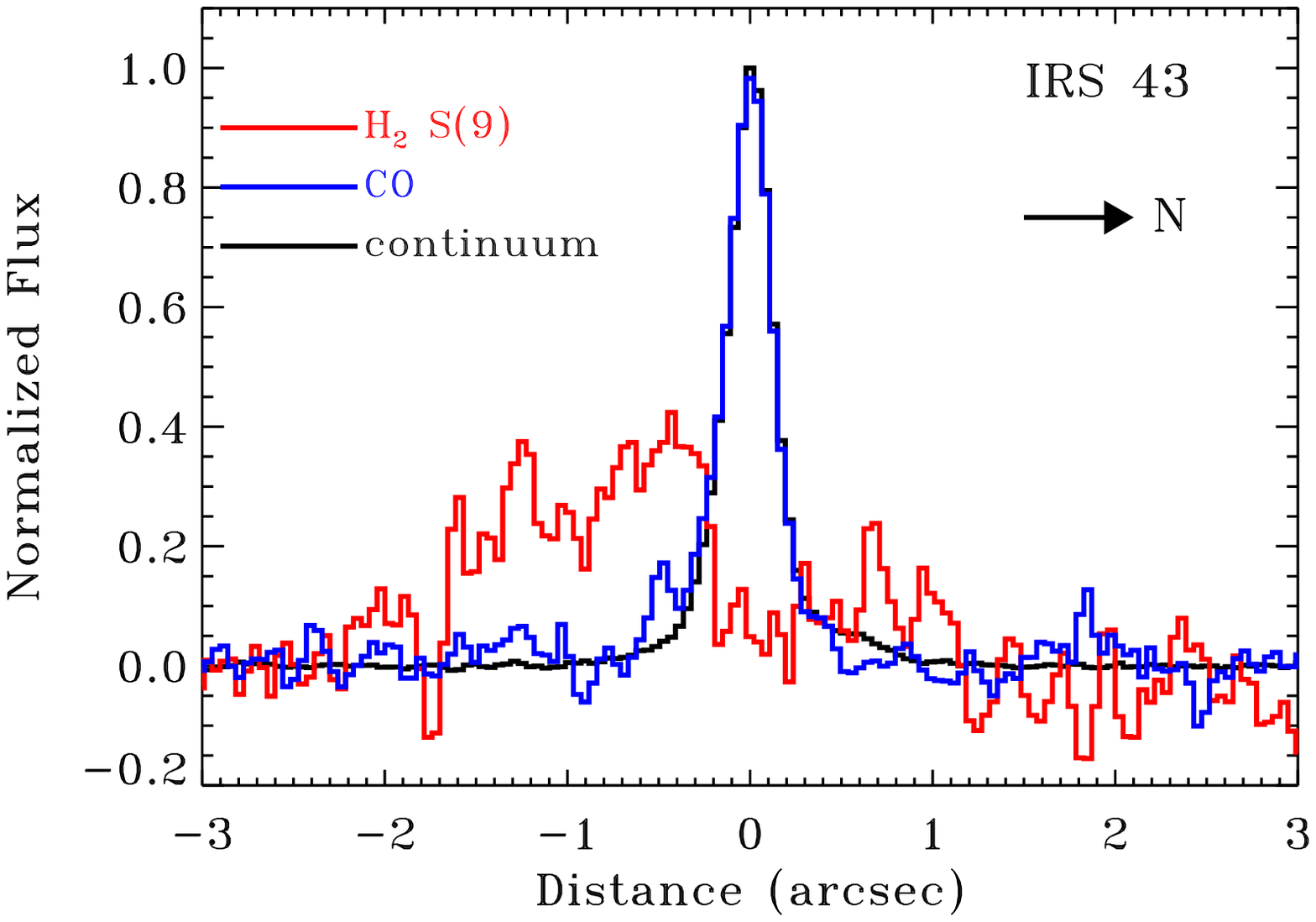}
\includegraphics[width=40mm]{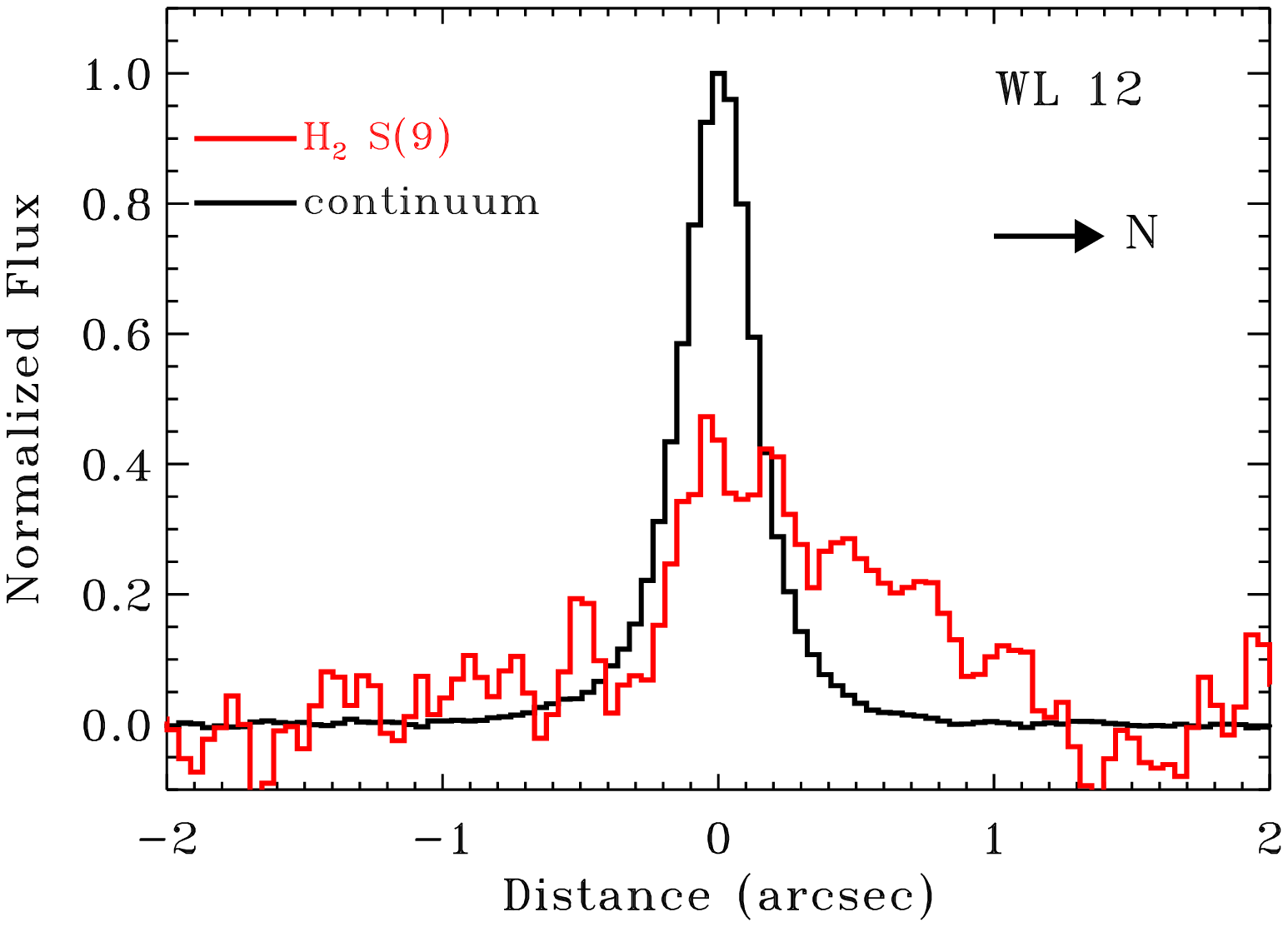}\includegraphics[width=40mm]{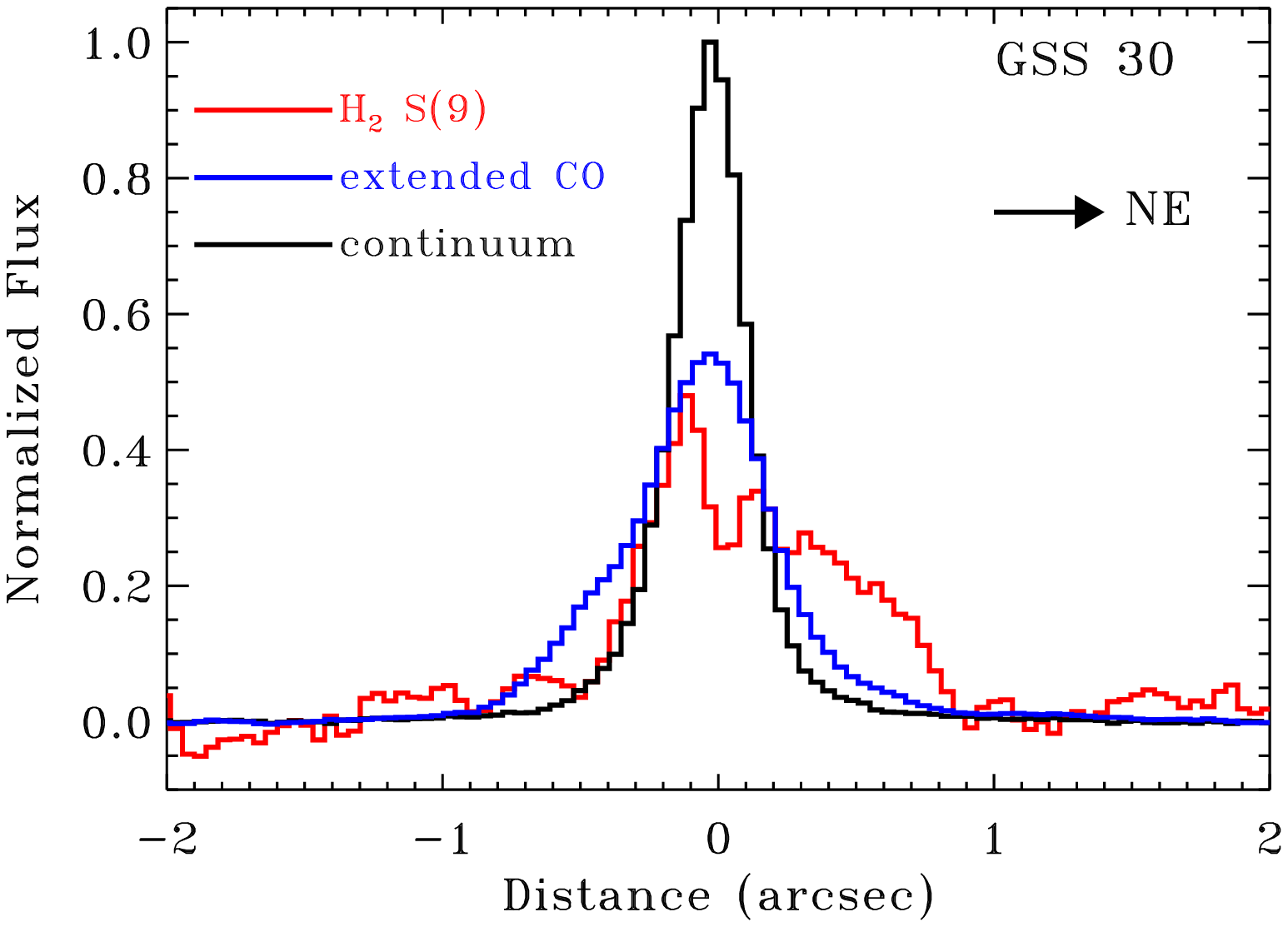}
\caption{Spatial profiles of M-band continuum emission, coadded $^{12}$CO $v=1-0$ line emission, and H$_2$ S(9) emission from four sources.  The H$_2$ emission from these four sources is much more extended than the CO and continuum emission, and is reminiscent of H$_2$ $v=1-0$ S(1) emission from outflows \citep{Beck08}.  The H$_2$ emission from Elias 29, L1551 IRS 5, IRS 63, and WL 6 is not extended and not shown here.  The CO emission from GSS 30 shown here is the blueshifted component between -20 and -10 \kms.  The CO emission on the line wings is not spatially extended beyond the continuum emission.}
\label{fig:spats}
\end{figure}

Emission in the H$_2$ S(9) line at 4.6947 $\mu$m is detected from 9 objects in our sample (see Figure~\ref{fig:h2specs} and Table~\ref{tab:h2specs}).  Most H$_2$ lines have small or negligible blueshifts and a FWHM of $\sim 20-30$ \kms.  
Figure~\ref{fig:spats} shows the spatial distribution of the H$_2$ emission in the cross-dispersion direction, along with the spatial distribution of continuum and CO emission.   The H$_2$ emission from five stars (IRS 63, Elias 29, SVS 20 S, L1551 IRS 5, and WL 6) is not extended along the slit direction and is consistent with a point source.  The H$_2$ emission is extended in the slit for the other 4 objects with detected emission (GSS 30, WL 12, IRS 43, IRS 44 W).  The observed H$_2$ emission from IRS 43 S is produced by molecular gas that extends $>1\farcs5$ ($>190$ AU) to the S of the central object.  The H$_2$ emission from WL 12 extends $1\farcs2$ ($\sim 150$ AU) to the S of the object but is not seen to the N.  

Some of the H$_2$ emission from IRS 44 is associated with IRS 44 W, similar to the case for CO emission.  Weak H$_2$ emission is extended by up to $\sim 1\farcs1$ ($\sim 140$ AU) to the E of IRS 44 W.  The equivalent width of H$_2$ emission is $\sim 50\%$ smaller on our night of good seeing than on our two nights with poor seeing and than in the ISAAC spectrum (Table \ref{tab:crires_isaac.tab}).  This result confirms that the S(9) emission from IRS 44 is spatially-extended because poor seeing increases the ratio of off-slit emission to on-slit emission.

\begin{table}
\caption{H$_2$ S(9) Line Detections}
\label{tab:h2specs}
{\footnotesize \begin{tabular}{lccccccccc}
\hline
Target        & $v_{H2}$$^a$ & FWHM & EW & Extended?\\
 & \kms & \kms  & \kms & \\
\hline
\multicolumn{5}{c}{Detected H$_2$ lines}\\
\hline 
IRS 44 E+W$^b$ &  -4.6(0.5) &  22(3)  & 9.4(2.0) & Y \\
IRS 44 E+W$^c$ &  -4.9 (0.9) &   23(5) & 9.9(2.4)  & Y \\
IRS 44 E+W$^d$    &  -6.6 (0.6)  &  25(3) & 6.7(1.3) & Y \\
Elias 29             & -9.5 (2.5)  &  26 (8) & 0.9 (0.2)  & N$^e$ \\
IRS 43 S+N              &  -5.1 (1.6)  & 18(4)   & 1.7(0.3) & Y \\
IRS 63$^f$               & -10.2 (3.5)  & 14(4)   & 0.7(0.3) & N \\
WL 12$^f$                & -17(1)       &  26(3)  & 16(2) & Y \\
L1551 IRS 5    & -17 (3) &  22(5) & 5.4(1.4)  & N$^e$ \\
GSS 30$^g$            &  -7.9 (0.7)   & 18(3.5) &  3.2(0.4) &Y  \\
\hline
\multicolumn{5}{c}{Broad H$_2$ lines: spurious detections?$^h$}\\
\hline
WL 6                 &  1.5(2.0)   & 89(10)  & 6.9(1.7) & N\\
SVS 20 S        & 19 (2)  & 70(14) & 5.3 (1.0) & N \\
\hline
\multicolumn{5}{c}{Non-detections}\\
\hline
Elias 32 &  \multicolumn{4}{c}{low S/N}\\
CrA IRS 2 &  \multicolumn{4}{c}{blended with CO}\\
HH 100 IRS  &  \multicolumn{4}{c}{No detection, telluric correction?}\\
HL Tau   & \multicolumn{4}{c}{No detection}\\
RNO 91   &  \multicolumn{4}{c}{No detection}\\
TMC 1A  & \multicolumn{4}{c}{No detection, bad telluric correction}\\
SVS 20 N & \multicolumn{4}{c}{low S/N}\\
\hline
\multicolumn{5}{l}{$^a$Relative to velocity of $^{13}$CO and C$^{18}$O absorption.}\\
\multicolumn{1}{l}{$^b$2008-04-27} & \multicolumn{2}{l}{$^c$2008-04-30} & \multicolumn{2}{l}{$^d$2008-08-07}\\
\multicolumn{5}{l}{$^e$Low FWHM in continuum for CRIRES observation}\\
\multicolumn{1}{l}{$^f$low S/N} &
\multicolumn{4}{l}{$^g$C$^{18}$O line detected at -30 \kms}\\
\multicolumn{5}{l}{$^h$Uncertain detection because of poor telluric correction}\\
\end{tabular}}
\end{table}

Some on-source non-detections are not significant.  
The CO 3--2 R(10) line at 4.6958 $\mu$m could mask any weak on-source H$_2$ emission from CrA IRS 2 and HH 100 IRS.  The non-detection of H$_2$ emission from Elias 32 is limited by poor S/N.
On the other hand, \citet{Bit08} detected S(9) emission from HL Tau at a level that should have been detectable in our spectrum$^6$.  The H$_2$ S(9) emission reported in \citet{Bit08} could be spatially-extended and not located within our slit, similar to the spatial distribution of H$_2$ 1-0 S(1) 2.12 $\mu$m emission from HL Tau \citep{Takami2007,Beck08}.  However, in our 2010 observation of HL Tau, the slit was aligned with the position angle of the outflow and the H$_2$ 1-0 S(1) 2.12 $\mu$m emission, yet no extended H$_2$ emission was detectable.
The \citet{Bit08} spectrum of HL Tau includes a stronger unidentified feature redward of the S(9) line, which may alternately point to a spurious detection in their spectrum.
\footnotetext[6]{Using a FWHM=12 \kms, as measured by \citet{Bit08}, for a Gaussian line centered at the systemic velocity, we measure an equivalent width upper limit of $\sim 0.1$ \kms, roughly 2.5 times weaker than that detected in the TEXES spectrum.  Some emission could be located at -10 to -30 \kms\ in the on-source spectrum but not be detectable because of a poor telluric correction.  Our March 2010 observation of HL Tau has a poor telluric correction at the H$_2$ line, presumably because of variable telluric absorption and is not usable for the analysis of on-source emission.  No off-source H$_2$ emission was detected in that observation.}

The broad, redshifted H$_2$ emission from SVS 20 S is not confirmed in the ISAAC spectrum and may be a spurious detection introduced by a poor telluric correction.  A strong telluric CO$_2$ absorption line is located at $+85$ \kms\ of the H$_2$ line (+118 \kms\ relative to the source velocity at the time of the observation).  The H$_2$ emission from WL 6 is also particularly broad, with a FWHM of 89 \kms.  The ISAAC spectrum includes a similar detection for WL 6, so this broad emission may be real.

The H$_2$ S(8) line at 5.0529  $\mu$m is included in a wavelength setting that was used to observe only GSS 30.  Any emission in this line is severely blended with strong emission in the CO $v=1-0$ P(37) line.  The H$_2$ S(8)/S(9) line flux ratio upper limit of $\sim 0.6$ indicates that the emission is produced in gas hotter than $\sim 700$ K.

\subsection{CO Wind Absorption}

\begin{figure}[!t]
\includegraphics[width=80mm]{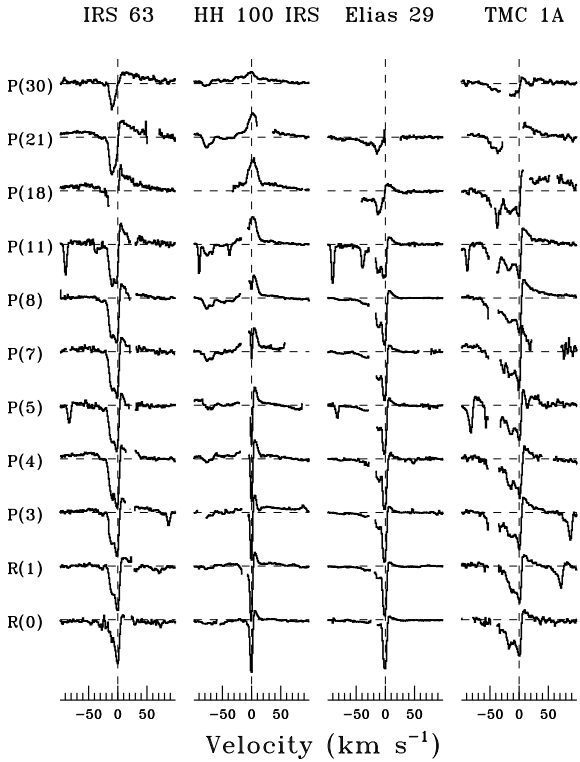}
\caption{Wind absorption towards four sources in $^{12}$CO 1--1 lines, with high- to low-$J$ from top to bottom.}
\label{fig:windplot}
\end{figure}

Within our sample, six objects (HH 100 IRS, IRS 63, WL 12, Elias 29, Elias 32, and TMC 1A) show wind absorption in CO $v=1-0$ lines (see Figs.~\ref{fig:gallery} and~\ref{fig:windplot}).  The wind to HH 100 IRS, TMC 1A, and Elias 29 is seen to velocities of $\sim 100$ \kms.  The wind absorption to Elias 32, IRS 63, and WL 12 is slower, with speeds up to $\sim50$ \kms.  These differences, and the lack of CO wind absorption in the majority of our sample, may be attributable to the different viewing inclinations, compositions, and wind velocities between sources.

\begin{figure*}[!t]
\includegraphics[width=65mm]{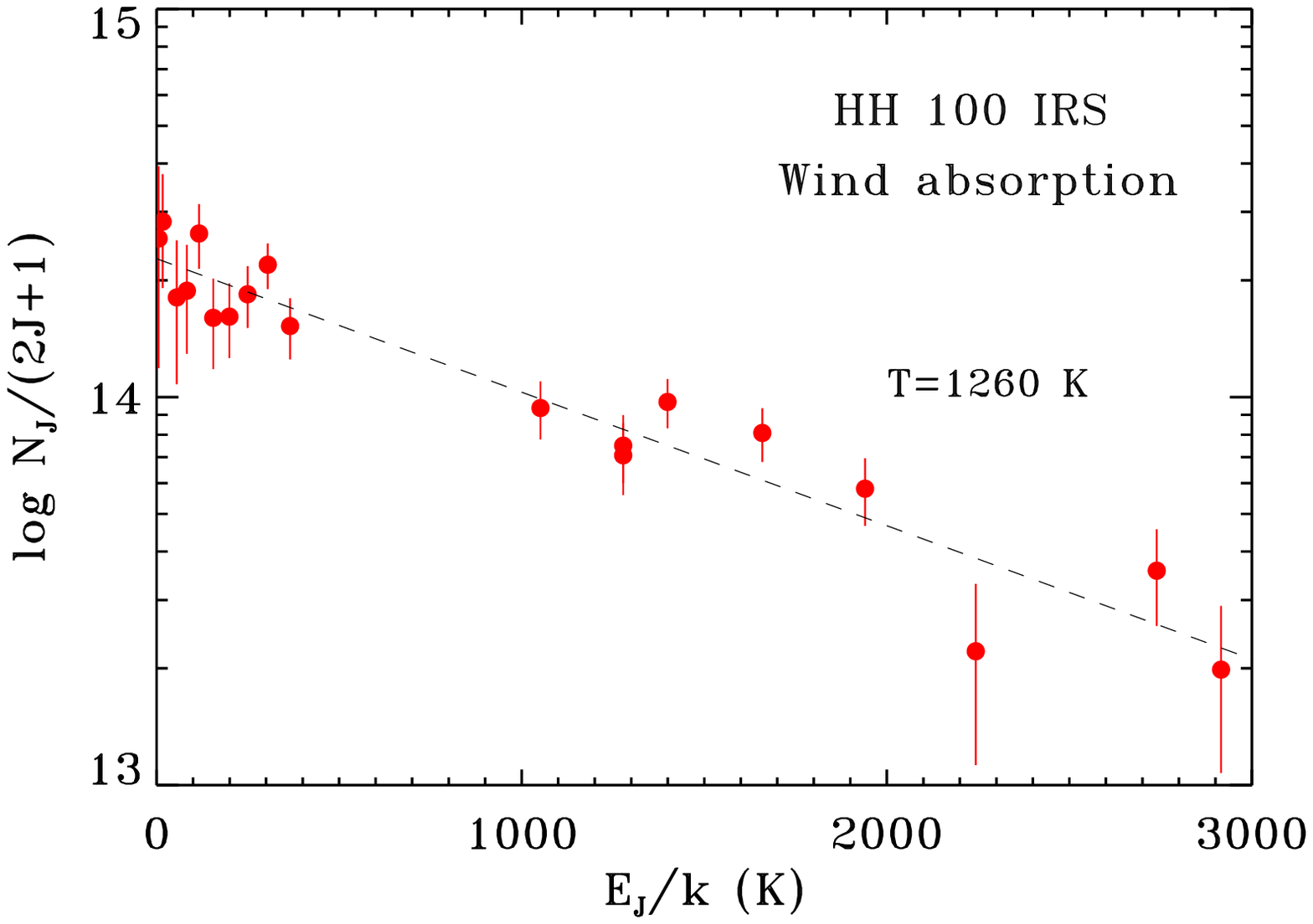}
\includegraphics[width=65mm]{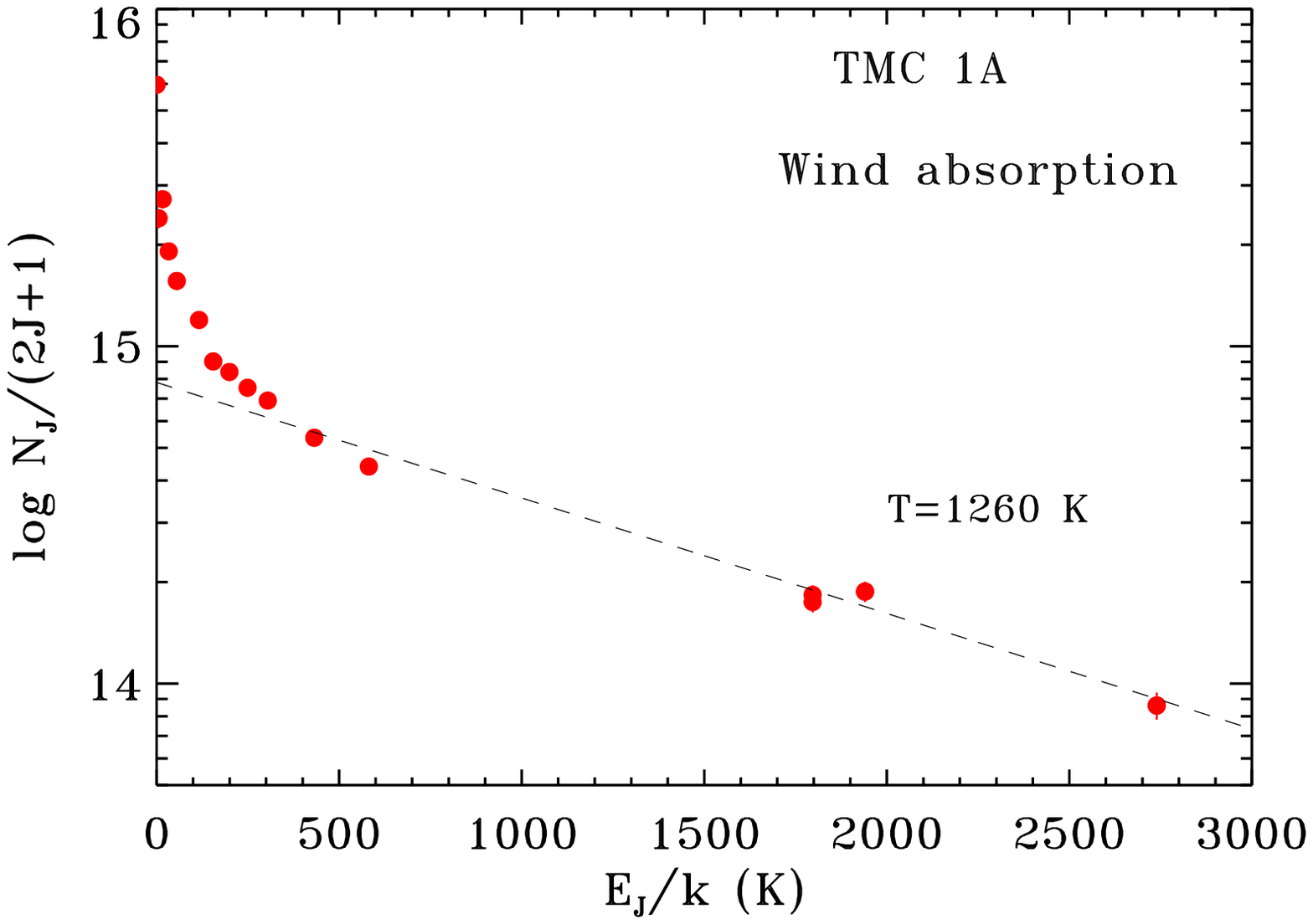}
\includegraphics[width=65mm]{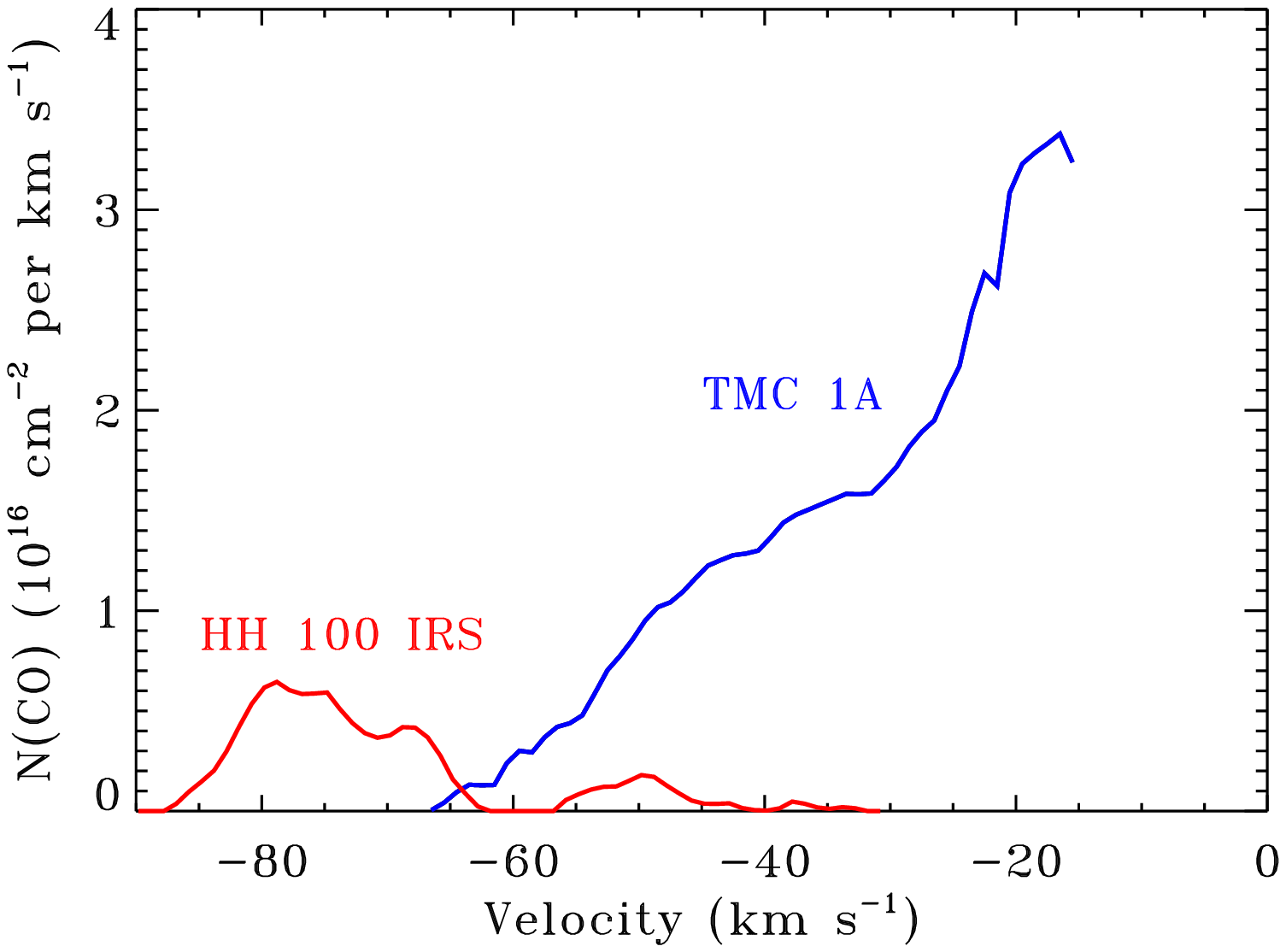}
\caption{Rotational diagram for wind absorption from HH 100 IRS (left) and TMC 1A (middle) covering the -55 to -80 \kms\ region that includes the double-peaked wind absorption.  The temperature of the absorbing gas for both stars is $1260 \pm 100$ K.  The right panel shows the column density of CO in 1 \kms\ bins.}
\label{fig:hh100_windtemp}
\end{figure*}

 The wind absorption is typically weakest at $J^{\prime\prime}=0$ and peaks in strength at $J^{\prime\prime}$=10 to 20.
For HH 100 IRS, the absorption is strongest in two distinct components at -65 to -85 \kms.  The absorption seen to TMC 1A is strongest at low velocity and becomes undetectable at $\lesssim -65$ \kms.  To calculate temperatures and column densities, we 
assume that the wind absorption is optically thin.  An alternate explanation, which applies to the slow wind seen from TMC 1A described below, is that the wind only attenuates a fraction of the M-band continuum emission.  In this case,  the absorption in low $J$ and high $J$ lines would still need to be optically thin because the absorption depth is largest for lines with $10<J<20$.  For TMC 1A, an optically-thick, low-velocity ($<20 \kms$) component is seen in low-$J$ lines and in $^{13}$CO lines.  Since the spectrum does not go to 0 at low velocities in the $^{12}$CO lines, the low-velocity absorption must obscure only about half of the M-band continuum emission produced by the inner disk.

The temperature of the wind for  HH 100 IRS and TMC 1A is $1260\pm100$ K and $1260 \pm40$ K, respectively (Fig.~\ref{fig:hh100_windtemp}).  Fig.~\ref{fig:hh100_windtemp} shows the column density of CO in 1 \kms\ bins in the winds seen to HH 100 IRS and TMC 1A.  The log of the total CO column density to TMC 1A, summed from 20--65 \kms, is 19.6, and for HH 100 IRS, summed from 60--90 \kms, is 18.4.

\begin{figure*}[!t]
\includegraphics[width=85mm]{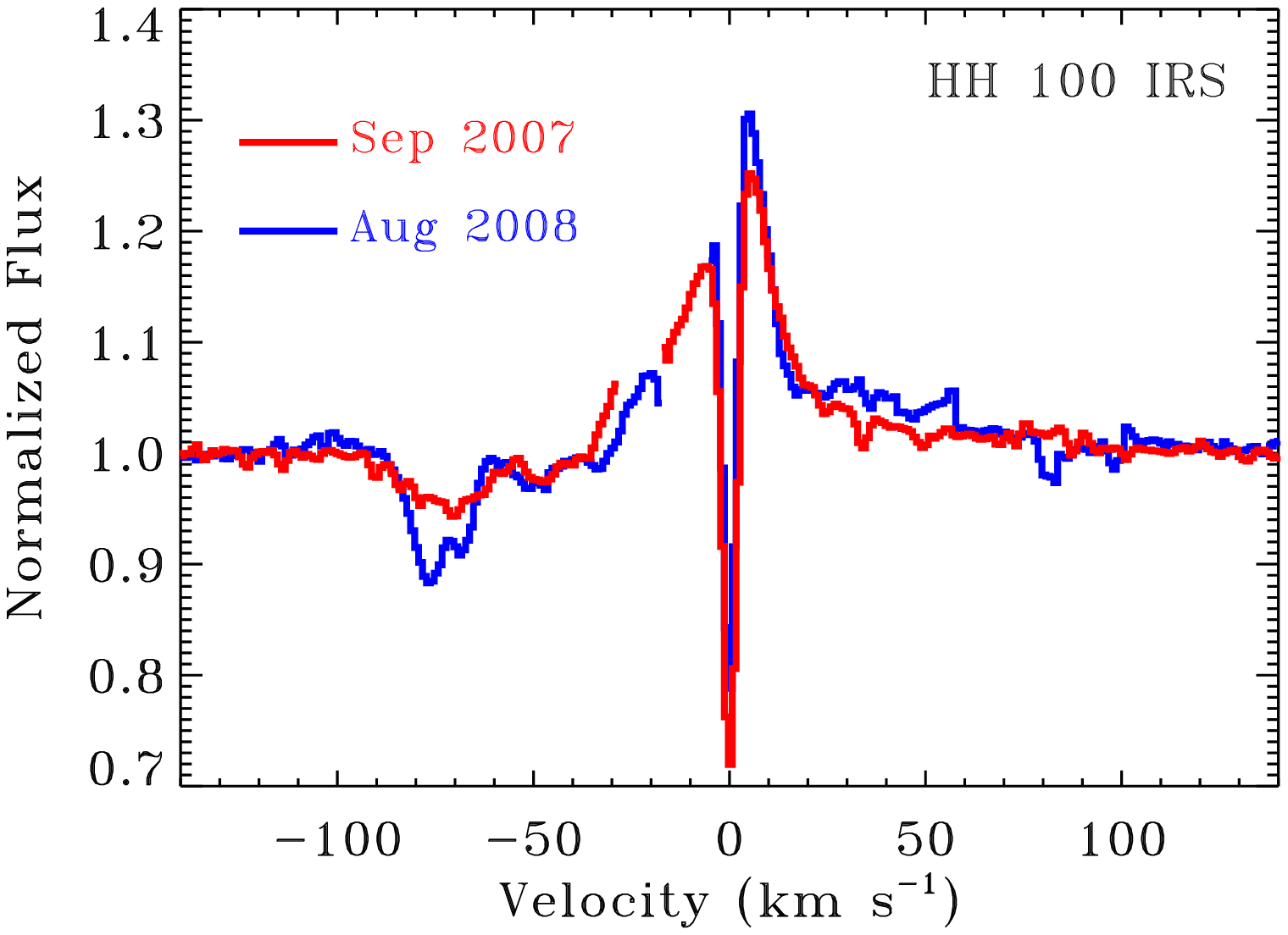}
\includegraphics[width=85mm]{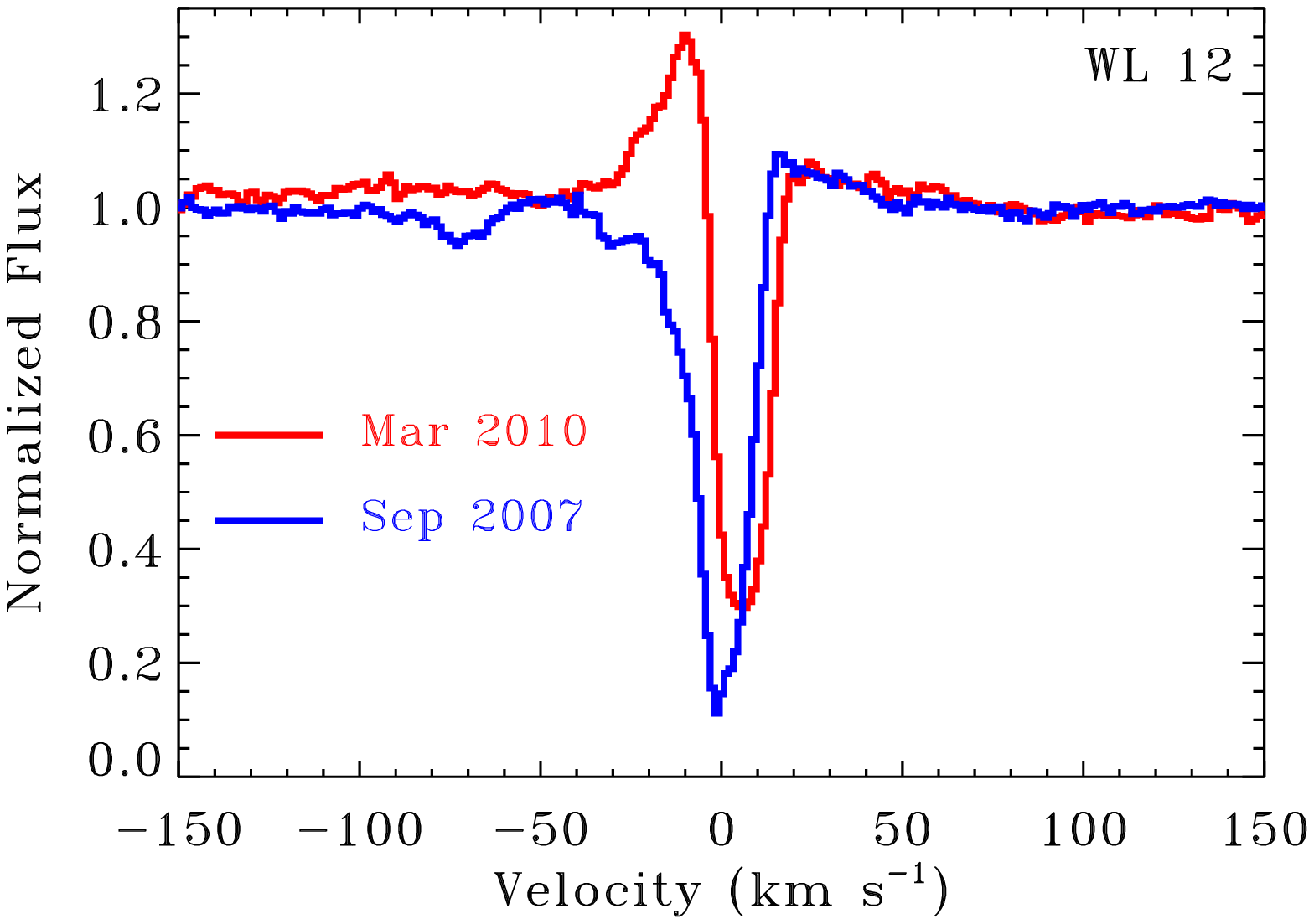}
\caption{{\bf Left:}  Comparing the CO fundamental line profile from HH 100 as seen in observations separated by 11 months.  The profiles shown here are the co-added lines of P(4), P(6), P(7), and P(8).  No significant difference is seen in the emission line profile between the two epocsh.  In August 2008 the wind absorption at -75 \kms\ was deeper than that seen in September 2007.  {\bf Right:}  Co-added CO emission from two observations of WL 12.  The observation of September 2007 covered low-J lines while the observation of March 2010 covered high-J lines, so the comparison of line profiles is not direct.  Some of the wind absorption in  2007 appears to go into emission in 2010.}
\label{fig:hhvar}
\end{figure*}

Figure~\ref{fig:hhvar} shows CO $v=1-0$ emission from observations of HH 100 IRS obtained 11 months apart, and of WL 12 obtained 3 years apart.  For HH 100 IRS the blueshifted CO absorption changes between the two observations, with stronger absorption at $\sim -70$ \kms\ in August 2008.  The shape and strength of the emission line profile remained the same, to within observational uncertainties.  For WL 12, some of the absorbing gas appears to have gone into emission, which could happen with a change in viewing angle of the wind or if the opacity of the wind decreased, allowing some emission to escape.

\section{DISCUSSION}

In \S 3, we describe four different components of CO rovibrational lines from embedded young stars: (1)  broad, vibrationally-excited emission (\S 3.1), (2) very narrow, spatially-extended emission (\S 3.3), (3) absorption from winds and envelopes (\S 3.6), and (4) narrow, optically-thick emission (\S 3.2).  In the following subsections, we discuss the physical interpretations of the different components of warm CO gas in the YSO.

\subsection{Disks in Broad CO Emission}

The broad CO emission seen from the embedded objects in our sample is similar to the warm CO emission seen from CTTSs.  Of the 12 CTTSs with detected CO $v=1-0$ emission in the \citet{Naj03} sample, nine are also detected in CO $v=2-1$ lines.  For these CTTSs, the FWHM range from 56--134 \kms\ and the line fluxes range from $(0.5-14)\times10^{-14}$ erg cm$^{-2}$ s$^{-1}$ ($0.3-10 \times 10^{-5}$ $L_\odot$ at 140 pc).
Similarly, in our sample most objects with broad CO $v=1-0$ line emission also show broad CO $v=2-1$ line emission, with FWHM that range from 55--140 \kms\ and line fluxes that range from $\sim (2-10) \times10^{-14}$ erg cm$^{-2}$ s$^{-1}$ ($2-10 \times 10^{-5}$ $L_\odot$ for a 120 pc distance and extinction $A_M=1$ mag.).  In the CTTS sample of \citet{Naj03}, $^{13}$CO emission was only detected to one source.  In our spectra, the broad component is not detected in $^{13}$CO lines.  The measured temperatures are roughly consistent with the CO temperatures in the CTTS sample in \citet{Salyk09}.    The centroids of the spectral line profile are consistent with the source velocities, and the emission is  centered on the source and not spatially extended beyond $\sim 10$ AU.    The most likely origin for this broad component of CO emission from embedded sources is the disk.

That vibrational levels $v=2,3,4$ are populated despite rotational temperatures of $\sim 1000$ K supports the disk interpretation for this emission, since large populations in the $v\geq 2$ level of CO typically requires FUV photoexcitation \citep{Krot80,vandis88,Bri07}.   Electronic (A-X) transitions of CO in the FUV are much stronger than the rovibrational transitions and become optically-thick at smaller optical depths.  As a consequence, gas that produces UV-excited emission from $v^\prime\geq 2$ will not be detectable in $^{13}$CO.  The lack of broad $^{13}$CO emission is therefore consistent with the presence of UV-excited gas and places a limit of $\log N$(CO)$<18$ in this inner disk region.  

Emission from the $v^\prime=2$ level has previously been detected from embedded young stars in $v=2-0$ overtone transitions from WL 16 \citep{Car93}, in overtone and fundamental emission from CTTSs \citep{Naj03,Bast10} and Herbig AeBe stars \citep{Bri07,Bri09,vanderPlas2009}.  That emission from $v^\prime>1$ traces the broad component is consistent with the far-UV field being strongest near the star. In contrast, far-UV-excited CO rovibrational emission from Herbig AeBe stars is usually produced at larger disk radii.
UV radiation fields from embedded young stars are not directly detectable but should be sufficiently strong to excite warm CO in the inner disk.

\begin{figure}
\includegraphics[width=90mm]{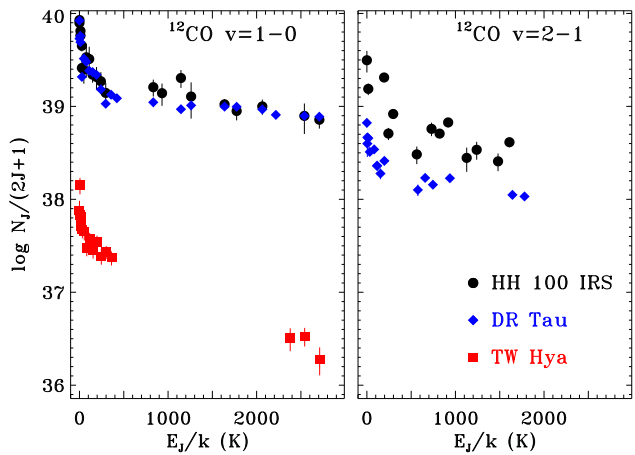}
\caption{A comparison of $^{12}$CO emission from the broad component of HH 100 IRS (black circles), DR Tau (blue diamonds), and TW Hya (red squares).  The emission in $^{12}$CO $v=1-0$ is very similar for HH 100 IRS and the CTTS DR Tau, with much weaker emission seen from the older CTTS TW Hya.  CO $v=2-1$ emission is stronger from HH 100 IRS than DR Tau.}
\label{fig:hhdrtw}
\end{figure}

Figure ~\ref{fig:hhdrtw} compares the rotational diagram of number of CO molecules for HH 100 IRS, DR Tau, and TW Hya.  The fluxes for TW Hya were measured by \citet{Sal07}.  The fluxes from DR Tau were measured from the CRIRES spectrum presented in \citet{Bast10}.  TW Hya is an $\sim8$ Myr star \citep{Mamajek2005} with a disk that has an inner dust hole of $\sim 4$ AU \citep{Calvet2002} and an accretion rate of $\sim 1.5\times10^{-9}$ M$_\odot$~yr$^{-1}$ \citep{Herczeg2008}.  
DR Tau is a younger star with an accretion rate of $3\times 10^{-7}$ M$_\odot$~yr$^{-1}$ \citep{Gullbring2000}, comparable to the $10^{-6}$ M$_\odot$~yr$^{-1}$ estimated for HH 100 IRS by \citet{Nis05}.  The strength of CO $v=1-0$ emission from DR Tau is remarkably similar to that from HH 100 IRS.  The inner disk structure of HH 100 IRS is probably somewhat similar to that of DR Tau and other CTTSs with high accretion rates.  The CO emission from both HH 100 IRS and DR Tau is much stronger than that from TW Hya, perhaps because TW Hya has a smaller accretion rate and therefore less viscous heating in the inner disk.

Double-peaked profiles are absent within our sample and are also somewhat rare for CTTSs.  
Of the many previously published samples of CO emission from CTTSs, only AA Tau, SR 21, GQ Lup, RNO 90, Elias 23 (this work, see Fig.~5), and DF Tau are resolved as double-peaked profiles$^7$ \citep{Naj03,Sal07,Pont08,Salyk09,Naj09,Hug09,Bast10,Brown11}.  Most CO lines from embedded objects are consistent with Gaussian profiles.  Some Gaussian profiles may be consistent with a Keplerian disk viewed face-on, as is the case for CO lines from several CTTSs \citep{Pont08,Goto2011}.
\footnotetext[7]{\citet{Naj03} suggest that the CO emission from RW Aur is double-peaked, but this inference was based on modeling a small spectral region that did not show a clear double-peaked line profile.  We find no evidence for a double-peaked CO emission profile in a CRIRES spectrum of RW Aur A (Brown et al., in preparation).}

To investigate the absence of double-peaked line profiles, \citet{Bast10} selected a subsample of 8 CTTSs with bright CO lines and high accretion rates.  In that sample, the combination of the narrow central peak and the lack of any spatially-extended emission, with upper limits of a few AU, indicates the presence of some gas near the star that is not in Keplerian rotation.   The subset of embedded objects that have low bolometric luminosities is  likely similar to the sample of Bast et al..  In both, the broad line widths indicate an origin near the where the inner disk is likely truncated.   The fluxes suggest an emitting area of $\sim 1$ AU$^2$.  \citet{Bast10} were unable to explain the line fluxes, profiles, and lack of any spatially-extended emission with an origin in the surface layers of a warm disk in Keplerian rotation and with a surface temperature described by a power law. One possible explanation attributes much of the line flux to a disk wind \citep{Pont10}.

Regardless of explanation, the large velocities in the line profile are likely explained by Keplerian broadening.  For the disk explanation, the Keplerian broadening is {\it in situ}, while for the disk wind explanation the Keplerian broadening would apply to the inner radius of the launch region.
Within the context of the disk/disk wind interpretation, the inner radii $R_{in} (\sin i)^2$ for the CO emission are calculated from the velocity of the best-fit Gaussian profile at 20\% the peak flux (Table~\ref{tab:inner.tab}) and a central mass $1$ $M_\odot$.  Although a direct measurement of this location in the line profile is typically not possible in our sample because of insufficient S/N, all broad line profiles are consistent with a Gaussian profile.  About 93\% of the warm CO emission is produced beyond this inner radius.  Alternately, we solve for inclination by setting $R_{in}=0.05$ AU for every source.  The listed inner radii and inclinations suffer from significant uncertainties.

\subsection{Molecules in Winds$^{8}$}

Molecular emission is often detected in winds from young stars \citep[e.g.][]{Shang2007,Beck08,Davis2010}.  As described in \citet{Panoglou11}, winds can have a  molecular component for one of the following reasons:  (i) entrainment of molecular gas in the envelope or cloud by the wind, (ii) molecular formation within the wind itself, or (iii) launching of molecular gas in a disk wind.  Typically, observations of  molecules in winds lack sufficient spatial resolution to discriminate between these possibilities.  Our data includes diagnostics of the wind near the launch region and also of the slow-moving molecular gas extended from the star.
\footnotetext[8]{For simplicity, the term {\it winds} used here is intended to incorporate all disk winds, MHD winds, and jets from a YSO.}

\subsubsection{Molecules launched in the wind}

The fast blueshifted CO absorption detected to HH 100 IRS, Elias 29, and TMC 1A suggests that the MHD wind itself is at least partially molecular when it is launched.  The slower CO absorption detected to Elias 32, WL 12, and IRS 63 could be explained by either an MHD disk wind or a slower thermal disk wind. 
When detected, wind absorption usually occurs close to the star because the density within the wind decreases with radius squared as the wind expands.  Molecular gas from the envelope that gets entrained in the outflow is highly unlikely to produce absorption close to the star and at such high velocities.  CO wind absorption in the M-band has also been detected from high-mass YSOs \citep{Mitchell1990,Thi2010}.  The presence of CO in the wind suggests that the wind from these sources favors launching from the disk rather than the stellar chromosphere or corona \citep{Matt2005,Ferreira2006}.  Although any significant mass loss from the star itself would have to be cooler than $10^4$ K \citep{Matt2007}, a chromospheric wind would likely be mostly neutral.  As an alternative to a wind that is molecular where it is launched, \citet{Glassgold1991} suggest that CO, but not H$_2$, may form within the wind itself.  However, in their models a high CO abundance requires a mass loss rate $>10^{-5}$ $M_\odot$ yr$^{-1}$, which is larger than the mass loss rates inferred for typical Stage 1 sources \citep{Bontemps1996}.

Models of MHD disk winds by \citet{Panoglou11} demonstrate that molecules can survive within the wind when the mass loss/accretion rate are sufficiently large ($M_{acc}\sim10^{-6}$ $M_\odot$ yr$^{-1}$), in material with an ionization fraction that is high enough to couple the molecular gas to the magnetic field.   The measured temperature of $1260\pm100$  K is roughly consistent with the predicted wind temperatures for objects with accretion rates of $10^{-7}-10^{-6}$ $\dot{M}$ yr$^{-1}$. 
Wind absorption in CO has not been detected previously to CTTSs.  Several CTTSs show on-source FUV H$_2$ emission with velocities up to $50-100$ \kms\ \citep{Her06}.  The frequency of CO absorption in winds from embedded objects and deficiency of CO absorption in winds from CTTSs is consistent with the survival of CO requiring large outflow rates to shield the outflow from irradiation by the central star.

That CO wind absorption is detected toward only some of the objects in our sample is consistent with the MHD wind being non-spherical and with a relatively wide opening angle near the star.   
In a survey of the He~I $\lambda10830$ line from CTTSs, \citet{Edwards2006} found that winds from stars seen pole-on tend to be faster and more optically-thick than winds from stars seen closer to edge-on.   Within our sample,  CO wind absorption is not detected to the two stars with the broadest CO emission lines, indicating that perhaps the same relation holds.  On the other hand, the narrow CO line emission from CrA IRS 2 indicates a low disk inclination, but no CO absorption is detected.  
The lack of CO wind absorption from HL Tau, with an inclination of 65--70$^\circ$ \citep{Close1997,Lucas2004}$^{9}$, also contrasts sharply with the deep, fast He~I absorption seen to the star \citep{Edwards2006}.  
 For both CrA IRS 2 and HL Tau, the likely interpretation is that the molecular fraction is low in the wind.  
For HL Tau, the discrepancy between the deep He I wind absorption and the lack of any detectable CO wind absorption could also be explained if 4.8 $\mu$m continuum is seen through a very different line-of-sight than the 1.1 $\mu$m continuum emission. 
\footnotetext[9]{\citep{Fur08} suggest an $i\sim15^\circ$ for the HL Tau disk from modelling the SED, but we consider the near-IR polarimetry a more reliable method to measure inclination because inclination is degenerate with many other parameters in broadband SED fitting and because the combination of high extinction and lack of ice absorption suggests disk attenuation rather than envelope extinction.}

\subsubsection{Spatially-extended molecular emission}

Powerful jets and winds from young stars can carve out cavities within the circumstellar envelope \citep[e.g.,][]{Whitney93, Woo01,Ybarra2006}.  
The interaction region between the cavity and the jet/wind entrains some cold gas, which is likely seen in  bipolar molecular outflows with velocities of a few \kms.  At the surface of the cavity wall, gas can be heated by shocks and UV photons from the central star \citep[e.g.][]{Spaans95,TvK09}.

Within our sample, H$_2$ 0--0 S(9) emission is detected in nine objects.  For four of these detections the H$_2$ emission is spatially-extended, always asymetrically about the star.
The spectral line profiles are typically centered within 10 \kms\ of the systemic velocity.  The extended component is likely produced in the envelope or nearby cloud material that is shocked by powerful outflows.  \citet{Greene2010} detected H$_2$ 1--0 S(0) and 1--0 S(1) emission from all 17 embedded young objects in their sample, of which 10 have emission that is that is spatially extended on 2--3$^{\prime\prime}$ scales.  
Compared with the H$_2$ 0-0 S(9) line reported here, the H$_2$ 1--0 S(0) and 1--0 S(1) emission lines have similar FWHM (20--30 \kms) and central velocities that differ by $\sim 5-10$ \kms.

Of the objects with extended H$_2$ emission, the slit PA was well-aligned (better than $20^\circ$) with the outflow axis only for GSS 30. 
The spatial distribution is likely analogous to the distribution of warm H$_2$ emission seen towards other embedded YSOs and environmentally-young CTTSs that drive powerful outflows \citep[e.g.][]{McCaughrean1997,Davis2001,Wal03,Beck08,Neufeld2008,Lah10,Greene2010}.  
The spatially unresolved H$_2$ emission could be produced by a disk, as is the case for weak H$_2$ rovibrational emission from a few CTTSs \citep{Bary2008}.  However, because the H$_2$ emission often includes a significant contribution from surrounding envelope/cloud material, a careful analysis would be required to use the line as a disk probe.  
In light of our results, the origin of H$_2$ emission from GSS 30, and perhaps Elias 29 and HL Tau, by \citet{Bit08} should be considered the wind/envelope interaction region rather than the disk.

Spatially-extended CO emission is also detected from two objects, GSS 30 and IRS 43, that show extended H$_2$ emission.  The extended CO line emission traces cooler gas and is spectrally more narrow than the H$_2$ line emission.  
That extended CO emission is not detected more frequently is likely explained by densities that are much lower than the 
critical density of $~5\times 10^{12}$ cm$^{-3}$ required to populate the $v=1$ levels needed to produce CO rovibrational emission \citep{Naj96}, although instead the CO/H$_2$ abundance ratio could be negligible.  On the other hand, the presence of extended CO emission from two objects may indicate high densities in the associated outflows.

\subsection{Disks and/or Outflows in Narrow CO Emission?}

Narrow CO emission is detected from 9 of 18 embedded objects within our sample.  The optical depth of this component is larger than that in the broad component, as evidenced by the high $^{13}$CO/$^{12}$CO line ratios.  Narrow line profiles have a wide range of properties, indicating that this classification is overly broad.  For GSS 30, IRS 44 W, WL 12, and SVS 20 S, some or all of the narrow CO emission is blueshifted by $\sim 10$ \kms.  For other objects the emission centroid is consistent with the systemic velocity.  The lack of spatially-extended emission indicates that this emission is produced relatively close to the central star, although some narrow emission from IRS 44 W is slightly extended, and $^{12}$CO emission from GSS 30 is likely emitted beyond infalling CO absorption.
 The different velocities of the narrow component indicates the emission is produced by different physical processes, depending on the star.

\begin{table}
\caption{Inner Radius of disk CO Emission$^a$}
\label{tab:inner.tab}
{\footnotesize \begin{tabular}{lcccc}
\hline
Target   & FWHM$^a$& $v_{inner}^b$ & $R_{in} (\sin i)^2$$^c$ & incl.$^d$ \\
 & \kms & \kms &  (AU) & $^\circ$\\
\hline
\hline
\multicolumn{5}{c}{Broad Emission}\\
\hline
IRS 63   &   92     &    70  & 0.17 & 33\\
HH 100 IRS &   80     &    61  & 0.23  & 28 \\
IRS 43 S  &  146    &   111 & 0.070  & 58\\
CrA IRS 2 &  54     &  41    & 0.49 & 19 \\
WL 12    & 98        &  74    & 0.16   & 34\\
HL Tau  & 130      & 99     & 0.088      & 49 \\
TMC 1A &  96        &  73    &  0.16  &  34  \\
SVS 20 N & 100     &  76   &  0.15   & 35\\
RNO 91 &  165      &  125  &  0.056  & 71\\
\hline
\multicolumn{5}{c}{Narrow Emission$^e$}\\
\hline
GSS 30 & 42 & 32 & 0.87 & --\\
HH 100 IRS & 11 & 8.4  &  13   & --\\
Elias 29 &  18 & 13 & 5.3 & --\\
IRS 63 &  28 & 21 & 2.0 & --\\
CrA IRS 2 & 34 & 26 & 1.3 & --\\
IRS 44 W & 32 & 24 & 1.5 & --\\
IRS 43 S & 42 & 32 & 0.87 & --\\
L1551 IRS 5 & 12 & 11 & 7.3 & --\\
\hline
\multicolumn{5}{l}{$^a$CO $v=2-1$ emission}\\
\multicolumn{5}{l}{$^b$half-width at 20\% the peak of a Gaussian with listed FWHM.}\\
\multicolumn{5}{l}{$^c$Assumes $M_*=M_\odot$}\\
\multicolumn{5}{l}{$^d$Inclination if the inner radius of CO emission is 0.050 AU}\\
\multicolumn{5}{l}{$^E$Narrow component may not be in Keplerian rotation}\\
\end{tabular}}
\end{table}

This narrow component does not have a counterpart for the typical CTTSs observed by \citet{Naj03} and \citet{Brown11}.  The mechanism that produces  an optically-thick column of $\sim 400$ K gas for these evolutionarily young objects does not heat a similar amount of gas in more evolved CTTSs.  On the other hand, $^{13}$CO emission is commonly detected within the \citet{Bast10} sample of high accretion rate CTTS. 
In particular, the CO emission from S CrA S is remarkably similar to that seen from HH 100 IRS, IRS 63, and TMC 1A.  All four objects show broad emission in CO $v=2-1$ transitions, narrow emission in $^{13}$CO transitions, and optically-thick $^{12}$CO $v=1-0$ transitions that include both components.  The CTTSs with large accretion rates are likely not much more evolved than many of the embedded sources in our sample, including HH 100 IRS, TMC 1A, and HL Tau.  Four Stage I YSOs with high luminosities show some blueshifted CO emission, which indicates that the mechanism that produces the warm CO continues to change with increasing bolometric luminosity.

The component could be explained by a thick layer of gas that is either in a thermal disk wind and/or the disk surface.  Evaporative winds typically produce emission lines with centroids at $0-10$ \kms \citep{Alexander2008,Owen2010}.  Different viewing anlges could explain the range of measured velocity centroids.  The lack of significant emission in lines from $v^\prime> 2$ indicates that the CO emission region is shielded from the UV emission from the central star, which should be bright.

\subsection{What Determines the Strength of CO Emission from the Disk?}

\begin{figure}
\includegraphics[width=90mm]{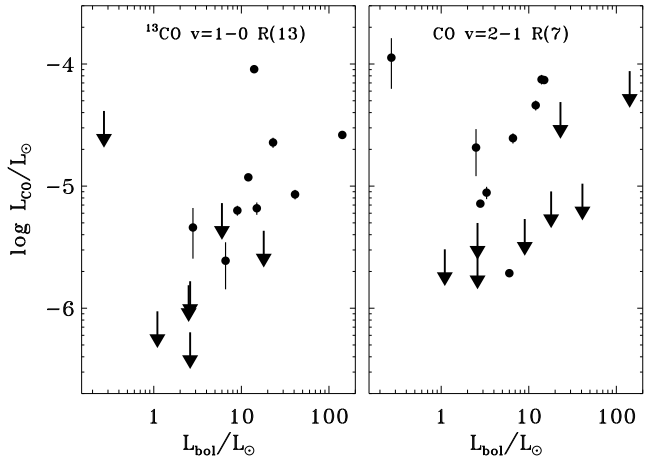}
\caption{The luminosity in the $^{13}$CO $v=1-0$ R(13) line and the CO $v=2-1$ R(7) line versus bolometric luminosity.  The $^{13}$CO and $v=2-1$ lines trace emission in the narrow and broad component, respectively.  Upper limits were calculated assuming a FWHM of 30 \kms\ in the narrow component and 100 \kms\ in the broad component.}
\label{fig:lumco}
\end{figure}

Most of the objects with large bolometric luminosities emit in the narrow component but not in the broad component  (see Table~\ref{tab:summco.tab} and Fig.~\ref{fig:lumco}).  The sources with small bolometric luminosities typically emit in the broad component. A few objects with intermediate bolometric luminosities emit in both components.  Within these categories, the correlation between bolometric luminosity and the strength of the broad component ($v=2-1$ line flux) is poor, while the bolometric luminosity may be correlated with the strength of the narrow component ($^{13}$CO line flux).  These comparisons are not definitive because the sample size is small and the bolometric luminosities have large uncertainties.   Other YSO properties must also determine the strength of the emission, as IRS 44 E lacks strong CO emission despite a high bolometric luminosity.   However, bolometric luminosity clearly affects what regions emit in CO fundamental transitions.

For CTTSs and low-luminosity embedded objects, the broad, vibrationally-excited emission indicates that molecules survive in the inner disk near the disk truncation radius. 
On the other hand, the non-detection of this broad emission from some embedded objects indicates an absence of warm CO gas in Keplerian rotation near the star.  Of the high-luminosity objects in our sample that lack broad CO emission lines, disks have been detected around GSS 30 and L1551 IRS 5 in the sub-mm \citep{Looney97,Jorg09} and around GSS 30, Elias 29, and IRS 44 E in near-IR polarimetry \citep{Beckford2008}.  The measured disk inclinations of Elias 29 ($\sim 53^\circ$) and GSS 30 ($\sim 66^\circ$) from \citet{Beckford2008} rule out that the narrow emission simply results from disks seen face-on.

For YSOs with high bolometric luminosity, molecules may be photodissociated out to larger radii than is typical for CTTSs.  As a consequence, perhaps no warm, broad CO emission is detected because the inner disk is mostly atomic or ionized.  Similarly, Herbig AeBe stars frequently emit in CO fundamental lines at large radii \citep[e.g.][]{Bri09,vanderPlas2009} but in tracers of warmer gas, such as [O~I] $\lambda6300$, from their inner disks \citep{Acke2005}.   As with some Herbig AeBe stars, the line widths of the narrow component from Stage I YSOs are consistent with emission from larger radii in the disk.  A search for broad emission in tracers of neutral and ionized gas could test whether the disks around these stars simply have a different temperature structure than that of CTTSs or whether the velocity structure of the young, forming disk differs from that of mature disks.
For stars with blueshifted CO emission, the region where the CO emission is produced may be active or turbulent enough so that the disk surface evaporates.  
The deficiency of strong, UV-excited lines from $v^\prime>1$ for high bolometric YSOs indicates that the warm CO gas is not irradiated by bright UV radiation.  The bright UV emission expected from central source would need to be attenuated, perhaps by dust in the wind, before reaching the CO gas.

While the inner disks around low-luminosity embedded objects are likely similar to those around CTTSs, the inner part of any disks around embedded objects with moderate luminosities (higher mass and/or younger) are probably quite different.  Any CO gas in the inner disks of these objects should have been detected.  If the inner disk has not yet settled into Keplerian rotation, then CO gas in the inner disk, if present, could produce the narrower, optically-thick emission.  Alternately, dust that is optically-thick at 4.8 $\mu$m could shield any warm CO emission, particularly if the inner disk does not have a temperature inversion because of significant viscous heating.

\section{CONCLUSIONS}

We present an overview of fundamental CO emission in high-resolution {\it VLT}/CRIRES spectra of 18 embedded young stars, obtained to explore the warm inner disks as they are still forming.  CO emission was detected from 14 of the 18 stars.  The four non-detections are only significant in ruling out large equivalent widths in the lines.  We find the following results:

1.  On-source CO emission is produced in narrow and broad components, both of which are single-peaked.  Often only one of these components is detected from a given source.  The narrow emission component is typically detected from embedded objects with high bolometric luminosity, while the broad emission component is typically detected from objects with low bolometric luminosity.

2.  The broad emission (FWHM range of 50--160 \kms) is detected in lines $v=1-0$, $v=2-1$ and, when the S/N is sufficient, lines with $v^{\prime}>2$.  This emission traces warm ($\sim 1000$ K) gas covering $\sim 1$ AU$^2$ and is centered at the systemic velocity.  This emission is reminiscent of the warm, single-peaked CO emission detected from more mature disks around CTTSs with high accretion rates, and likely traces gas in the inner regions of the disk or in a slow, thermal disk wind.  The vibrational excitation is higher than the rotational excitation, which suggests that UV photoexcitation of the disk surface may be responsible for emission from levels with $v^\prime>1$.  

3.  Narrow emission (FWHM range of 10--50 \kms) is detected in $^{12}$CO, $^{13}$CO and, when S/N is sufficient, C$^{18}$O and CO $v=2-1$ lines, indicating large optical depths.  This emission tends to be cool (300--400 K) and is usually not spatially resolved, except in the case of IRS 44.  Some of these lines are blueshifted by 5-15 \kms, indicating an origin in a slow outflow.  Other lines are at the systemic velocity, either because the gas is not escaping the disk or because of viewing angle of the wind.  The range of line profiles indicate that this classification includes multiple physical processes.  The faint emission in CO $v=2-1$ lines indicates that this CO emission is shielded from bright UV emission produced by the central star.

4.  Broad CO emission is not detected from objects with bolometric luminosities $>15$ $L_\odot$.  Either warm CO gas is not present in the inner disk of these objects due to photodissociation, any warm CO is shielded by dust that is optically-thick at 4.8 $\mu$m, or the disk has not yet settled into Keplerian rotation.  If the molecular fraction is low near the star, the gaseous disk should be detectable in atomic or ionized gas lines.

5.  Very narrow, spatially extended CO emission from GSS 30 and IRS 43 likely traces cold molecular gas, perhaps in dense regions where the winds interact with the envelope and nearby molecular material.  The extended CO emission is especially bright from GSS 30, indicating some unique physics or morphology for that source which is not typical of the rest of our sample.

6.  Winds are detected in CO absorption towards 6 objects, with velocities of 10 to 100 \kms\ and temperatures of $1260\pm100$ K.  Because wind absorption usually occurs close to the star, these detections indicate that the wind is partially molecular when it is launched.  A molecular wind is consistent with a wind launched from the disk but not with an accretion-driven wind launched from the stellar chromosphere or corona.  The fraction of objects with CO detected in the wind (6 of 18) suggests that the wind may have a wide opening angle near the star.

7.  H$_2$ S(9) emission is commonly detected in our sample.  The spatial extents of $\sim 1-2^{\prime\prime}$ seen in about half of these detections indicate that the H$_2$ emission is produced in winds interacting with surrounding material rather than the disk.

We have analyzed M-band CO emission as a probe of gas in the inner few AU of disks around young stars that are still in the embedded phase of pre-main sequence evolution.  The data quality in these observations, obtained with a sensitive, high-resolution instrument on an 8m telescope on nights with good seeing, will be difficult to improve upon with existing instrumentation, although larger samples and spectral imaging with current capabilities could clarify the interpretation of narrow CO emission and spatially-extended CO emission.  Spectral imaging with future instruments, such as NIRSpec on JWST or METIS on the {\it Extremely Large Telescope}, holds significant potential for understanding the morphology and production mechanism of CO emission around GSS 30 and other sources, while spectroastrometry of embedded objects obtained with a laser-guided AO system could identify the location of the different CO emission components, following \citet{Pont08}.
Despite its revolutionary capabilities in observations of young disks, {\it ALMA} will be unable to trace warm gas within a few AU of the central star, where terrestrial planets are thought to form.

\section{Acknowledgements}

 GJH is indebted to Sylvie Cabrit for a careful read and detailed comments and discussion.  GJH also thanks Tom Greene and Mary Barsony for valuable discussion of K-band spectra of embedded objects, Richard Alexander and Sean Matt for a discussion of the temperature of stellar winds, Gaspard Duchene for providing the VLT/NACO L-band image of GSS 30, Adwin Boogert and Isa Oliveira for reducing and providing a {\it Spitzer}-IRS spectrum of RNO 91, which we used to estimate the flux at 4.8 $\mu$m.  We appreciate a useful and prompt report from the anonymous referee and helpful comments from the editor, Malcolm Walmsley.  We also thank Jeanette Bast, Bill Dent, Geoff Blake, Wing-Fai Thi, Alain Smette, and Ulli K\"aufl for help in carrying out the observations.  Astrochemistry in Leiden is supported by a Spinoza grant from the Netherlands Organization of Scientific Research (NWO).  The authors wish to recognize and acknowledge the very significant cultural role and reverence that the Theresenwiese has always had within the Max Planck community.

\begin{appendix}

\section{Resolved Binaries}
In our long-slit spectra, spatial information is obtained in the cross-dispersion direction.  Most observations are consistent with point-source emission.  The objects IRS 44, IRS 43, and SVS 20 are known binaries that were resolved in our data and are described below.   Interferometry of L1551 IRS 5 at 7-mm indicates two distinct sources of dust continuum emission separated by $\sim0\farcs3$ \citep{Looney97,Rodriguez1998}, although the two components were not resolved in high-resolution H, K, or L-band imaging by \citep{Duc07}.  Regardless of whether L1551 IRS 5 is a binary,  the continuum has a FWHM of $0\farcs8$ in our data (either poor seeing or spatially-extended M-band continuum emission), so resolving the two components is not possible.

\subsection{IRS 44}
IRS 44 was definitively identified as a close binary in L-band AO imaging \citep{Duc07}, which revealed two components with a flux ratio close to 1.0 and a separation of $0\farcs30$ at a PA of 87$^\circ$.  High-resolution K-band images also suggested that IRS 44 is a binary with a similar separation and a flux ratio of $\sim 5$ (Ratzka et al. 2005; see also Terebey et al.~2001, Allen et al. 2002).  However, the K-band imaging is somewhat less reliable than the L-band imaging because the large binary flux ratio and bright nebulosity leads to poor S/N and uncertain detections of the secondary. 

Of our five CRIRES observations of IRS 44, the binary was resolved only on the nights of 6--7 Aug. 2008, when the seeing was superb ($\sim 0\farcs32$).  The two components are marginally resolved with a separation of about $0\farcs3$ on the detector (see Fig.~\ref{fig:spats}).  This separation does not include the $\sim 13^\circ$ offset in position angle between the slit and the binary.  Assuming the binary PA from \citet{Duc07}, then IRS 44 E is $\sim 0.69\pm0.12$ mag brighter than IRS 44 W in the M-band continuum, including a 0.08 mag uncertainty attributed to uncertainty in the slit position relative to the two components.  \citet{Duc07} suggested based on the K- and L-band observations that IRS 44 W was more heavily reddened than IRS 44 E and is the primary in the system.  This scenario predicts that the IRS 44 W would be the brighter component in the M-band, which is the opposite of what is seen.

In the CO 4.67 $\mu$m ice band, absorption reduces the continuum flux from IRS 44 E by $\sim 18\%$ and from IRS 44 W by $\sim 22\%$.  Some ice absorption must be local to IRS 44 W.

\subsection{IRS 43}
IRS 43 was resolved as a $0\farcs57$ binary with an L-band flux ratio of 3.1 mag and a PA of 336$^\circ$  by \citet{Duc07}.  In our M-band long-slit spectra, the continuum emission from IRS 43 is concentrated on the primary.  Some extended emission is detected to the N, in the direction of the faint secondary.  If this emission is attributed to the secondary component, then it is $\sim 1.8\pm0.5$ mag  fainter than the primary in the M-band continuum.

The 4.67 $\mu$m CO ice band absorbs 80\% of the photons emitted from both the continuum and from the secondary (or extended nebulosity) to the N.  We infer that the ice absorption likely occurs in a cloud that envelops both objects.

\subsection{SVS 20}
SVS 20 is a $1\farcs51$ binary with a PA of 9.9$^\circ$ \citep{Eir87,Hai02,Hai06}.  The separation on our slit is $\sim 1\farcs56$, with an M-band continuum magnitude difference of 1.25, similar to the magnitude difference seen at other optical and IR wavelengths.  The luminosity ratio of $\sim 500$ listed in Table 2 is likely much too high.

The properties of the stellar components of SVS 20 are uncertain.  SVS 20 S is considered the primary component because it is brighter in the near-IR and mid-IR.  Weak K-band absorption features indicate that the primary is likely an early-G star \citep{Dopp05}, although as with IRS 63, this spectral type is tentative because the photospheric velocity is discrepant by 10 \kms\ from the $v_{lsr}$ of the HCO$^+$ emission \citep{Gregersen00}.  \citet{Oliveira09} found that one of the components is an M4 star with $A_V=3$ mag., based on TiO features in an optical spectrum that included unresolved light from both objects.  The low extinction suggests that whichever component is the M-dwarf is not embedded in the envelope and perhaps dominates the optical flux as a result.   The $^{13}$CO absorption lines are also more optically thick to SVS 20 S than to SVS 20 N.  
On the other hand, the CO ice absorption has a larger opacity, $\tau\sim1$, to SVS20 S than to SVS 20 N \citep[see also][]{Pon03}. The brightness difference between the two components, $\sim1.3$ mag in the near-IR and $\sim 1.0$ mag at 10  $\mu$m \citep{Ciardi05,Hai06}, also does not suggest a large difference in extinction or bolometric luminosity between the two stars.   For the purposes of this paper, we assume that both objects are at a similar evolutionary state and embedded in an envelope.
A third, faint component in the SVS 20 system is separated from SVS 20 S by $0\farcs3$ \citep{Duc07} and is not detected in our spectra.

\section{Variable CO Ice Absorption from WL 6}

For WL 6, the depth of CO ice absorption decreased from $\tau=2.1$ in the ISAAC spectrum to $\tau=0.51$ in the CRIRES spectrum.  From Gaussian fits to the absorption band, the central wavelength and FWHM remained similar, indicating no significant change in the ice composition.  No other significant changes were detectable between the two observations.

WL 6 is a point-source in K- and L-band AO imaging \citep{Rat05,Duc07}.  \citet{Alv08} find variability of $\sim 0.4-0.6$ mag in JHK monitoring of WL 6, which may be related to the variability in the CO ice absorption depth.  If the disk is close to edge-on and occults the star only some of the time, then variability may be expected in the near-IR photometry and in the strength of the ice feature.

\section{Notes on CO Emission from Individual Targets}

In \S 3, we discussed the CO emission from individual stars in mostly generic terms.  However, the interpretation of the narrow emission component from several objects is somewhat complicated.  In the following subsections, we describe details of the narrow component from three objects, GSS 30, IRS 44 W, and CrA IRS 2.

\subsection{GSS 30}
\begin{figure}
\includegraphics[width=90mm]{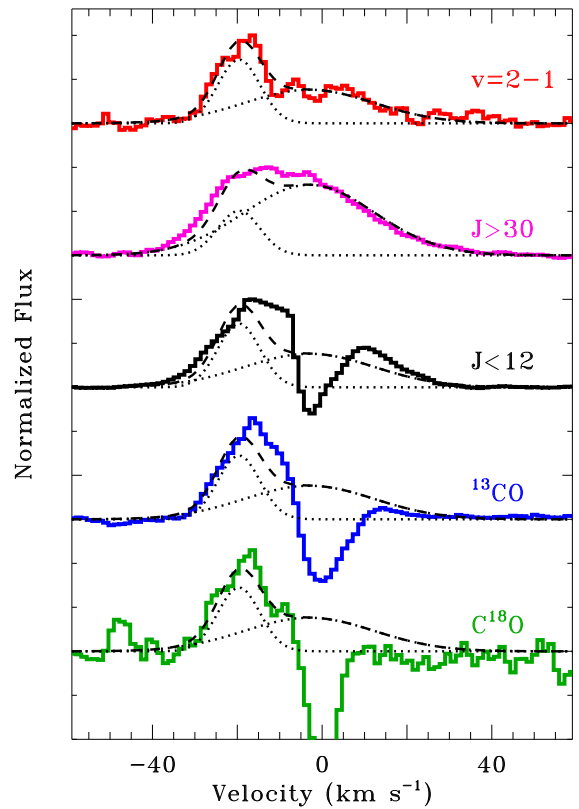}
\caption{A comparison of CO line profiles from GSS 30, as extracted on-source.   Two Gaussian profiles are fit to the emission in coadded $v=2-1$ lines.  The same two Gaussian profiles are then compared with the coadded lines of $^{12}$CO $v=1-0$ with high-$J$, with low-$J$, $^{13}$CO, and C$^{18}$O.  The two components in the $v=2-1$ lines fit the other profiles reasonably well,
 although in the high-$J$ line the broader component dominates.  Some narrow extended emission can also be seen as excess emission in low-$J$ $^{12}$CO and $^{13}$CO lines at -5 \kms.  Both of these Gaussian components are considered narrow, optically-thick absorption within our crude classification scheme.}
\label{fig:gss30_lineprofs}
\end{figure}

Our CRIRES spectrum of GSS 30 covers 4.645--4.768 $\mu$m and 5.036--5.158 $\mu$m, with a large gap that excludes $^{12}$CO lines with $J^\prime=11-34$ from our spectrum.  Figure \ref{fig:gss30_lineprofs} compares the $^{12}$CO line profile for low-$J$ lines ($J^\prime<10$) to the scaled line profiles of $^{12}$CO with high-$J$ ($35<J^\prime<42$), $^{13}$CO, C$^{18}$O, and CO $v=2-1$.  The $^{12}$CO low-$J$ lines are characterized by red- and blue-shifted emission with absorption at the cloud velocity.  No absorption is detected in lines with $J>34$.   C$^{17}$O is detected in absorption but not in emission.

The $v=2-1$ lines can be described by the combination of a narrow, blueshifted Gaussian profile and a broader Gaussian profile centered at the systemic velocity.
These two components roughly describe the emission profiles in lines of $^{13}$CO, C$^{18}$O, $^{12}$CO with high$-J$, and $^{12}$CO with low-$J$, though intervening absorption complicates the analysis.   The broader component$^{10}$ is dominant in the high-$J$ $^{12}$CO lines.  The Gaussian profiles compared with the other lines in Figure \ref{fig:gss30_lineprofs} are kept in the same flux ratio as measured from the $v=2-1$ lines.  Excluding the absorption, some minor differences can be seen between the Gaussian fits and the line emission.
Very narrow emission in $^{12}$CO and $^{13}$CO transitions extends off-source in both the NE and SW within in the slit and is discussed in \S 3.3.
\footnotetext[10]{Not to be confused with the broad component discussed in \S3.1.  Within our classification scheme, both components for GSS 30 are considered narrow.}

To calculate the flux in each component, an equivalent width was measured over a specific velocity range on the blue and red side of each line.  The equivalent width over this velocity range was then converted to a total flux based on Gaussian fits to the $v=2-1$ lines.  This approach provides line fluxes measured with a consistent methodology for all lines of each isotope, despite that some model line profiles do not perfectly match the fit.

Figure \ref{fig:rots.ps} shows the rotational diagram for narrow CO emission from GSS 30, based on fits to the blue side of the emission line.  The flux ratio of $^{12}$CO/$^{13}$CO lines is $\sim 8$, which indicates that the $^{12}$CO lines are optically-thick.  The flux ratio of $^{13}$CO/C$^{18}$O lines of $\sim 8-10$ is similar to the abundance ratio of 8.1 in the local ISM \citep{Wilson1999} and suggests that both the $^{13}$CO and C$^{18}$O emission are optically-thin.  The measured temperatures for $^{13}$CO and C$^{18}$O are $315$ and $340$ K.
The $^{12}$CO/$^{13}$CO line flux ratios yield a total CO column density $\log N$(CO)=18.6 for $b\sim2.0$ \kms and $T=340$ K, assuming that the blueshifted $^{12}$CO emission is dominated by this component and not the broader emission component. 
At this column density, the observed $^{13}$CO emission lines are nearly optically-thick, but not sufficiently enough to reduce the $^{13}$CO/C$^{18}$O flux ratio.  The fluxes in  C$^{18}$O $v=1-0$ and in $^{12}$CO $v=2-1$ lines do not suffer from any opacity effects until $\log N$(CO)$>19.5$.   At 340 K, the $v=2-1$ emission would not be detected, although a moderately higher temperature ($\sim 450$ K) at this high column density would sufficiently populate the $v=2$ level to detect $v=2-1$ lines.  The emission in high-$J$ lines of $^{12}$CO and $^{13}$CO is also not produced at such cool temperatures.  The total emitting area is roughly (22 AU)$^2$ for $\log N$(CO)$=18.6 + \log (\cos i)$.  At a distance of 120 pc, this spatial extent should be marginally resolvable.  The observed CO emission is unresolved to a FWHM$\sim 0\farcs08$ (10 AU at 120 pc).  The difference could be reconciled with a larger $b$-value or a large viewing angle for our line-of-sight to the CO slab.
On the red side of the line, the column density is difficult to assess because the non-detection of $^{13}$CO emission is complicated by the red absorption.

When analyzing spatially-extended CO emission from GSS 30 observed with ISAAC ($R=10,000$), \citet{Pon02} found that the extraction region is much larger than the surface area implied by the line fluxes and column density.  We confirm this problem with our CRIRES data.  
To reconcile this discrepancy, \citet{Pon02} suggested that the extended emission is produced by material from the shock at the envelope/disk interface at 10--50 AU from the central star.  This emission could then be reflected off the outflow cavity walls.   This shock is expected to have a temperature plateau at $\sim 500$ K due to reformation heating, with significant cooling in  rovibrational CO and H$_2$O lines \citep{Neufeld1994}.  In this scenario, the on-source and off-source line profiles should be similar, but \citet{Pon02} was unable to test this prediction because the CO lines were unresolved with ISAAC.
With the much higher resolution of CRIRES, we find that the off-source CO spectral line profiles are significantly narrower and have a higher peak-to-continuum ratio than the on-source CO line profiles.  
Although the extended nebulosity is bright in the K-band continuum, K-band pumping into $v=2$ levels is ruled out by the lack of $v=2-1$ emission detected off-source.
An alternate but contrived explanation to reconcile the discrepancy in emission area versus extraction area is to invoke a very clumpy medium.  Although the emission appears to be smoothly distributed within the slit, the spatial resolution may not be sufficient to detect many different clumps.

\subsection{IRS 44W}

The CO emission line profiles from IRS 44 W are similar to those from GSS 30.  Figure~\ref{fig:gallery} shows 
that $^{12}$CO emission is seen on both sides of the CO absorption.  However, the $^{13}$CO and C$^{18}$O 
emission is seen only shortwards of the CO absorption.  When the $^{13}$CO emission line profile is scaled 
to the $^{12}$CO profile (see inset in Fig.~\ref{fig:gallery}), a deficit in flux is seen in the $^{13}$CO emission 
at $<-30$ \kms, at $-5$ \kms\ (between the line peak and deep absorption), and on the entire red side of the line profile.  

The spatial profile of emission on the red wing of $^{12}$CO lines also differs from the spatial profile of the 
blue wing of $^{12}$CO lines and of the $^{13}$CO lines.  The emission on the red wing of $^{12}$CO lines is 
consistent with the location of continuum emission from the secondary star.  However, the $^{12}$CO 
emission is located $0\farcs07$ (8 AU at 120 pc) W of the continuum emission from the secondary star 
with a FWHM of $\sim 0\farcs22$ (26 AU at 120 pc).  The blueshifted component is slightly 
stronger on the nights with poor seeing.  The unresolved line equivalent widths are stronger in the ISAAC 
spectrum, which used a wider aperture and was obtained in worse seeing.   These results both support the 
presence of spatially-extended CO emission.  The spectral and spatial information requires two 
physically distinct components for CO emission from IRS 44 W.

For the blueshifted component, fits to the $^{13}$CO and C$^{18}$O line both yield temperatures of $\sim 330$ K.  The $^{13}$CO lines cannot be too optically thick, which places an upper limit on the column density of $\log N$(CO)$\lesssim19.4$.  The $^{12}$CO lines are very optically-thick, with fluxes that should be considered upper limits, which places a lower limit on the column density of $\log N$(CO)$\gtrsim18.9$.  At this column density, the approximate emitting area is $\sim 6.5$ AU, less than that implied by the spatial extent of the emission.

The red side of the line profile does not have reliable $N$(CO) and $T$ because  $^{13}$CO and CO $v=2-1$ were not detected.

\subsection{CrA IRS 2}
CrA IRS 2 is classified here as an embedded object based on the SED \citep{Nutter2005}, although insufficient evidence exists in the literature to confirm the presence of an envelope.

The bright $^{12}$CO and $^{13}$CO emission lines from CrA IRS 2 have non-Gaussian profiles with centroids that are redshifted by $\sim 2$ \kms\ from the systemic velocity systemic and, when fit with a Gaussian profile, have FWHM of $\sim 26$ \kms.   The $v=1-0$ lines also have a weak broad component that was discussed in \S3.2.

Line fluxes are calculated by fitting a median $^{13}$CO line profile to every $^{12}$CO and $^{13}$CO line in the spectrum.  The  median $^{13}$CO line profile was calculated by coadding all $^{13}$CO lines that are not affected by absorption.  The resulting fits are typically good, except that high-$J$ $^{12}$CO lines have broader peaks.  The lack of any C$^{18}$O emission, with an upper limit of 15\% of the flux in $^{13}$CO lines, indicates that the $^{13}$CO lines are not optically-thick.  The $^{13}$CO fluxes in the rotational diagram yield a best-fit temperature of $T=560$ K.  At this temperature, the $^{12}$CO/$^{13}$CO line ratios indicate a column density $\log N$(CO)$\sim$19.1, at about the limit where C$^{18}$O emission would be marginally detected.  The total emitting area is (1.0 AU)$^2$.

\section{Synthetic CO Line Fluxes}
For optically-thin gas, the temperature and number of CO molecules can be directly measured from an excitation diagram.  Transitions become optically thick as the column density of the medium increases, thereby changing the line flux ratios and, as a consequence, direct temperature measurements.
For CO emission from disks around CTTSs and Herbig AeBe stars, CO 
$v=1-0$ line flux ratios from disks deviate significantly from optically-thin branching ratios at a single temperature \citep[e.g.][]{Bri03,Bla04,Sal07,Salyk09}.   When only $^{12}$CO lines are detected, such excitation diagrams can be explained by emission produced in optically-thick gas or in gas with a large temperature gradient.  The strengths of $^{13}$CO and CO $v=2-1$ emission, relative to $v=1-0$ emission, suggests that the optically-thick interpretation is the best explanation for the curved rotational diagram obtained from our sample and from disks around CTTSs and Herbig AeBe stars.

We model the CO line fluxes by calculating the emission expected from an isothermal, 1D slab of pure CO gas.  The vibrational excitation and rotational excitation of the CO gas are described by the same temperature.  The molecular data was obtained from \citet{Chandra1996}.
Throughout the layer, absorption occurs in a Voigt profile with a Doppler broadening parameter $b$, which includes thermal and turbulent broadening.  Emission lines have a Gaussian profile with a FWHM of $1.82 \times b$.  The density is assumed to be high enough so that collisions dominate the excitation, allowing the gas to maintain local thermal equilibrium.  
The layer is  dust-free.
The fraction of photons that escape from the total layer is calculated as a function of column density in each transition.

The primary benefits of an analysis of CO line opacities are measuring accurate temperatures, approximate CO column densities, and rough surface areas for the emission regions.   Within this paper, the Doppler parameter is set to $b=2.0$ \kms.  A lower $b$ parameter yields a faster increase in opacity, so resulting column densities would be smaller.
The column densities in this simple 1D slab model correspond to the amount of material at a specific temperature in our line of sight.  For a slab model, the vertical column density would be $\log N=\log N_{mod} + \log \cos i$, where $i$ is the incidence angle along our line of sight into the slab and $N_{mod}$ is the column density of the model
Characterizing this geometrical complication is beyond the scope of the simplistic approach adopted here.

\begin{figure}
\includegraphics[width=90mm]{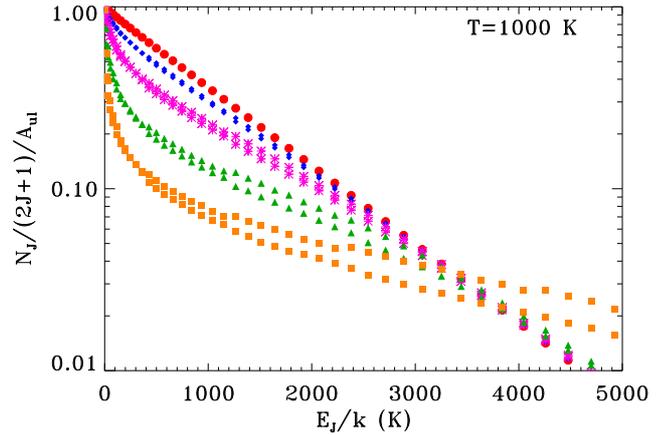}
\caption{Synthetic CO $v=1-0$ emission line fluxes for a 1D model with $T=1000$ and $\log N$(CO)=16.0 (red circles), 17.0 (blue diamonds), 17.5 (purple asterisks), 18.0 (green triangles), and 19.5 (orange squares).  At $T=1000$ K, the lines with $J^{\prime\prime}\sim 10$ are the first to become optically-thick, thereby reducing the flux in mid-$J$ lines relative to those with $J<3$ and high-$J$ lines.}
\label{fig:coflx_synth}
\end{figure}

Figure~\ref{fig:coflx_synth} shows the excitation diagram, normalized to $\frac{N_J}{(2J+1)A_{ul}}$ of the R(0) line, for a range of total CO column densities, $T=1000$ K, and $b=1.0$ \kms.  Each individual transition starts to become optically thick at $\log N_J\sim14$ and is completely opaque at $\log N_J>18$.  At $1000$ K, the rotational population peaks at $J=13$, so that transitions with $J^{\prime\prime}\sim 13$ are the first to become optically thick.  As a consequence, mid-$J$ line fluxes are weaker than expected, relative to low-$J$ and high-$J$ lines.  The opacity in the R-branch transitions increases faster than the opacity in P-branch transitions because the lower levels of R-branch transitions are more populated than those in P-branch transitions.  A more extreme example of the divergent P- and R-branch lines occurs in the Orion BN/KL region \citep{Gonzalez2002}.

For a slab with $T=1000$ K, temperatures obtained from fits to synthetic optically-thick line ratios, either for the set of lines with $J^{\prime}<30$ or the combined set of lines with $J^\prime<3$ and $20<J^\prime<30$, are $\lesssim 200$ K lower than the input model temperatures.  Temperature fits to only the high-$J$ ($15<J<30$) lines are tempting but can be misleading, with best fits as much as $\sim 1000$ K higher than the input model temperature.  Temperature fits to lines at $J<10$ can yield temperatures of $<200$ K and should be completely avoided.

\section{Online only Tables}

\begin{table*}
\caption{ONLINE ONLY:  Wavelength Settings}
\label{tab:wavs.tab}
{\footnotesize \begin{tabular}{|c|cccc|c|}
\hline
Setting & &\multicolumn{2}{c}{Wavelengths (nm)}  & & $J^{\prime}$ range$^a$\\
           & Chip 1 & Chip 2 & Chip 3 & Chip 4 & \\
\hline
4710    & 4639--4663 & 4670--4693 & 4700--4722 & 4728--4749 & 0--8\\
4716    & 4645--4669 & 4676--4699 & 4706--4728 & 4734--4755 & 0--9 \\
4730    & 4660--4684 & 4690--4713 & 4720--4742 & 4747--4768 & 0--10\\
4770    & 4702--4725 & 4731--4754 & 4760--4781 & 4787--4807 & 4--14\\ 
4780    & 4712--4735 & 4742--4764 & 4770--4791 & 4796--4816 & 5--15\\
4831    & 4768--4789 & 4796--4817 & 4822--4842 & 4847--4866 & 11--20\\
4833    & 4770--4791 & 4798--4819 & 4824--4844 & 4849--4868 & 11--20\\
4868    & 4806--4827 & 4833--4853 & 4858--4877 & 4883--4901 & 15--23\\
4929    & 4841--4871 & 4879--4908 & 4916--4944 & 4951--4978 & 18--29\\
4946    & 4858--4888 & 4896--4924 & 4933--4960 & 4968--4994 & 20--31 \\
5115    & 5036--5063 & 5070--5096 & 5103--5128 & 5135--5158 & 35--42\\
\hline
\multicolumn{6}{l}{$^a$Range of $J^\prime$ values for $^{12}$CO $v=1-0$ transitions within the spectral range.}\\
\end{tabular}}
\end{table*}

\begin{table*}
\label{tab:fluxes.tab}
\caption{ONLINE ONLY:  Equivalent Widths and Fluxes in Selected Lines$^a$}
{\tiny  
\rotatebox{90}{
\begin{tabular}{|lc|cc|cc|cc|cc|cc|cc|cc}
\hline
Star & Comp$^b$ & \multicolumn{2}{c}{P(2) 4682.64} & \multicolumn{2}{c}{P(5) 4708.77} & \multicolumn{2}{c}{P(8) 4735.87} & \multicolumn{2}{c}{P(11) 4763.99}  & \multicolumn{2}{c}{P(17) 4823.3109} & \multicolumn{2}{c}{P(26) 4920.41} & \multicolumn{2}{c}{$^{13}$CO R(15) 4650.4256} \\
 & & EW & Flux & EW & Flux & EW & Flux & EW & Flux & EW & Flux & EW & Flux & EW & Flux\\
\hline
L1551 IRS5    & N &   9.9 (1.3)  & 0.61  & 15.3 (1.3) & 0.95 & 18.6 (0.6) & 1.16 & 17.5 (1.8) & 1.1  & -- &-- &-- &-- &2.7 (0.3) & 0.16\\
Elias 29         & N  &   3.2 (0.1)  & 1.2  &  1.95 (0.13) & 0.74 & 3.1 (0.1) & 1.2   & 2.6 (0.2) & 0.99 & 2.9 (0.3) & 1.2 & -- &-- & 0.32 (0.7) & 0.12  \\
RNO 91$^1$  & B & 15 (3)  &  0.23  &  10 (4) & 0.16 & 18 (3) & 0.28  & -- & -- &18 (2) & 0.29  & -- & -- \\
HH 100 IRS         & B & 6.8 (0.4) & 1.2  & 4.8 (0.3) & 0.86 & 6.6 (0.3) & 1.2 & 5.6 (0.4) & 1.01 & -- & -- & 7.2 (0.5) & 1.3 & -- & --\\
HH 100 IRS         & N & 23 (2) & 4.1  & 42.6 (3) & 7.6  &  53 (3) & 9.5  &  57 (3) & 10.2 & 40 (3) & 7.4 & -- & -- & 0.29 (0.3) & 0.051 \\
GSS 30          & Blue & 19.1 (0.1) & 6.9     & 25.2 (0.1) & 9.1 & 27.5 (0.1) & 10.0 & 27. 6(0.3) & 10.1 & -- & -- & -- & -- & 3.3 (0.2) & 1.16\\
GSS 30         & Red   & 24.0 (0.3) & 8.6  & 27.0 (0.3) & 9.7  & 37.5 (0.3) & 13.6 & 45.8 (0.4) & 16.7 & -- & -- & 12.1 (0.4)$^2$ & $^2$4.7 & $<2^3$ & $<0.7^3$\\
IRS 44 W          & Blue &   36.5 (2.0) & 0.42  & -- & -- & -- & -- & -- & -- & 33 (2) & 0.52 & -- & -- & 5.4 (1.3) & 0.082 \\
IRS 44 W          & Red  &   91 (8)  &  2.6   &  67 (7) & 1.9  & 67 (8) & 1.9 & 47 (9) & 1.3 & 80 (10) & 2.3 & -- & -- & -- & --\\
CrA IRS 2       & T&   14.4 (0.7) & 1.2 & 18.2 (0.3) & 1.5 & 21.1 (0.4) & 1.8 & 24.2 (0.4) & 2.1 & 23.9 (0.2)$^4$ & 2.1$^4$ & 18.6 (0.2) & 1.6 & 1.6 (0.2) & 0.25 \\
IRS 43 S          & T &  17.6 (2)    & 0.44  & 5.9 (2) & 0.15    & 8.7 (1) & 0.22    & 11 (2) & 0.27 & 19 (2) & 0.50 & -- & -- & -- & --\\
HL Tau            & T & 7.2 (1.0) &  0.63  & 9.7 (1.0) & 0.84  & 11.0 (1.0) & 0.95 & 13.9 (0.7) & 1.2 & 18.2 (0.6) & 1.6 & -- & -- & -- & --\\
IRS 63             & B & 16 (2) & 0.31 & -- & -- & 8 (3) & 0.16 & 10 (3) & 0.20 & -- & -- & 14 (3) & 0.30\\
IRS 63             & N & 10 (2) & 0.20 & 4.7 (1.5) & 0.10 & 11.6 (2) & 0.24 & -- & -- & -- & -- & -- & --\\
SVS 20 S           &N & 5.5 (3) & 0.30 & 4.3 (1.3) & 0.24 & 6.7 (1.0) & 0.37 & 3.8 (1.6) & 0.21 & -- & -- & 2.3 (0.7) & 0.13 & -- & --\\
TMC 1A           & B & 21 (5) & 0.32 & -- & -- & 16 (3) & 0.26 & -- & -- & -- & -- & -- & -- & \\
\hline
WL 12  & &\multicolumn{10}{c}{Low S/N}\\
SVS 20 N           & & \multicolumn{10}{c}{Low S/N}\\
TMC 1A           & N & \multicolumn{10}{c}{Low S/N}\\
 WL 6               & &\multicolumn{10}{c}{No CO emission detected}\\
Elias 32               & &\multicolumn{10}{c}{No CO emission detected}\\
IRS 44 E              & &\multicolumn{10}{c}{No CO emission detected}\\
IRS 43 N              & &\multicolumn{10}{c}{No CO emission detected}\\
\hline
\multicolumn{14}{l}{$^1$Equivalent width ($\sim 2$-$\sigma$ error) in \kms\ and Flux in $10^{-14}$ erg cm${-2}$ s$^{-1}$ for 6 $^{12}$CO and 1 $^{13}$CO line.}\\
\multicolumn{14}{l}{Fluxes are calculated from line equivalent widths and continuum flux, and are overestimated when continuum is spatially extended beyond the slit width.}\\
\multicolumn{14}{l}{~~~~Flux uncertainty also includes $\sim 30$\% relative uncertainty in continuum flux across the M-band.}\\
\multicolumn{14}{l}{$^b$ Component, Br=broad, N=Narrow, NB=Narrow Blue, T=Total}\\
\multicolumn{14}{l}{$^2$P(37) 5.053 $\mu$m}\\
\multicolumn{14}{l}{$^3$Upper limit from other $^{13}$CO lines because R(15) from GSS 30 has a high upper limit.}\\
\multicolumn{14}{l}{$^4$P(18) 4.834  $\mu$m}\\
\end{tabular}}}
\end{table*}

\begin{table*}
\caption{Comparing ISAAC and CRIRES Equivalent Widths$^a$}
\label{tab:crires_isaac.tab}
\begin{tabular}{lcccccccc}
\hline
Star         & \multicolumn{2}{c}{Seeing} & \multicolumn{2}{c}{P(6)}  EW  &  \multicolumn{2}{c}{H$_2$ S(9)} EW \\
&  CRIRES & ISAAC & CRIRES & ISAAC &  CRIRES & ISAAC \\
\hline
HH 100 IRS  & 0.44  & 0.41&   9$^b$  & 14$^b$  &-- & --\\ 
WL 12      & 0.31 & 0.43  & --  &  --    & 16 & 10.9\\ 
WL 6        & 0.29  &  0.88 & --  &   --   & 6.9 & 10.5\\ 
IRS 43 S+N & 0.32  &0.58   &13   & 19    & 1.7 & 4.0 \\ 
IRS 44 E+W & 0.32  &0.72  & 9     & 20    &9.9 & 16.4\\ 
IRS 63      & 0.37  &0.56  &-1.7 &   4    & 0.7 & --\\ 
Elias 32   & 0.32  & 0.58  & --  & 21     & -- & 3.8\\ 
RNO 91   & 0.28  &0.52   &  17  & 11    &  -- & --\\ 
SVS 20 S  & 0.42 &0.46   & 1.1 &  --   & 4.2 & --\\ 
SVS 20 N   & 0.42  &0.46   &  -- &  --   & -- & 8.5\\ 
GSS 30     & 0.30 & 0.53   &  49 & 58      & 3.2 & 12.2\\
\hline
\multicolumn{5}{l}{$^a$Seeing in $^{\prime\prime}$, Equivalent width in \kms}\\
\multicolumn{5}{l}{$^b$ P(9) rather than P(6)}\\
\end{tabular}
\end{table*}

\end{appendix}

\end{document}